\documentclass[5p,times,number,sort&compress,preprint]{elsarticle}
\journal{Engineering Applications of Artificial Intelligence}

\usepackage{textcomp}

\usepackage{amsmath,amsfonts}
\usepackage{amsthm}
\usepackage{amsmath}

\usepackage{amstext}
\usepackage{amssymb}
\usepackage{mathtools}
\usepackage{bm}  %
\usepackage{siunitx}
\usepackage{physics}

\DeclareMathOperator*{\argmin}{arg\,min}

\newtheorem{theorem}{Theorem}
\newtheorem{lemma}{Lemma}
\newtheorem{remark}{Remark}

\usepackage{array}
\usepackage{color, soul} %
\usepackage{ulem} %

\usepackage{graphicx}
\graphicspath{%
}
\usepackage{tikz}
\usepackage[list=true,labelformat=simple]{subcaption}
\usepackage{stfloats}

\usepackage{url}
\usepackage{verbatim}
\usepackage{cite}

\usepackage{xcolor}

\usepackage[inline]{enumitem}
\usepackage{times}
\usepackage{algorithm,algpseudocode,float}
\usepackage{setspace} %

\algnewcommand{\algorithmicand}{\textbf{ and }}
\algnewcommand{\algorithmicor}{\textbf{ or }}
\algnewcommand{\AlgAnd}{\algorithmicand}
\algnewcommand{\AlgOr}{\algorithmicor}

\newsavebox{\myparbox}
\newlength{\myparboxwidth}

\providecommand{\norm}[1]{\lVert#1\rVert}

\newcommand{\defeq}{{\, \stackrel{\rm def}{=} \,}}

\newcommand\Angle[1]{\setbox0=\hbox{$\mskip 7mu minus 4mu#1$}%
  \raise.21ex\hbox{$/$}\hskip-0.95ex\underline{\raise\dp0\hbox{\box0}}}

\ExplSyntaxOn

\NewDocumentCommand \vect { s o m }
 {
  \IfBooleanTF {#1}
   { \vectaux*{#3} }
   { \IfValueTF {#2} { \vectaux[#2]{#3} } { \vectaux{#3} } }
 }

\DeclarePairedDelimiterX \vectaux [1] {\lbrack} {\rbrack}
 { \, \dbacc_vect:n { #1 } \, }

\cs_new_protected:Npn \dbacc_vect:n #1
 {
  \seq_set_split:Nnn \l_tmpa_seq { , } { #1 }
  \seq_use:Nn \l_tmpa_seq { \enspace }
 }
\ExplSyntaxOff

\usepackage[firstpageonly=true]{draftwatermark}

\usepackage{hyperref}
\hypersetup{
    colorlinks=true,
    linkcolor=blue,
    filecolor=magenta,
    urlcolor=cyan,
    bookmarks=true,
}
\usepackage[noabbrev]{cleveref}  %
\Crefname{figure}{Fig.}{Figs.}

\begin{document}

\begin{frontmatter}
\title{Real-Time Measurement-Driven Reinforcement Learning Control Approach for Uncertain Nonlinear Systems \tnoteref{t1}}
\tnotetext[t1]{This work was partially supported by NSERC Grant~EGP~537568-2018.}

\author[1]{Mohamed Abouheaf}
\ead{mabouhe@bgsu.edu}

\author[2]{Derek Boase}
\ead{dboas065@uottawa.ca}

\author[2]{Wail Gueaieb\corref{cor1}}
\ead{wgueaieb@uottawa.ca}

\author[3]{Davide Spinello}
\ead{dspinell@uottawa.ca}

\author[4]{Salah Al-Sharhan}
\ead{salah27@ieee.org}

\cortext[cor1]{Corresponding author}

\affiliation[1]{%
  organization={College of Technology, Architecture and Applied Engineering},%
  addressline={Bowling Green State University},%
  city={Bowling Green, OH},%
  postcode={43403-0001},%
  country={USA}%
}

\affiliation[2]{%
  organization={School of Electrical Engineering and Computer Science, University of Ottawa},%
  addressline={800 King Edward Avenue},%
  city={Ottawa, ON},%
  postcode={K1N~6N5},%
  country={Canada},%
}

\affiliation[3]{%
  organization={Department of Mechanical Engineering},%
  addressline={161 Louis Pasteur},%
  city={Ottawa, ON},%
  postcode={K1N~6N5},%
  country={Canada},%
}

\affiliation[4]{%
  organization={Machine Intelligence Research Labs},%
  addressline={},%
  city={Auburn, WA},%
  postcode={98071-2259},%
  country={USA},%
}

\begin{abstract}

The paper introduces an interactive machine learning mechanism to process the measurements of an uncertain, nonlinear dynamic process and hence advise an actuation strategy in real-time. For concept demonstration, a trajectory-following optimization problem of a Kinova robotic arm is solved using an integral reinforcement learning approach with guaranteed stability for slowly varying dynamics. The solution is implemented using a model-free value iteration process to solve the integral temporal difference equations of the problem. The performance of the proposed technique is benchmarked against that of another model-free high-order approach and is validated for dynamic payload and disturbances. Unlike its benchmark, the proposed adaptive strategy is capable of handling extreme process variations. This is experimentally demonstrated by introducing static and time-varying payloads close to the rated maximum payload capacity of the manipulator arm. The comparison algorithm exhibited up to a seven-fold percent overshoot compared to the proposed integral reinforcement learning solution. The robustness of the algorithm is further validated by disturbing the real-time adapted strategy gains with a white noise of a standard deviation as high as 5\%.
\end{abstract}

\end{frontmatter}

\DraftwatermarkOptions{%
angle=0,
hpos=0.5\paperwidth,
vpos=0.97\paperheight,
fontsize=0.012\paperwidth,
color={[gray]{0.2}},
text={
  \parbox{0.99\textwidth}{%
    \copyright\ 2023. This manuscript version is made available under the CC-BY-NC-ND 4.0 license\\ \href{https://creativecommons.org/licenses/by-nc-nd/4.0/}{https://creativecommons.org/licenses/by-nc-nd/4.0/}\\
    This is the accepted version of the manuscript. Published manuscript DOI: \href{https://doi.org/10.1016/j.engappai.2023.106029}{10.1016/j.engappai.2023.106029}%
  }},
}

\section{Introduction}
\label{sec:Intro}
Measurement-driven solutions based on adaptive learning concepts are challenged by many factors, such as the need to incorporate the dynamics of the process explicitly into the underlying strategies~\citep{aastrom2013adaptive,Vech22}. Many adaptive approaches  have been designed offline and lack the ability to capture high-order model-following dynamics~\citep{Xueyun23,MOD20,Com22,FWA22}. Hence, many adaptive learning approaches employ either complex or computationally expensive algorithms~\citep{TRA21,slid21,Com22,FWA22}. This gets more challenging when adaptive mechanisms are adopted for coupled regulation and optimization missions, where the dimensions of the state and action spaces grow significantly~\citep{Sutton_1998,MA21,Gra21}.

Machine Learning approaches have been employed in many instrumentation applications, such as quality monitoring of laser cladding~\citep{Kao2020}, vortex flow-meter design~\citep{Thummar2021}, measurement of residual Oxygen concentration~\citep{Jianjun2020}, vision-based measurement systems~\citep{Fedullo2021}, state-of-charge prediction~\citep{Yichun2021}, localization of faults and network status detection~\citep{Mohammed2021}, Parkinson’s disease diagnostics~\citep{Talitckii2021}, machine health monitoring~\citep{Chaoyi2021}, real-time aging prediction of integrated circuits~\citep{Huang2019}, vision systems calibration in welding robots~\citep{Zou2020}, and sleep apnea analysis~\citep{Bahrami2022}.
Other machine learning forms have adopted Recurrent Neural Network (RNN) and Long Short-Term Memory (LSTM) networks to solve various optimization problems~\citep{Sherstinsky2020}. A maneuvering system is developed for Unmanned Aerial Vehicles (UAVs) using dynamic inversion and LSTM network approaches~\citep{Cao2022}.
Another adaptive cruise mechanism adopted a transfer learning idea that is based on LSTM networks~\citep{Zhou2022}. The Hierarchical Temporal Memory (HTM) and LSTM network approaches have been adopted to predict short-term arterial traffic flow~\citep{Mackenzie2019}. A deep neural network that employs a feedback control concept is adopted to solve an intelligent structural control problem~\citep{Rahmani2019}. It makes use of a state-selector function to avoid forgetting key states by the neural-networks and hence improve the overall performance.
An LSTM-based deep learning approach has been used to predict the vapor mass quantity in the adsorption bed~\citep{Skrobek2020}. Other machine learning mechanisms based on LSTM, Bidirectional Long Short-Term Memory (BiLSTM), and Gated Recurrent Unit (GRU) have been employed to improve the efficiency of adsorption cooling systems~\citep{Skrobek2022}.

Reinforcement learning (RL) is a class of Machine Learning that offers a structured approach to learn the best strategy-to-follow~\citep{sut92,Zolfpour14,Vladimir22}. It leverages feedback from an agent's interactions with its environment to either reward or penalize the agent's actions through the utility of a value function~\citep{AlMahamid22}. The goal of the agent is to maximize a cumulative sum of the rewards~\citep{Bertsekas1996}. This class of adaptive systems uses two-step techniques known as policy iteration (PI) and value iteration (VI)~\citep{Liu22,Wasala20,Bertsekas1996,Busoniu2010}. For nonlinear applications, Integral Reinforcement Learning (IRL) approaches are adopted to solve optimal control problems~\citep{Abouheaf2019}. The means of adaptive critics are employed to implement the RL solutions using two neural network structures, namely the actor and critic networks~\citep{Bertsekas1996}. The adaptive critics approximate the strategy-to-follow using an actor neural network; while the value of applying a certain strategy is approximated by a critic neural network. These approaches have been used to solve cooperative control problems for multi-agent systems communicating over graphs~\citep{AbouheafAuto14,AbouheafCTT2015,AbouheafICRA19}. An adaptive Fuzzy-RL mechanism is adopted to control flocking motion of a swarm of robots in~\citep{ICRA21}. Regression models such as iterative and batch least squares are employed to implement the PI solutions~\citep{Busoniu2010,Srivastava2019}.
The adaptive approaches are adopted to control underactuated vehicles and distributed generation sources~\citep{AbouheafTrans20,AbouheafIRL2019}.

Linear Quadratic Tracking Regulators (LQT) provide offline control strategies that solve the optimal tracking control problems. This requires a knowledge of the system dynamics, {where the optimal control gains} are then applied to the forward evolution of the state~\citep{Lewis2012}. This problem is ubiquitous in modern control applications, namely in intelligent control systems~\citep{aastrom2013adaptive}. In order to develop robust adaptive control solutions, it is often desirable to develop a model for the plant, or at least an approximate dynamic model. Although this approach has certain benefits, modeling the dynamics of a system can require assumptions that may narrow the applicability of the model and introduce uncertainty about the dynamic system parameters. Applying model approximations techniques, such as linearization for instance, can lead to a loss of generality. In cases where a dynamic model is available, certain model-reference adaptive control approaches may be considered. These involve backstepping, sliding mode control, and Lyapunov methods, for example~\citep{Byrne1995,Moore2014,Vempaty2016,BenAmor2017,Zuo2017,SHI2017,hu2010distributed,Chen2021}. Given the dependency of such methods on the dynamics the process, the control strategies inherit such limitations. This can be seen in~\citep{Vempaty2016}, for example, where the lateral motion of a 5-DoF trailer system is stabilized with a model reference adaptive system (MRAS) using Lyapunov theory~\citep{Vempaty2016}.

The above mentioned challenges are tackled in this work using an integral adaptive learning approach.
Herein, another form of MRAS is proposed for the online control of unknown nonlinear systems. It is then validated using a 6-DoF Kinova robotic arm. The control gains are adapted to reflect the variations in the dynamics of the robotic application. The adaptive learning algorithm actuates the joints following interactive reference-trajectories. It employs a data-driven scheme to determine the variations in the control strategies needed to move the end-effector between the desired positions. The controller relies on an online IRL mechanism with guaranteed stability for slowly varying dynamics.
The work contributes an online measurement-driven adaptive learning mechanism that (i)~adopts incremental learning capabilities to improve the control strategy in real-time, (ii)~avoids incorporating the process and the reference-model dynamics explicitly into the underlying control strategy, (iii)~provides a flexible feedback mechanism in terms of the order of the model-following dynamics, and (iv)~allows for online approximate solutions for a class of optimal tracking problems.
It builds on the contributions of~\citep{AbouheafIRL2021} to develop an online IRL control mechanism for nonlinear model-following systems with uncertain dynamics. The work presented herein supports the theoretical findings of~\citep{AbouheafIRL2021} with solid data-processing and practical evidence. A data-driven approach is developed for the real-time control of a 6-DoF Kinova robotic arm, as a highly nonlinear dynamic system. It adopts incremental learning features to adapt the strategies without estimating the dynamics of the robotic arm or explicitly expressing the strategies in terms of the reference-trajectory dynamics. In addition, the control structure is modified to accommodate incremental learning capabilities associated with the control strategy (i.e.,~provide data-driven features to the underlying strategies allowing for online adaptations). The value of using the temporal difference equation in rejecting high-frequency noise is also explored and analyzed in details. Finally, the proposed solution is benchmarked against another model-free approach for dynamic payload and disturbances.

The rest of the paper is organized as follows: The objectives of the adaptive learning problem are highlighted in \Cref{sec:arm_env}. \Cref{sec:Prob} introduces a description of the trajectory-tracking optimization problem. \Cref{sec:Opt} lays out the mathematical foundation of the temporal difference solution, including a stability analysis of that solution. The IRL solution and its implementation using an adaptive critics technique are introduced in \Cref{sec:IRL}. The practical experimentation setup and results are discussed in \Cref{sec:Exp_Set,sec:Anl}, respectively. Finally, main concluding remarks are highlighted in \Cref{sec:Con}.

\section{Instrumentation Setup of the Kinova Robotic Arm}\label{sec:arm_env}
The data-driven approach presented herein is applicable to a large-class of nonlinear systems. It is applied to solve the model-following problem of a Kinova JACO manipulator, where each joint is controlled independently of the other.
Note that in addition to the nonlinear nature of the dynamics of each joint, it is also time-dependent due to the simultaneous motion of the joints, which adds to the complexity of the control problem.
The inherited features in the IRL solution enable the direct use of joint-measurements without estimating the dynamics of the robotic arm. Furthermore, the temporal difference structure of the online IRL solution (i.e., integral temporal difference or Bellman equation) enables robust filtering characteristics for the high-frequency content of the processed signals. This will be highlighted later when presenting the results in \Cref{sec:Anl}.

\begin{figure}[!ht]
    \centering
    \includegraphics[scale=0.11]{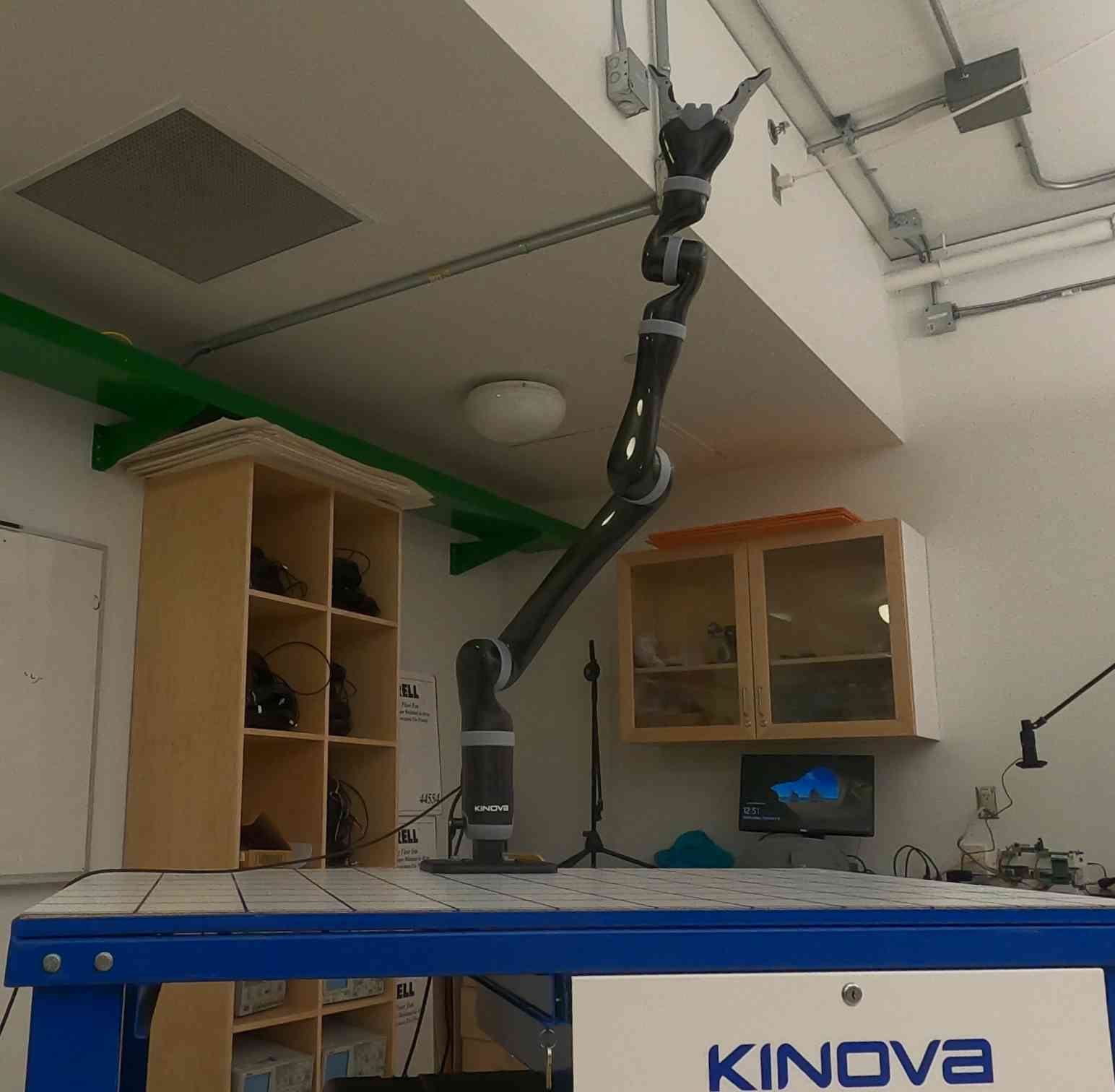}
    \caption{Kinova JACO manipulator}
    \label{fig:kinova_arm}
\end{figure}

The Kinova JACO manipulator is a 6-DoF robot equipped with a two-finger gripper, as shown in \Cref{fig:kinova_arm}. The actuators are interfaced through  commercial RS-485 communication protocol providing a data rate of \SI{12}{Mbps} and a high-level control frequency of \SI{100}{\Hz}. Their low-level control loops operate at \SI{500}{\MHz}. Each servometer is equipped with a position sensor with a resolution of ${3,686,400}/{\text{turn}}$. Herein, the robot is controlled through the Robot Operating System (ROS).

Each control signal {$u_i(t)\in\mathbb{R}$} actuates the motion of joint \textit{i} (i.e., the position angle $\theta_i(t)$), and the angular velocity is calculated by $\dot{\theta}_i(t)= {\left(\theta_i(t) - \theta_i(t-\nu)\right)}/{\nu}$, where $t$ is a time-index, $\nu$ is a sampling-time, and $i \in \{1, 2, 3, 4\}$ represents the joint that is being controlled. Hereafter, index \textit{i} is omitted for simplicity, since the algorithm is the same for each joint.

\section{Problem Formulation}
\label{sec:Prob}

This section introduces the mathematical foundation of the  trajectory-tracking optimization problem. The goal is to manipulate the joints of the robotic arm simultaneously to follow a desired trajectory in real-time. The learning scheme decides on the actuation signals of the joints online through adapted strategies to regulate the trajectory tracking-errors between the desired and measured angular positions ${\theta}^d(t)$ and ${\theta}(t)$, respectively. Hence, the dynamic model of each joint is given by
\begin{align}
  \label{eq:Sys}
 {\dot \theta} & = f\left({\theta}(t),{u}(t) \right).
\end{align}
It is assumed here that the drift dynamics of the joint are embedded in unknown function $f$. The adapted strategy does not require to incorporate such dynamics explicitly in its structure. Herein, the trajectory-tracking problem is formulated as an optimization problem, where the main objective is to select an actuation signal in real-time (i.e., a control strategy that is decided based on a machine learning process) to regulate the trajectory-tracking errors ${\varepsilon}(t)={\theta}^d(t)-{\theta}(t)$  (i.e.,~ideally, the aim is to achieve $\lim_{t \to \infty} \norm{ {\varepsilon}(t) } = 0$). The control signal due to an attempted strategy $\pi$ is given by  ${u}^{\pi}(t)={u}^{\pi}(t-\nu)+{\eta}^{\pi}(t)$, where the correction in the actuation signal is represented by ${\eta}^{\pi}(t)$ such that ${\eta}^{\pi}(t)=\omega^{\pi}_0\,\varepsilon(t)+\omega^{\pi}_\nu\,\varepsilon(t-\nu)+\omega^{\pi}_{2\nu}\,\varepsilon(t-2\nu)$. The notation $\nu$ refers to the desired sampling interval needed to collect the real-time measurements. Further, the tuple $\{\omega^{\pi}_0,\omega^{\pi}_\nu,\omega^{\pi}_{2\nu}\}$ defines the control gains of a policy $\pi$, which will be determined using the online RL solution. The structure of the correction signal ${\eta}^{\pi}(t)$ could vary to reflect high-order error dynamics. This is done by increasing the number of employed error samples (i.e., $\varepsilon(t-o\,\nu), o=0,1,\dots, \mathcal{L}$). Herein, a second-order tracking error system is considered (i.e., $o=2$), which proved to be sufficient to achieve the optimization goals, as shall be demonstrated later. The signal ${\eta}^{\pi}(t)$ decides on the relative angular deviations to apply without any information on the dynamics of the robotic arm.
The choice of $o=2$ means that, this approach can represent as double as the error dynamics of a model-following problem.

\begin{figure}[htb]
  \centering
  \includegraphics[width=1\linewidth]{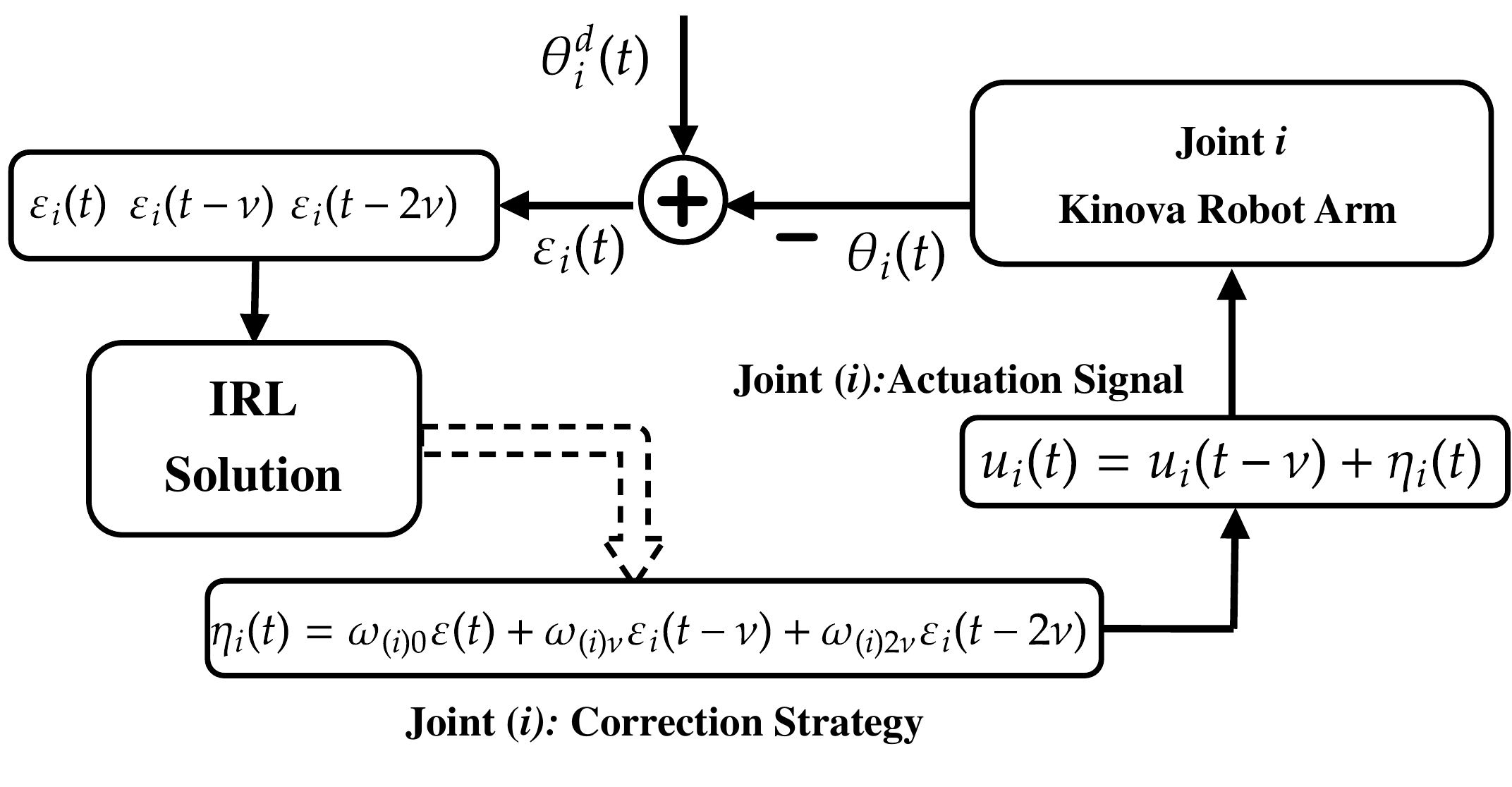}
  \caption{Trajectory-tracking system}
  \label{fig:control}
\end{figure}

The trajectory-tracking problems are mostly solved using adaptive learning approaches, since it is difficult to implement the optimal tracking solutions for complex high-order error dynamical systems~\citep{aastrom2013adaptive}. Typically, optimal tracking control problems are solved offline based on the knowledge of the system dynamics. Further, such solutions often start by solving a subset of coupled differential equations offline before solving the remaining subset online~\citep{Lewis2012}. The dynamics of the reference-model must be explicitly embedded in the structure of such solutions, which makes their complexities dependent on the number of controlled joints. Moreover, the solutions do not ensure robustness against varying dynamics of the robotic arm. All of this urges for an innovative model-free adaptive solution that can be realized in real-time. Herein, the proposed IRL scheme is divided into two parts. First, an optimal tracking control setup is considered to develop a temporal difference structure that can be solved online. Then, an adaptive learning scheme based on IRL is considered to solve the trajectory-tracking optimization problem. The solution is realized by sampling and processing the tracking errors of each joint, as illustrated in \Cref{fig:control}.

\section{Optimal Trajectory Tracking}
\label{sec:Opt}
This section introduces the mathematical foundation of the online adaptive learning solution. It relies on an optimal control setup rather than employing an explicit optimal tracking control one. This is done to relax the challenges associated with solving the optimal tracking control problems~\citep{Lewis2012}. However, this strategy needs a proper process to ensure that it does not require information on the robot dynamics. The solution is realized in a measurement-driven manner, where a moving finite storage of tracking-error samples is considered. As highlighted earlier, the number of error samples defines the order of the error dynamics. This structure exhibits an advantage compared to low-order error dynamical systems employed in classical adaptive systems~\citep{Lewis2012,Chen2018}. The goal is to select an optimal strategy ${\pi*}$ which yields an optimal correction signal ${\eta^{\pi*}}$ for each joint independently, to follow a desired trajectory. The error samples are stored in a vector
$\bm{X}(t) = \vect{ \varepsilon(t) , \varepsilon(t-\nu) , \varepsilon(t-2\nu) }^T \in \mathbb{R}^3$, which signifies the state vector in an optimal control setup. A convex performance index $P$, that is inspired by the linear quadratic regulator structure, is employed such that $P^{\pi}(t)=\int_{t}^{\infty} {\cal{U}}(\bm{X}(\xi),{\eta}^\pi(\xi)) \, d \xi,
$ where the cost function ${\cal{U}}(\dots)$ incorporates $\bm{X}$ and ${\eta}^\pi$ such that
${\cal{U}}(\bm{X}(\xi),{\eta}^\pi(\xi))=\frac{1}{2}\left(\bm{X}^T(\xi) \,\bm{Q}\, \bm{X}(\xi)+{{\eta}^{\pi}}^T(\xi) \,{R} \, {\eta}^\pi(\xi)\right)$. The positive definite matrices $\bm{0} < \bm{Q}\in\mathbb{R}^{3\times3}$ and $0 < {R}\in\mathbb{R}$ are weighting structures. In this particular case, $R$ is a real scalar. 

The performance index $P$ evaluates a strategy $\eta^\pi(t)$ over a finite interval. The following developments explain
\begin{enumerate*}[label=(\roman*), before=\unskip{: }, itemjoin={{; }}, itemjoin*={{; and }}]
\item the setup of the implicit optimal control solution
\item the conditions needed to develop a candidate solution
\item the temporal difference structure employed by the adaptive learning solution
\item the model-free structure of the optimal control strategy.
\end{enumerate*}
These developments combine findings from the adaptive and optimal control theories.

\begin{theorem}
  \label{thm:CTHJB}
  Let a kernel solution function $S \left( \bm{X}(t),{\eta}^\pi(t) \right)$ be non-negative and $S \left( \bm{0}\right)=0$. Thus,  
  \begin{enumerate}[leftmargin=*, widest*=9]
  \item $S^*\left(\bm{X}(t),{\eta}^\pi(t)\right)$ represents an optimal solution for the Hamilton-Jacobi-Bellman equation (HJB) equation \newline$H( \bm{X}(t), \allowbreak \grad{S}^*\left(\bm{X}(t),{\eta}^{\pi*}(t)\right), \allowbreak {\eta}^{\pi*}(t) )=0$.
  \item $S \left(\bm{X}(t),{\eta}^\pi(t)\right)$ represents a Lyapunov function.
  \end{enumerate}	
\end{theorem}

\begin{proof}%
\begin{enumerate}[leftmargin=*, widest*=9]
\item The Hamiltonian for the trajectory-tracking optimization problem decides the optimal policy along the trajectory of the error dynamics $\dot{\bm{V}}^\pi(t)=0$ such that
\begin{equation*}
H(\bm{X}(t),\bm{\mu}(t), {\eta}^\pi(t))=\bm{\mu}^T(t) \, \dot{\bm{V}}^\pi(t)+{\cal{U}}\left(\bm{X}(t),{\eta}^\pi(t)\right),
\end{equation*}
where $\bm{V}^\pi(t)=\vect{ \bm{X}^T(t) , {{\eta}^\pi(t)}^T }^T$ and $\bm{\mu}$ denotes a Lagrange multiplier associated with the constraints $\dot{\bm{V}}^\pi(t)$.

The kernel solution structure $S(\dots)$ is selected to be convex in vector $\bm{V}^\pi(t)$ such that
\begin{equation}
  S\left(\bm{X}(t),{\eta}^\pi(t)\right)
  =\frac{1}{2}
  {\bm{V}^\pi}^T(t)\,
  \bm{\mathcal{H}}\,
  \bm{V}^\pi(t),
  \label{val}
\end{equation}
where
$\displaystyle  \bm{0}<
\bm{\mathcal{H}}
\equiv
\begin{bmatrix*}[l]
  \bm{\mathcal{H}}_{\bm{X}\bm{X}}
  & 
  \bm{\mathcal{H}}_{\bm{X}{\eta}}
  \\
  \bm{\mathcal{H}}_{{\eta}\bm{X}}
  &
  \bm{\mathcal{H}}_{{\eta}{\eta}}
\end{bmatrix*}
\in \mathbb{R}^{4\times4} ,
$
$\boldsymbol{H}_{{\eta}\bm{X}} \in \mathbb{R}^{1 \times 3}$ and $\boldsymbol{H}_{{\eta}{\eta}} \in \mathbb{R}$.

\par The relation between the kernel solution and $\bm{\mu}$ is explained by the Hamilton-Jacobi (HJ) theory~\citep{Lewis2012}. Herein, the Lagrange multiplier is found to be $\bm{\mu} \equiv\grad{S} \left(\bm{X}(t),{\eta}^\pi(t)\right)={\partial S\left(\bm{X}(t),{\eta}^\pi(t)\right)}/{\partial \bm{V}^\pi(t)}$. Substituting $\grad{S} \left(\bm{X}(t),{\eta}^\pi(t)\right)$ into the Hamiltonian $\mathcal{H} (\dots)$, yields a Bellman equation given by
\begin{equation}
  \grad{S} \left(\bm{X}(t),{\eta}^\pi(t)\right)^T \, \dot{\bm{V}}^\pi(t)+{\cal{U}}\left(\bm{X}(t),{\eta}^\pi(t)\right)=0. 
  \label{CTBell}
\end{equation}
It can be noted that this expression is an infinitesimal representation of $P^{\pi}(t)\defeq S\left(\bm{X}(t),{\eta}^\pi(t)\right)=\int_{t}^{\infty} {\cal{U}}(\bm{X}(\xi),{\eta}^\pi(\xi)) \, d \xi$. Therefore, \eqref{CTBell} can be restructured such that
\begin{multline}
  H(\bm{X}(t),\grad{S}\left(\bm{X}(t),{\eta}^{\pi}(t)\right), {\eta}^{\pi}(t))= \\
  \dot S\left(\bm{X}(t),{\eta}^\pi(t)\right)+{\cal{U}}\left(\bm{X}(t),{\eta}^\pi(t)\right). 
  \label{CTBelldt}
\end{multline}
Solving $H(\bm{X}(t), \allowbreak \grad{S}\left(\bm{X}(t),{\eta}^{\pi}(t)\right), {\eta}^{\pi}(t))=0,$ while applying the optimal strategy yields the HJB equation. The optimal signal ${\eta^{\pi*}}$ is derived by applying Bellman's optimality conditions such that ${\eta^{\pi*}(t)}=\argmin_{\eta^\pi(t)}H(\bm{X}(t),\grad{S}\left(\bm{X}(t), {\eta}^{\pi}(t)\right), \allowbreak {\eta}^{\pi}(t))$. The optimal solution $S^{*}$ is found by solving the HJB equation given by
\begin{multline}
  H(\bm{X}(t),\grad{S}^*\left(\bm{X}(t),{\eta}^{\pi*}(t)\right), {\eta}^{\pi*}(t))= \\
  \dot S^{*}\left(\bm{X}(t),{\eta}^{\pi*}(t)\right)+{\cal{U}}\left(\bm{X}(t),{\eta}^{\pi*}(t)\right)=0. 
  \label{CTHJB}
\end{multline}
\item Since the kernel solution function $S$ is quadratic and convex, then it represents a Lyapunov candidate function. Using~\eqref{CTBelldt} and taking the time-derivative of the kernel solution $S$ yield $\dot S\left(\bm{X}(t),{\eta}^\pi(t)\right)=-{\cal{U}}\left(\bm{X}(t),{\eta}^\pi(t)\right) \le 0$.
Therefore, $S\left(\bm{X}(t),{\eta}^\pi(t)\right)$ fulfills the conditions of a Lyapunov function.
\end{enumerate}
\end{proof}

This mathematical setup solves the optimal tracking control problem through finding the optimal kernel solution $S^*$. However, in order to realize the solution in real-time, a temporal difference structure and an explicit model-free form of the optimal strategy are required. Hence, \Cref{thm:IRL-Bellman} builds on the results of \Cref{thm:CTHJB} to develop a temporal difference structure that can be employed by the reinforcement learning solution.

\begin{theorem}
  \label{thm:IRL-Bellman}
  The kernel solution $S^*\left(\bm{X}(t),{\eta}^{\pi*}(t)\right)$ satisfies an integral Bellman optimality equation that is given by
  \begin{multline}
    \label{IRLOpt}
    S^*\left(\bm{X}(t),{\eta}^{\pi*}(t)\right)=\int_{t}^{t+\nu}{\cal{U}}\left(\bm{X}(\xi),{\eta}^{\pi*}(\xi)\right)\,d\xi
    \\
    +
    S^*\left(\bm{X}(t+\nu),{\eta}^{\pi*}(t+\nu)\right),
  \end{multline}
  where the optimal policy $\pi*\defeq\{\omega^*_0,\omega^*_\nu,\omega^*_{2\nu}\}$ satisfies the stationarity condition of the optimization problem.
\end{theorem}

\begin{proof}%
Applying Euler approach on the Bellman equation~\eqref{CTBell}, yields a temporal difference structure given by
\begin{equation*}
  \frac{S\left(\bm{X}(t),{\eta}^{\pi}(t)\right)-S\left(\bm{X}(t+\nu),{\eta}^{\pi}(t+\nu)\right)}{\nu}
  ={\cal{U}}\left(\bm{X}(t),{\eta}^{\pi}(t)\right).
\end{equation*}
Equivalently, this could be written as 
\begin{multline} S\left(\bm{X}(t),{\eta}^{\pi}(t)\right)=\int_{t}^{t+\nu}{\cal{U}}\left(\bm{X}(\xi),{\eta}^{\pi}(\xi)\right)\,d\xi
  \\
  +
  S\left(\bm{X}(t+\nu),{\eta}^{\pi}(t+\nu)\right).
  \label{IRLBell}
\end{multline}
This represents a temporal difference form, where Bellman's optimality conditions can be applied to find the optimal strategy. Hence,  ${\eta}^{\pi*}(t)=\argmin_{{\eta}^{\pi}(t)} \allowbreak S\left(\bm{X}(t),{\eta}^{\pi}(t)\right)$. This and~\eqref{val} yield
\begin{equation}
{\eta}^{\pi*}(t)=- {\cal{\bm H}}_{{\eta}{\eta}}^{-1}  {\cal{\bm H}}_{{\eta}\bm{X}} \, \bm{X}(t).
\label{optpol}
\end{equation}
This optimal policy structure ${\pi*}=- {\cal{\bm H}}_{{\eta}{\eta}}^{-1}  {\cal{\bm H}}_{{\eta}\bm{X}},$ is model-free and relies only on the kernel solution matrix ${\cal{\bm H}}$. Therefore, this matrix is adapted in real-time using the tracking error measurements. Employing this optimal strategy into~\eqref{IRLBell}, yields the Integral Bellman optimality equation~\eqref{IRLOpt}. The optimal correction signal is given by $\eta^{\pi*}(t)=\bm{\omega}^*\bm{X}(t)$ where the optimal control gains follow $\bm{\omega}^*= \vect{ \omega^*_0 , \omega^*_\nu , \omega^*_{2\nu} } = - {\cal{\bm H}}_{{\eta}{\eta}}^{-1}  {\cal{\bm H}}_{{\eta}\bm{X}}$
\end{proof}

The next result discusses the stability of the trajectory-tracking error system following the solution of the Integral Bellman optimality equation \eqref{IRLOpt} using the optimal strategy~\eqref{optpol}.

\begin{lemma}
  \label{thm:stability}
  Let the value of the initial kernel solution ${S}$ be bounded such that  ${S} \left(\bm{X}(0),{\eta}^\pi(0)\right)\le\cal{T}$. Given a bounded independent command trajectory ${\theta}^d(t)$, the trajectory tracking error dynamical system is asymptotically stable with $\lim_{t\rightarrow  \infty} \varepsilon(t) = 0$ and $\lim_{t\rightarrow \infty} \dot S\left(\bm{X}(t),{\eta}^\pi(t)\right)=0$.
\end{lemma} 

\begin{proof}%
According to \Cref{thm:CTHJB}, the solution  
$S\left(\bm{X}(t),{\eta}^\pi(t)\right)$ is shown to be a Lyapunov function. Then, the inequality given by $S\left(\bm{X}(t),{\eta}^\pi(t)\right)\le S\left(\bm{X}(0),{\eta}^\pi(0)\right)\le\mathcal{T}, \forall t$ holds. This yields, $S\left(\bm{X}(t),{\eta}^\pi(t)\right)\in L_\infty$. So, $\varepsilon(t),\varepsilon(t-\nu),\varepsilon(t-2\nu) \in L_\infty$ and $\bm{\mathcal{H}}$ (implicitly signifies a vector of control gains $\bm{\omega}^*$) $\in L_{\infty}$.
Since $S$ is proved to be a Lyapunov function and the integral Bellman equation can be formulated as $\int_{0}^{t}\dot S \left(\bm{X}(\vartheta),{\eta}^\pi(\vartheta)\right) \, d\vartheta=S \left(\bm{X}(t),{\eta}^\pi(t)\right)-S \left(\bm{X}(0),{\eta}^\pi(0)\right)$. Then, $-\int_{0}^{t}\dot S\left(\bm{X}(\vartheta),{\eta}^\pi(\vartheta)\right) \, d\vartheta \le S\left(\bm{X}(0),{\eta}^\pi(0)\right)$.
This shows that $\dot S\left(\bm{X}(t),{\eta}^\pi(t)\right) \in L_\infty$, and consequently implies that $\dot \varepsilon(t), \, \dot \varepsilon(t-\nu), \, \dot \varepsilon(t-2\nu) \in L_\infty$. Using~\eqref{CTHJB} and~\eqref{optpol}, a conclusion can be made such that $\int_{0}^{t} \bm{X}^T(\vartheta) \,  \left(\bm{Q}+\bm{\omega}^{*T}{R}\bm{\omega}^*\right)\, \allowbreak \bm{X}(\vartheta) \, d\vartheta \le S \left(\bm{X}(0),{\eta}^\pi(0)\right)$, which reveals that $\varepsilon(t), \, \varepsilon(t-\nu), \, \varepsilon(t-2\nu) \in L_2$ and $\dot S \left(\bm{X}(t),{\eta}^\pi(t)\right) \in L_2$. Applying Barbalat's Lemma~\citep{aastrom2013adaptive}, yields $\lim_{t\rightarrow \infty} \dot S\left(\bm{X}(t),{\eta}^\pi(t)\right)=0$. Thus, the tracking error dynamic system is asymptotically stable and $\lim_{t\rightarrow \infty} \varepsilon(t)=0$.
\end{proof}

\begin{remark}
Adaptive control solutions with time-delays adopt direct and indirect parameter estimation schemes in real-time control strategies. In the model-reference adaptive solution proposed here, the algorithm is based on multi-step time sampling of the tracking error, where the number of steps is left as a user-defined parameter. The time shifting induced by the multi-step sampling naturally incorporates a time-delay into the underlying strategy without requiring additional parameter estimation approaches.
\end{remark}

The above results provide a temporal difference solution framework that uses a model-free strategy to solve an optimal trajectory tracking problem in real-time. It relies on kernel structure~\eqref{val} and employs a model-free strategy~\eqref{optpol} to solve the integral Bellman optimality equation~\eqref{IRLOpt}. This optimal solution cannot be realized analytically. Hence, a reinforcement learning solution scheme is considered next to realize the solution.
\section{Reinforcement Learning Adaptive Solution}
\label{sec:IRL}
The Bellman optimality equation~\eqref{IRLOpt} and the optimal model-free control strategy~\eqref{optpol} will be used to develop an adaptive learning solution adopting the heuristic form of IRL to control the joints of the Kinova obotic arm, simultaneously in real-time. A value iteration mechanism will be considered to realize the online IRL solution. Then, an adaptive critics structure is employed to approximate the IRL solution in real-time using a gradient-descent technique.

\subsection{Online Value Iteration Solution}
A simplified value iteration procedure for each joint of the robotic arm is described as follows:
\begin{enumerate}
	\item Initialize the kernel solution matrix ${\cal{\bm H}}^0$, control signal $u(0)$, correction signal $\eta(0)$, and error vector $\bm{X}(0)$.
	\item Start an iterative process $r=0,1,2,\dots,N$, where $r$ refers to a sequence of adapted or updated policies.
	\begin{enumerate}[leftmargin=*, widest*=9]
	\item Solve for the new kernel solution ${\cal{\bm H}}^{r+1}$ such that
	\begin{multline}	S^{r+1}\left(\bm{X}(t),{\eta}^{r}(t)\right)=\int_{t}^{t+\nu}{\cal{U}}\left(\bm{X}(\xi),{\eta}^{r}(\xi)\right)\,d\xi
		\\
		+
		S^r\left(\bm{X}(t+\nu),{\eta}^{r}(t+\nu)\right).
		\label{Val_crt}
	\end{multline}
	\item Improve the correction strategy as
	\begin{equation}
		{\eta}^{r+1}(t)=- \left[{\cal{\bm H}}_{{\eta}{\eta}}^{-1}  {\cal{\bm H}}_{{\eta}\bm{X}}\right]^{r+1} \, \bm{X}(t).
		\label{Val_act}
^{}	\end{equation}
	\end{enumerate}
\item Upon convergence of $\norm{{\cal{\bm H}}^r}$
 terminate the adaptation.
\end{enumerate}

This online IRL solution is based on a value iteration mechanism that solves the temporal difference equation~\eqref{Val_crt} and updates the control strategy \eqref{Val_act} in the above mentioned designated order.
\Cref{thm:convergence} below verifies the convergence of the adapted kernel solution following this iterative procedure.

\begin{theorem}
  \label{thm:convergence}
  Let the value iteration solution update the kernel matrix ${\cal{\bm H}}^r$ and the optimal strategy ${\cal{\bm \omega}}^r$ using~\eqref{Val_crt}~and~\eqref{Val_act}, respectively. Then,
  \begin{enumerate}[leftmargin=*, widest*=9]
  \item The value iteration process yields a sequence $0 \le S^0 \le S^1 \le S^2 \le \dots \le S^*$ that converges to the optimal solution of~\eqref{IRLOpt}.
  \item The strategies advised by \eqref{Val_act} are stabilizing ones.
  \end{enumerate}
\end{theorem}
        
\begin{proof}%
\begin{enumerate}[leftmargin=*, widest*=9]
\item The initial kernel solution is bounded such that $0 < S^0\left(\bm{X}(\bm{0}),{\eta}^\pi(\bm{0})\right)\le\mathcal{T}$. Thus, according to \eqref{Val_crt}, the equality
$
        S^{r+1}\left(\bm{X}(t),{\eta}^{\pi}(t)\right)
        =
        \sum_{i=0}^{r}S^{1}\left(\bm{X}(t+i\,\nu),{\eta}^{\pi}(t+i\, \nu)\right)
        -\sum_{i=1}^{r} \allowbreak S^{0}\left(\bm{X}(t+i\,\nu),{\eta}^{\pi}(t+i\, \nu)\right)
$ holds. This leads to a non-decreasing sequence (i.e.,~$0 \le S^0\le S^1\le \dots \le S^r\le S^{r+1},\,\forall r$).
The trajectory-tracking error dynamical system is shown to be asymptotically  stable. Therefore, the cumulative cost is bounded (i.e., $0<\int_{0}^{\infty} {\cal{U}}\left(\bm{X}(\xi),{\eta}^{\pi}(\xi)\right) d\xi \le \bar{\mathcal{T}}$). This leads to a converging sequence such that $0 \le S^0\le S^1\le \dots \le S^r\le S^{r+1} \le \mathcal{T+\bar T},\,\forall r$. Accordingly, the value iteration procedure results in a sequence $ 0 \le S^0\le S^1\le \dots \le S^*$, where $S^*$ is the solution of Bellman optimlaity equation~\eqref{IRLOpt}.
\item The integral Bellman optimality equation $S^{r}\left(\bm{X}(t),{\eta}^{r}(t)\right)-S^{r}\left(\bm{X}(t+\,\nu),{\eta}^{r}(t+\, \nu)\right)=\int_{t}^{t+\nu} {\cal{U}}\left(\bm{X}(\xi),{\eta}^{r}(\xi)\right) d\xi
$ employs the optimal strategies ${\eta}^r$, $\forall r, \nu$ given by \eqref{Val_act}. This and the stability results yield
\begin{align*}
	\small
& \lim_{\xi\rightarrow \infty}S^{r}\left(\bm{X}(\xi),{\eta}^{r}(\xi)\right)=0
\le
\dots
\le
S^{r}\left(\bm{X}(t+2\,\nu),{\eta}^{r}(t+2\, \nu)\right)
\le
\\&S^{r}\left(\bm{X}(t+\,\nu),{\eta}^{r}(t+\, \nu)\right)
\le
S^{r}\left(\bm{X}(t),{\eta}^{r}(t)\right).
\end{align*}
Therefore, the strategies $\mu^r,\forall r$ are stabilizing and the resulting sequence of updated strategies can be written as
\begin{align*}
	\small
&S^{0}\left(\bm{X}(t),{\eta}^{0}(t)\right)
\le
S^{1}\left(\bm{X}(t+\,\nu),{\eta}^{1}(t+\, \nu)\right)
\le\\
&S^{2}\left(\bm{X}(t+2\nu),{\eta}^{2}(t+2\nu)\right)
\le 
\dots 
\le
S^{r}\left(\bm{X}(t+r\nu),{\eta}^{r}(t+r\nu)\right).
\end{align*}
\end{enumerate}
\end{proof}

\begin{remark}
Generally, Lyapunov solutions for model-following problems require the knowledge of the process dynamics and of the desired trajectory dynamics. IRL is adopted in optimization problems written in a temporal difference form, as for example induced by the integral Bellman equation in a discrete-time form. Temporal difference forms with function approximation are the core of families of model-free solution methods for optimization problems. Herein, the simultaneous solution of~\eqref{Val_crt} and~\eqref{Val_act} provides a novel algorithmic approach to the optimal tracking problem based on the computational form of the IRL.
\end{remark}

\subsection{Adaptive Critics Implementation}

Neural networks are adopted to implement the value iteration solution. This is done using critic and actor neural networks to approximate the kernel function and associated optimal strategy. These structures are adapted in real-time using data measured along the trajectory of the robotic arm system. The kernel solution value function is approximated such that
\begin{equation*}
\hat S\left(\bm{X}(t),\hat{{\eta}}(t)\right)
=\frac{1}{2}
{\bm{V}}^T(t)\,
\bm{\Omega}_c\,
\bm{V}(t),
\end{equation*}
where $\bm{\Omega}_c$ represents the critic weights, ${\hat{{\eta}}(t)}$ is the approximation of correction control signal, and $\bm{V}^\pi(t)= \vect{ \bm{X}^T(t) , \hat{{\eta}}(t) }^T$. 
The critic weights $\bm{\Omega}_c > \bm{0}$ approximate the kernel solution matrix $\bm{\mathcal{H}} > \bm{0}$ which is adapted by the value iteration process. This structure  is inspired by that of the kernel matrix $\bm{\mathcal{H}}$.

The structure of the actor neural network follows the form of the optimal strategy~\eqref{Val_act} such that
\begin{equation*}
\hat{{\eta}}(t)=\bm{\Omega}_a \, \bm{X}(t),
\end{equation*}
where the actor neural network weights $\bm{\Omega}_a$ approximate the optimal control gains $\omega^*= \vect{ \omega^*_0 , \omega^*_\nu , \omega^*_{2\nu} }$.

Herein, a gradient-descent approach is employed to adapt the actor and critic weights. Thus, the adaptation error structures of the critic and actor are inspired by the value iteration procedure given by~\eqref{Val_crt} and~\eqref{Val_act}, respectively. The tuning error associated with adapting the critic weights is given by $E_{Critic}= \frac{1}{2}\left(\hat S\left(\bm{X}(t),\hat{{\eta}}(t)\right)-\tilde S(t)\right)^2,$ where $\tilde S(t)$ is defined by $\tilde S(t)={\cal{U}}\left(\bm{X}(t),\hat{{\eta}}(t)\right)+\hat S\left(\bm{X}(t+\nu),\hat{{\eta}}(t+\nu)\right)$. Therefore, the critic tuning law follows
\begin{equation}
\bm{\Omega}^{(r+1)}_c=\bm{\Omega}^{(r)}_c-\alpha_c \, E_{Critic}^{(r)} \bm{V}^\pi \,{\bm{V}^\pi}^T,
\label{Crt_Law}
\end{equation}
where $0<\alpha_c<1$ is an adaptation rate for the critic weights.
Similarly, the tuning error associated with adapting the actor weights is formulated as $E_{Actor}= \frac{1}{2}\left(\hat{{\eta}}-\tilde u\right)^2$,
where $\tilde u= - \boldsymbol{\Omega}_{c\hat{{\eta}}\hat{{\eta}}}^{-1} \, \boldsymbol{\Omega}_{c\hat{{\eta}}\bm{X}}$. Thus, the resulting actor adaption law is given by
 \begin{equation}
 \bm{\Omega}^{(r+1)}_a=\bm{\Omega}^{(r)}_a-\alpha_a \, E_{Actor}^{(r)}  \,{\bm{X}}^T,
 \label{Act_Law}
\end{equation}
where $0<\alpha_a<1$ is the learning rate of the actor weights.
The detailed steps of the online adaptive critics implementation of the IRL solution are listed in~\Cref{alg:alg1}. 

\begin{remark}
Herein, the problem is formulated as deterministic, consistently with the literature on reinforcement learning solutions of optimal control problem with model-free setups, where optimal control policies are learned by relying on measurements along a specific system's trajectory, eliminating the necessity of the system's drift dynamics/model which may be unknown or highly uncertain \citep{Lewis2012}. Additive noise is used as disturbance in various scenarios below to test and illustrate the robustness of the adaptive learning solutions.
\end{remark}

\begin{algorithm}[htb!]
  \setstretch{1} %
  \caption{Online Adaptive Critics Solution}\label{alg:alg1}
  \begin{algorithmic}[1] %
    \Require
    \Statex Total number of adaption steps $N$
    \Statex Sampling interval ${\nu}$
    \Statex Adaptation rates $\alpha_a$ and $\alpha_c$
    \Statex Weighting matrices $\boldsymbol{Q}$ and ${R}$
    \Statex Convergence threshold $\sigma$ and a time-window of length $L$ to check convergence
    \Ensure
    \Statex Tuned actor and critic weights (i.e., $\bm{\Omega}^{(t+\ell\nu)}_a$ and $\bm{\Omega}^{(t+\ell\nu)}_c,\quad \ell=0,1,\dots, {N}$)
    \Statex
     \State Define the desired trajectory $\theta^{d}(t), \forall t$
    \State Initialize the angular position $\theta(0)$, control signal $u(0)$, actor weights $\bm{\Omega}^{(0)}_a$, and critic weights $\bm{\Omega}^{(0)}_c$ %
    \State Calculate the error vector $\bm{X}(0)$

    \State $\ell \gets 0$ 
    \State Convergence\_Condition $\gets$ False
    \While {$\ell <N$ \AlgAnd Convergence\_Condition $=$ False}
    \State Compute the correction control signal $\hat{{\eta}}(\ell\nu)$
    \State Adjust the control signal $u(\ell\nu)+\hat{{\eta}}(\ell\nu)$ and then actuate each joint
    \State Observe $\theta((\ell+1)\, \nu)$ 
    \State Use $\theta^{d}((\ell+1)\, \nu)$ to find $\varepsilon(t+(\ell+1)\,\nu)$
    \State Calculate ${\cal{U}}(\bm{X}(\ell\,\nu),\hat{{\eta}}(\ell\nu))$ and $S(\bm{X}(\ell\,\nu),\hat{{\eta}}(\ell\nu))$ 
    \State Find $\bm{X}((\ell+1)\,\nu)$ and $\hat{{\eta}}((\ell+1)\,\nu)$. Hence, get $S(\bm{X}((\ell+1)\,\nu,\hat{{\eta}}((\ell+1)\nu))$
    \State Find the target values of the critic and actor neural network approximators $\tilde S(\ell\,\nu)$ and $\tilde{{\eta}}(\ell\nu)$
    \State Adapt the critic weights  $\Theta_c^{(\ell+1)}$\Comment{use tuning law~\eqref{Crt_Law}}
    \State Adapt the actor weights $\Theta_a^{(\ell+1)}$\Comment{use tuning law~\eqref{Act_Law}}
    \If{$\ell > {L}$\AlgAnd $\norm{\bm{\Omega}_c^{ (\ell+1-l)}-\bm{\Omega}_c^{(\ell-l)}} \le \sigma,$ $\forall l \in \{0,1,\ldots,{L}\}$,}
    \State $\bm{\Omega}_c^{(*)} \gets \bm{\Omega}_c^{(\ell+1)}$ 
    \State Convergence\_Condition $\gets$ True
    \EndIf
    \State $\ell \leftarrow \ell+1$
    \EndWhile
    \Statex \Return $\bm{\Omega}^{(t+\ell\nu)}_a$ and $\bm{\Omega}^{(t+\ell\nu)}_c$, for $\ell=0,1,\dots, {N}$
  \end{algorithmic}
\end{algorithm}

\section{Experimental Setup}
\label{sec:Exp_Set}
Five experiments are carried out to evaluate the controller's performance under various conditions.
In the first four experiments, the control scheme is applied to simultaneously track the nominal trajectories of the four joints (base, shoulder, elbow, and wrist) under four different conditions. In these experiments, the initial joint positions are taken as $\bm{\theta}(0)= \vect{0, 90, 180, -135, 180, 0}^T$ (degrees).
Note that the last two joints of the robot are not controlled.
The first experiment implements the control algorithm in an ideal case where there is no payload and no dynamic disturbances. The response to this experiment is used as a baseline for the comparison with the other responses, along with the nominal trajectories. To test the algorithm under a different payload, the second experiment is conducted while a gripper, mounted at the robot's end-effector, holds a constant payload of 3~lb, which represents approximately $91\%$ of the arm's total payload capacity. The purpose of the third experiment is to assess the controller's performance in the face of sudden payload changes. It is conducted by starting with no payload and then abruptly adding a 2.5~lb  sandbag to the gripper around the halfway mark of the experiment. To evaluate the disturbance rejection capacity of the proposed solution, a fourth experiment is carried out where the actor weights are synthetically disturbed by additive noise drawn from the normally distributed random variable $M\sim N(0, \sigma)$ during the beginning of the experiment. This experiment is repeated while varying the duration and standard deviation of the weight disturbances. The values used are tabulated in \Cref{tab:disturbance_params}. Such a scenario, demonstrates the robustness of the algorithm to variations in initial weights and shows a strong ability to adapt the weights as needed. Furthermore, the experiment mimics measurement noise that may be imposed on or passed to the actuation signals. It is important to note that this synthetic disturbance is superimposed to the actual measurement noise intrinsic to the instrumentation of the manipulator's arm.
The fifth experiment tests the controller's ability to compensate for a time-varying payload by controlling a single-joint (the elbow) while the gripper gradually lifts a sandbag off the top of a table so that the payload carried by the robot gradually increases to the full-weight of the sandbag (2.5~lb). The initial joint positions in this time-varying payload experiment are set to $\bm{\theta}(0)= \vect{90, 143, -45, -135, 180, 0}^T$ (degrees).
A brief summary of the five experiments is provided in \Cref{tab:exp_summary}. Experiment~4 is the only one that was not conducted on the real robot. It is rather run on ROS/Gazebo simulation platform to protect the robot from any erratic behaviors that may arise due to the injected noise.

	\begin{table}[!ht]
		\caption{Disturbance experiment parameters}
		\centering
		\begin{tabular}{c|c|c|c|c}
			Duration (\%)	&
      20 & 40 & 60 & 100 \\
			\hline
      Variance, $\sigma^{2}$ &
			0.025 &  0.025 & 0.020 & 0.0125
		\end{tabular}
		\label{tab:disturbance_params}
	\end{table}

\begin{table}[!ht]
  \caption{Summary of the experiments}
  \centering
  \small
  \begin{tabular}{c|p{2.6in}}
    \hline
    Experiment	&	Description\\
    \hline\hline
    1		&	No payload, no disturbance\\
    2		&	Constant payload of 3~lb throughout the experiment\\
    3	        &	Abrupt payload of 2.5~lb halfway throughout the experiment\\
    4 		&	Random distrurbance of actor weights at the beginning of the experiment\\
    5		& 	Time-varying payload\\
    \hline
  \end{tabular} 
  \label{tab:exp_summary}
\end{table}

\begin{figure}[!ht]
  \centering
  \subcaptionbox{Joint~1%
  \label{fig:nominal_trajectories:j1}}%
{%
  \includegraphics[width=0.45\columnwidth]{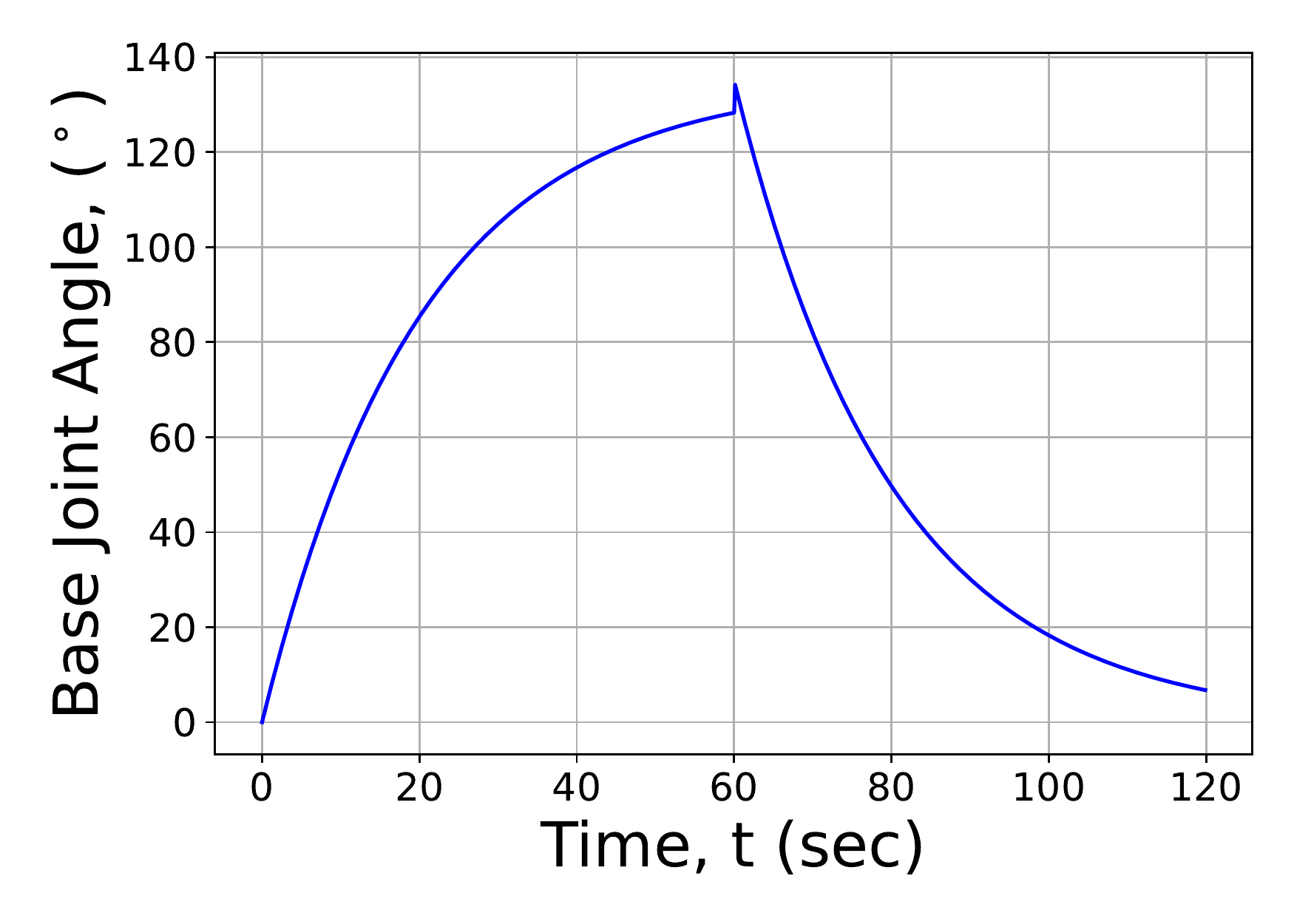}%
}
\hfill%
\subcaptionbox{Joint~2%
  \label{fig:nominal_trajectories:j2}}%
{%
  \includegraphics[width=0.45\columnwidth]{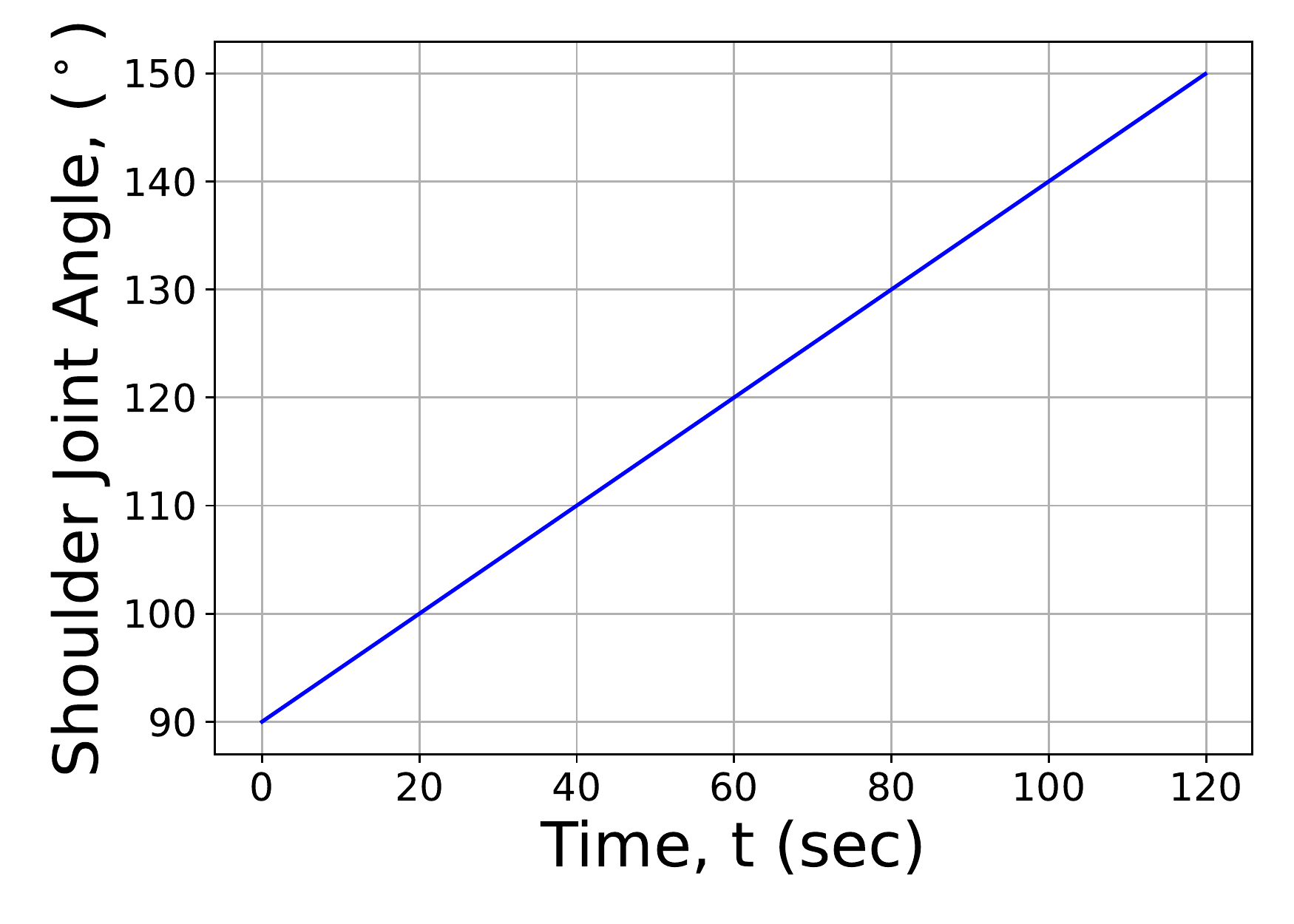}%
}
\\[2ex]
\subcaptionbox{Joint~3%
  \label{fig:nominal_trajectories:j3}}%
{%
  \includegraphics[width=0.45\columnwidth]{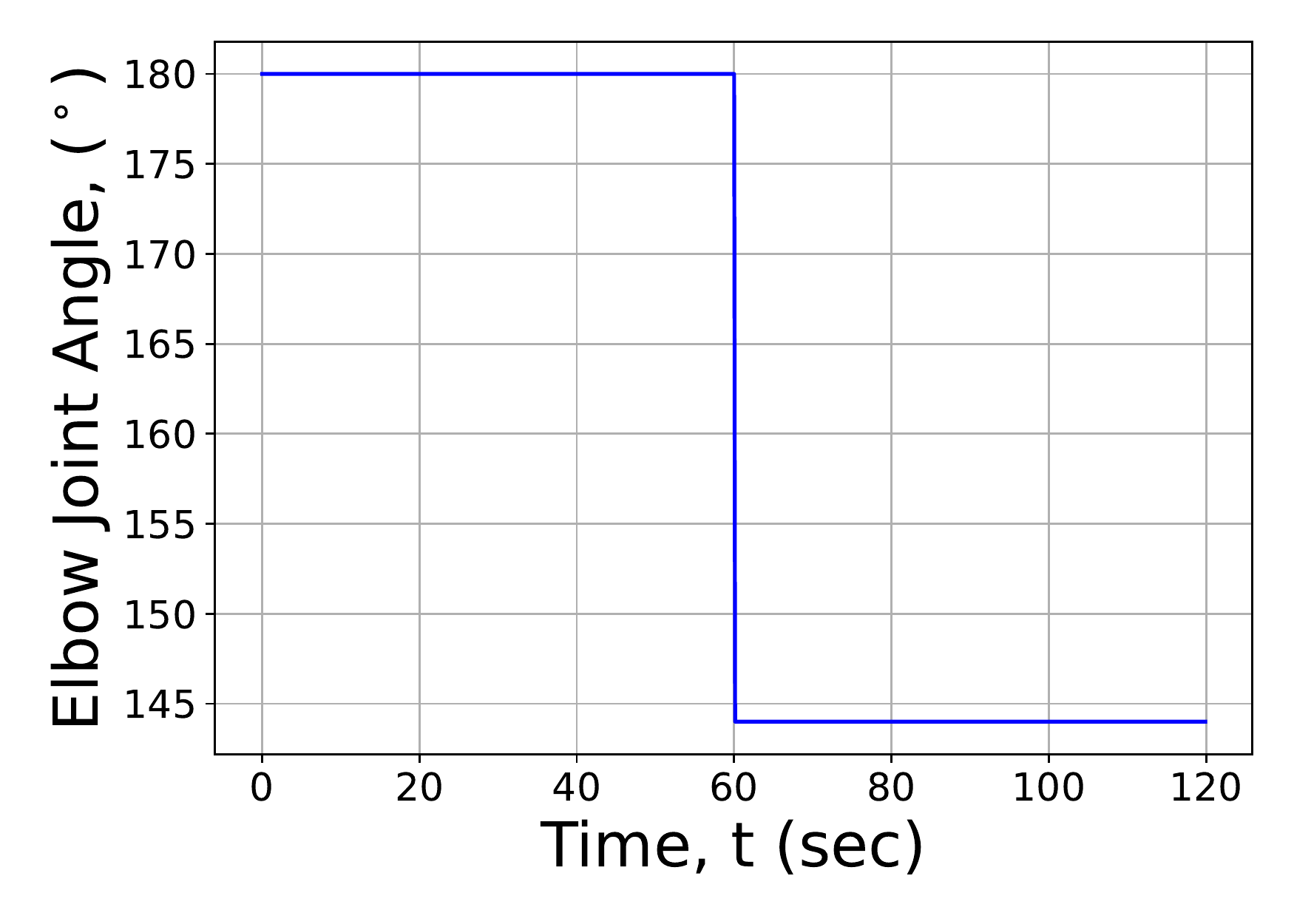}%
}
\hfill%
\subcaptionbox{Joint~4%
  \label{fig:nominal_trajectories:j4}}%
{%
  \includegraphics[width=0.45\columnwidth]{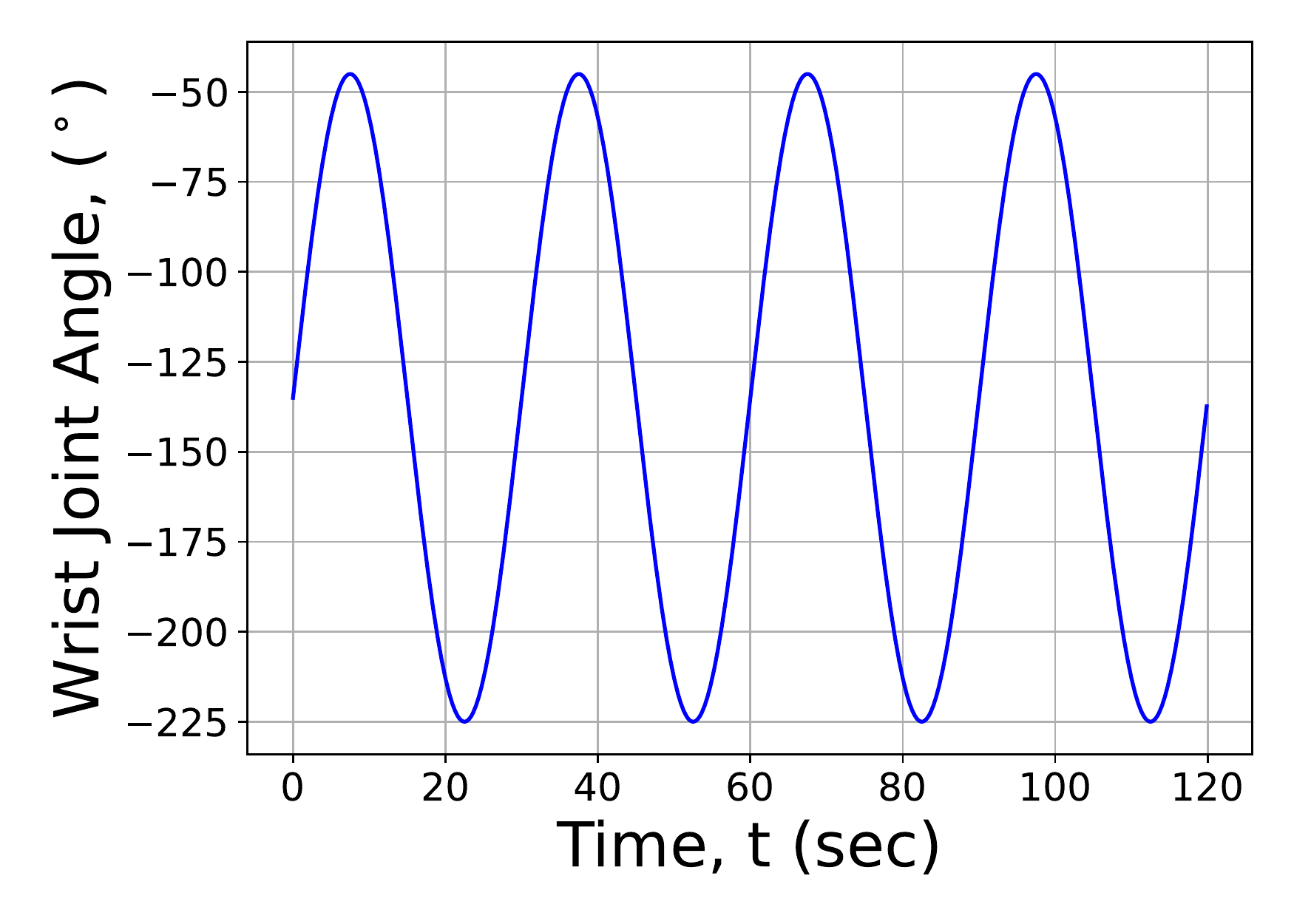}%
}
\\[2ex]
\subcaptionbox{Joint~2%
  \label{fig:nominal_trajectories:j2-time-varying}}%
{%
  \includegraphics[width=0.45\columnwidth]{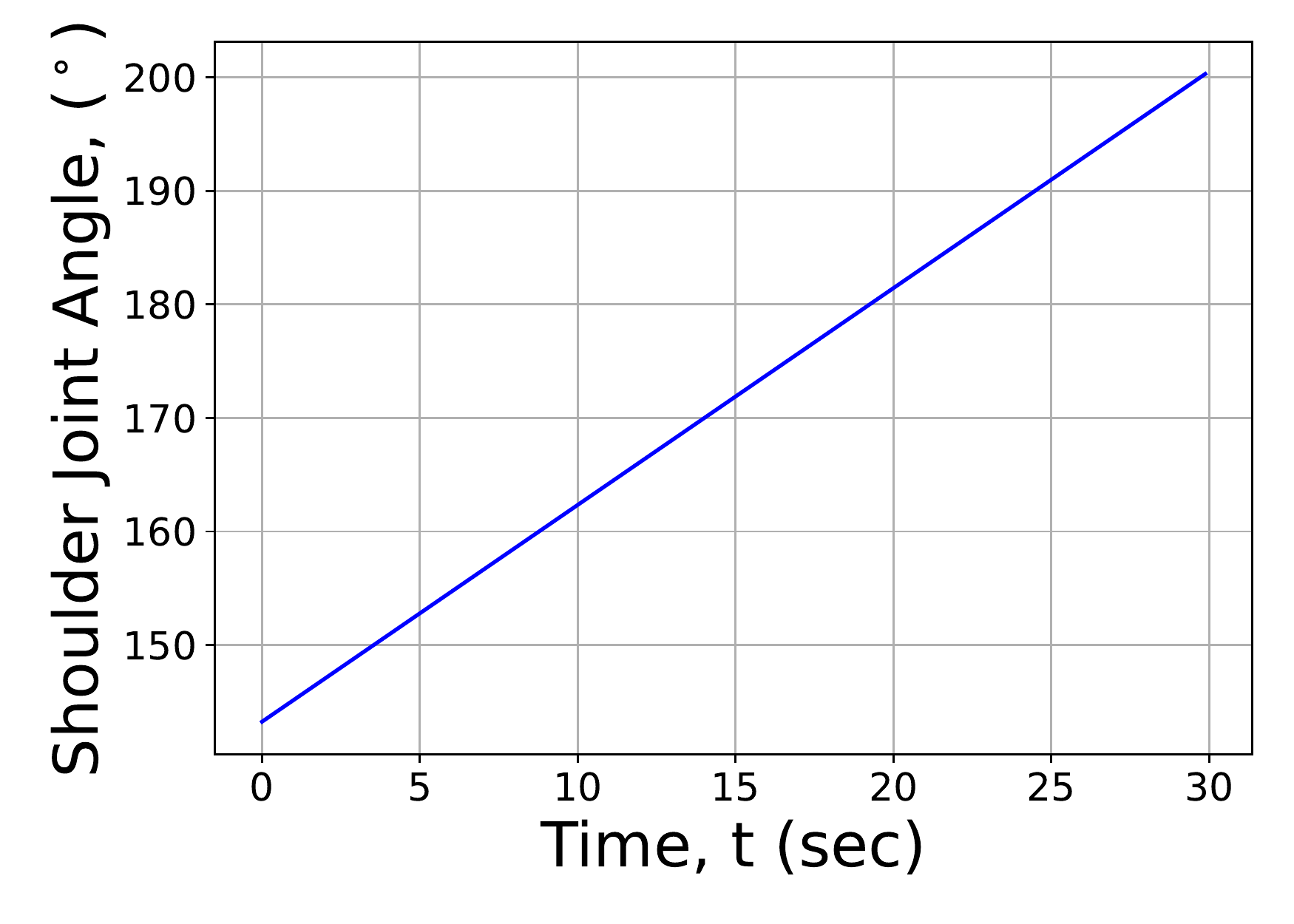}%
}
\caption{\subref{fig:nominal_trajectories:j1}--\subref{fig:nominal_trajectories:j4}: Nominal trajectories for experiments 1 to 4. \subref{fig:nominal_trajectories:j2-time-varying}: Elbow's reference trajectory for experiment~5.}
\label{fig:nominal_trajectories}
\end{figure}

The nominal joint trajectories for the first four experiments are shown in \Cref{fig:nominal_trajectories:j1,fig:nominal_trajectories:j2,fig:nominal_trajectories:j3,fig:nominal_trajectories:j4}, while that of the fifth experiment is shown in \Cref{fig:nominal_trajectories:j2-time-varying}.
This collection of reference trajectories provides variety of model-following dynamics to test, including slow dynamics and faster dynamics (the latter represented by the high frequency sinusoidal signal). Further, sharp and nonlinear forms of trajectories are considered. The goal is to test the robustness of the IRL solution under co-exiting interacting dynamic trajectories.
The base joint trajectory in \Cref{fig:nominal_trajectories:j1} represents an exponential growth for a predefined duration of three time-constants followed by an exponential decay for another duration of three time-constants. This trajectory is used to observe the system's ability to follow a path with time-varying position and velocity. The linear reference trajectory of the shoulder shown in \Cref{fig:nominal_trajectories:j2} probes the system's ability to follow a time-varying angular position while maintaining a constant velocity. The elbow's target trajectory in \Cref{fig:nominal_trajectories:j3} aims to analyze the ability of the system to maintain a constant angular position for a period of time before reacting to a step change. A notable difference between the elbow's trajectory and the other trajectories is the discontinuity of the former. The step change in the desired elbow's angular position should force the IRL algorithm to extrapolate and deliberately choose points that are not specified by the given values. Finally, the reference trajectory of the wrist shown in \Cref{fig:nominal_trajectories:j4} investigates the algorithm's ability to track sinusoidal signals.

The high-level control loop operates at a rate of \SI{8}{\Hz} for the IRL algorithm with $N = 960$ and $\nu = \SI{0.125}{\color{blue}{\second}}$, where here $N$ and $\nu$ represent the discrete-time index and the cycle period of the actuation system, respectively. In all test cases, the adaptation rates for the actor and critic weights are selected to be $ \alpha_a = 0.01$ and $\alpha_{c} =0.05$, respectively. The initial critic matrices are randomly generated for the first test case and then held constant throughout the remainder of the test cases for consistency purposes. Since the joint encoders sampling frequency and the actuation frequency are \SI{50}{\hertz} and \SI{8}{\hertz}, respectively, the corresponding time delay is negligible. The high-level ROS control between the external computer and the Kinova manipulator is secured through a USB-B connection while the embedded low-level actuator controls leverage the RS-485 protocol.

The critic weighting matrices for the four joints are set to
{\allowdisplaybreaks
  \begin{align*}\small
\bm{\Omega}_{c_{1}}
  & =
    \begin{bmatrix*}[l]
      0.80350 & 0.30937 & 0.84494 & 0.71454\\
      0.30937 & 0.21330 & 0.31157 & 0.36979\\
      0.84494 & 0.31157 & 1.09927 & 0.53435\\
      0.71454 & 0.36979 & 0.53435 & 0.92855
    \end{bmatrix*},
\\
  \bm{\Omega}_{c_{2}}
  & =
    \begin{bmatrix*}[l]
      0.42471 & 0.49183 & 0.54214 & 0.52932\\
      0.49183 & 0.66047 & 0.31157 & 0.59427\\
      0.54214 & 0.31157 & 1.15278 & 0.89470\\
      0.52932 & 0.59427 & 0.89470 & 1.05989
    \end{bmatrix*},
\\    
  \bm{\Omega}_{c_{3}}
  & =
    \begin{bmatrix*}[l]
      0.55195 & 0.39823 & 0.37661 & 0.25486\\
      0.39823 & 0.51538 & 0.42652 & 0.33598\\
      0.37661 & 0.42652 & 0.40267 & 0.35263\\
      0.25486 & 0.33598 & 0.35263 & 0.48076
    \end{bmatrix*},
\\
\bm{\Omega}_{c_{4}}
  & =
    \begin{bmatrix*}[l]
      1.52075 & 0.78373 & 1.29889 & 1.07203\\
      0.78373 & 0.93637 & 0.54343 & 0.28335\\
      1.29889 & 0.54343 & 1.41173 & 0.91172\\
      1.07203 & 0.28335 & 0.91172 &  1.2790
    \end{bmatrix*}.
\end{align*}}%
The initial actor weights are decided using the above critic matrices such that
$\Omega_a^0 = -{\cal{\bm H}}_{{\eta}{\eta}}^{-1}  {\cal{\bm H}}_{{\eta}\bm{X}} + Y$, where $Y \sim N(0, 0.1)$ . The random variable is added to the base, shoulder, and elbow joints, to show robustness to the initial conditions of the actor weights and to more readily demonstrate the adaptability of the IRL.
The values of $\bm{Q}$ and $R$ are quoted below for completeness.
\begin{align*}
	\small
  \bm{Q}_{1}
  &=
    \begin{bmatrix*}[l]
      0.51503 & 0.25789 & 0.06581\\
      0.25789 & 0.19214 & 0.07471\\
      0.06581 & 0.07471 & 0.03784
    \end{bmatrix*},
      &
  R_1
  & =
    0.07451,
\\
  \bm{Q}_{2}
  &=
    \begin{bmatrix*}[l]
      0.85038 & 0.59431 & 0.47996\\
      0.59431 & 0.51992 & 0.24568\\
      0.47996 & 0.24568 & 0.38590
    \end{bmatrix*},
      &
  R_2
  & =
    0.00876,
\\
  \bm{Q}_{3}
  &=
    \begin{bmatrix*}[l]
      0.74237 & 0.51836 & 0.63923\\
      0.51836 & 0.41698 & 0.46758\\
      0.63923 & 0.46758 & 0.60572
    \end{bmatrix*},
      &
  R_3
  & =
    0.49363.
\\
  \bm{Q}_{4}
  &=
    \begin{bmatrix*}[l]
      0.56665 & 0.36332 & 0.53481\\
      0.36332 & 0.47523 & 0.37991\\
      0.53481 & 0.37991 &  0.5266
    \end{bmatrix*},
      &
  R_4
  & =
    0.019686.
\end{align*}

To put the performance of the IRL algorithm in perspective, we compare it with a high-order model-free adaptive control (HOMFAC) scheme presented in~\citep{Xu2021}, which is an improved version of the algorithm proposed in~\citep{Chi-Hou-Jin-Huang-2018}.
The adaptive learning solution differs in the structure when compared with that of the HOMFAC approach. The adaptive solution employs adaptable strategies unlike the HOMFAC approach, where fixed control gains are considered. Furthermore, the reinforcement learning strategy captures explicit high-order error dynamics, while the order of error dynamics is controllable. This is unlike the HOMFAC case, where explicit zero-order error dynamics are utilized into the control strategy. This means that the adaptive learning solution has more capacity to react to the variations
in the error patterns. The simulation cases in \Cref{sec:Anl} highlight the impact of such differences.
Just like the IRL algorithm, the HOMFAC technique is applied to each of the four joints. Both algorithms are implemented in Python~2.7 and interfaced with the robot through ROS Melodic. The parameter values used for the HOMFAC algorithm are the same as those presented in~\citep{Xu2021} and are tabulated in \Cref{tab:params_HOMFAC}. Parameters $\alpha$, $\eta$, $\lambda$, $\mu$, and $\rho$, dictate the adaptation properties of the estimator $\phi(t)$ and calculation of the control action $u(t)$. Parameters $\phi^0_i$, $i =1, 2, 3, 4$, represent the initial conditions of the joint estimators. These values are chosen experimentally.
In order to successfully implement the HOMFAC algorithm on the Kinova JACO arm, the frequency is reduced to \SI{5}{\hertz}. Attempts to implement the HOMFAC algorithm with higher frequencies induced non-convergent responses.

\begin{table}
		\caption{Parameters used for the HOMFAC algorithm}
		\centering
		\small
	\begin{tabular}{c|c|c|c}
		\hline
		Parameter	&	Value & Parameter	&	Value\\
		\hline\hline
		$\alpha$ 		&	$\vect{1/2, 1/4, 1/8, 1/8}$ &
		$\eta$		&	0.8 \\
		$\lambda$	&	0.1&
		$\mu$ 		&	0.01\\
		$\rho$ 		& 	0.8&
		$\phi^0_{1}$&	15\\
		$\phi^0_{2}$&	15&
		$\phi^0_{3}$&	25\\
		$\phi^0_{4}$&	25 &\\	
		\hline
	\end{tabular} 
	\label{tab:params_HOMFAC}
\end{table}

\section{Results and Discussion}
\label{sec:Anl}
\begin{figure}[!ht]
  \centering
  \subcaptionbox{Experiment 1%
   \label{fig:actual_trajectories:j1_nm}}%
  {%
    \includegraphics[width=0.49\columnwidth]{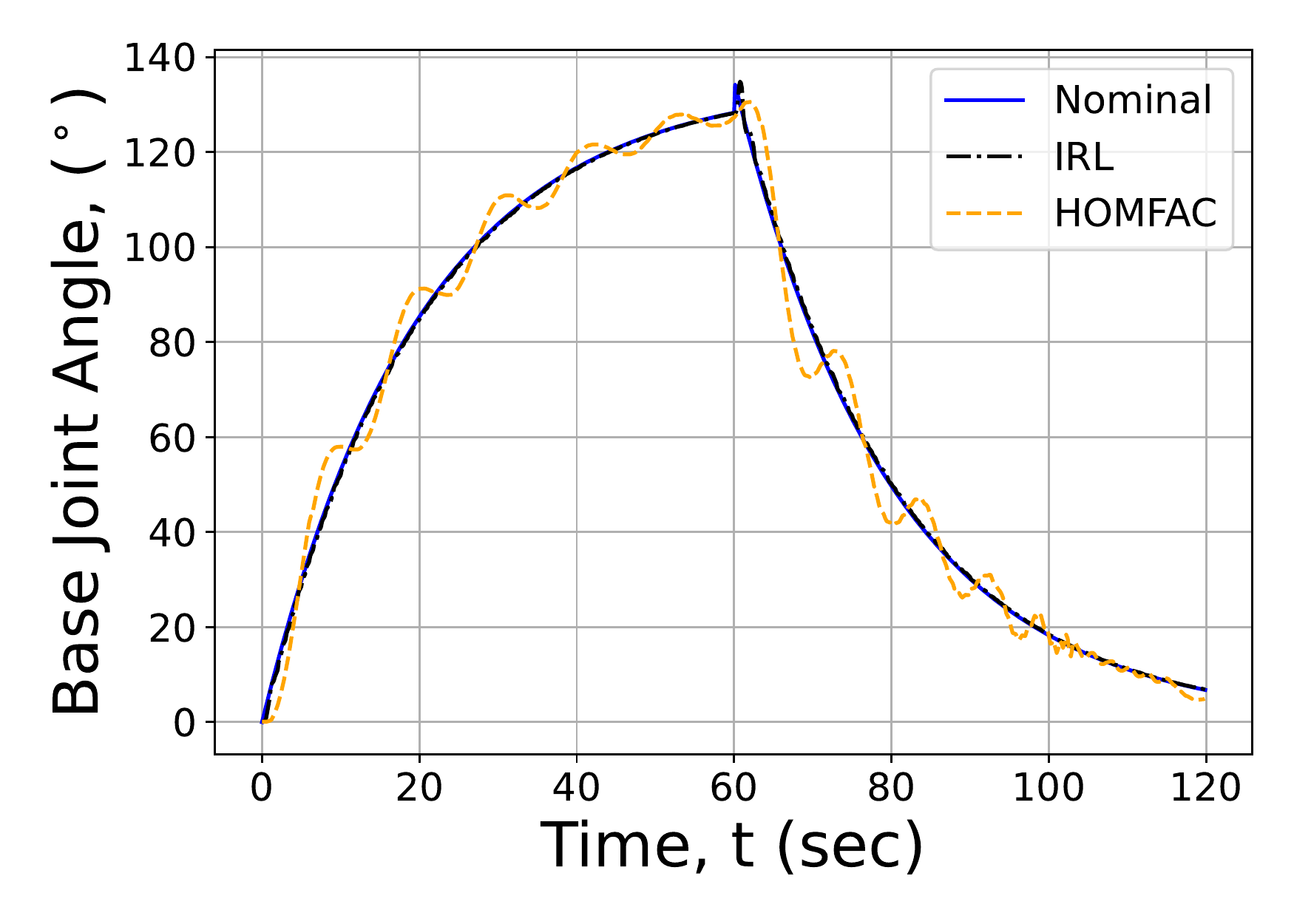}%
  } 
  \hfill%
  \subcaptionbox{Experiment 2%
   \label{fig:actual_trajectories:j1_3lb}}%
  {%
    \includegraphics[width=0.49\columnwidth]{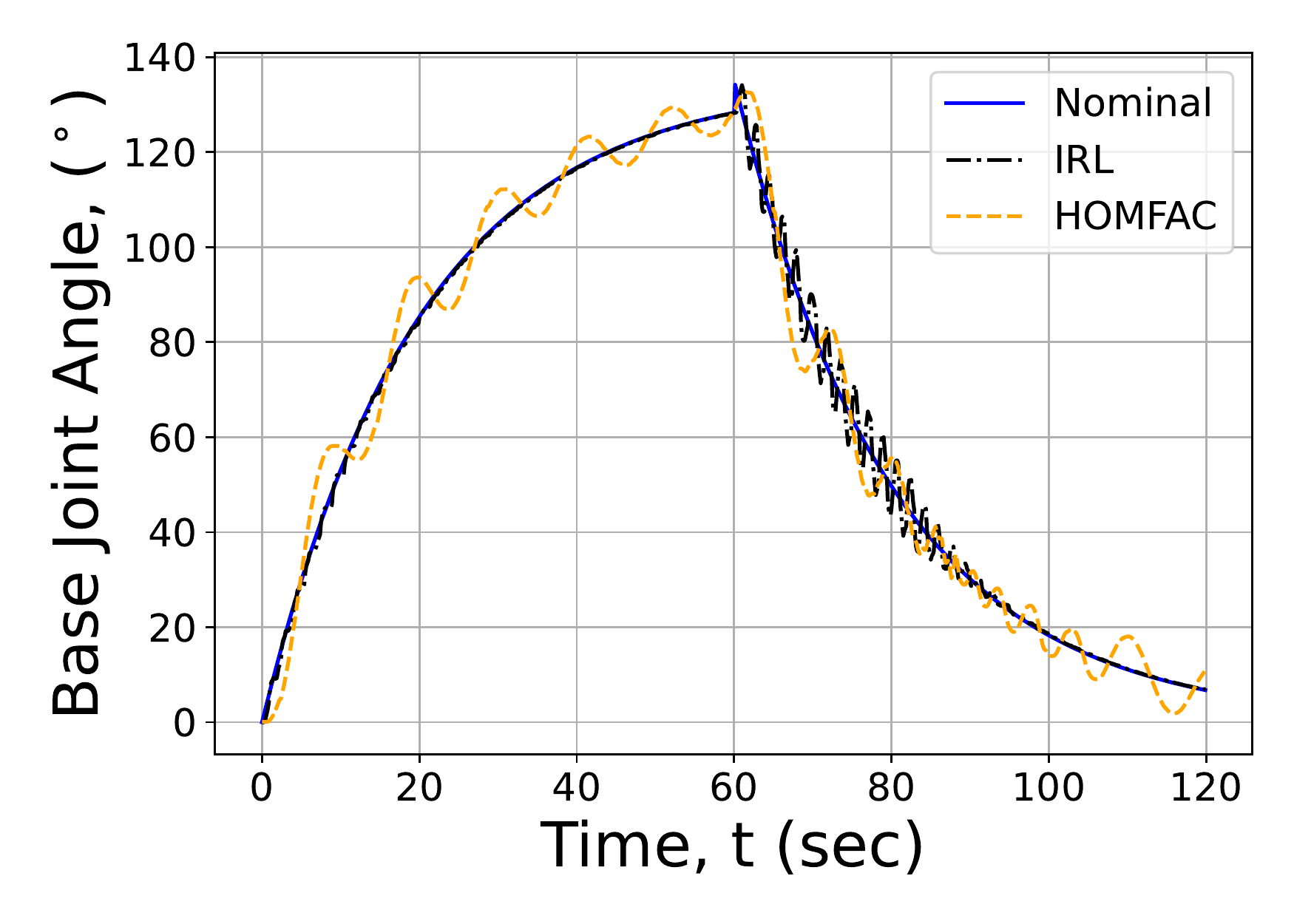}%
  }
  \\[2ex]
  \mbox{}\hfill
  \subcaptionbox{Experiment 3%
    \label{fig:actual_trajectories:j1_disturbed}}%
  {%
    \includegraphics[width=0.49\columnwidth]{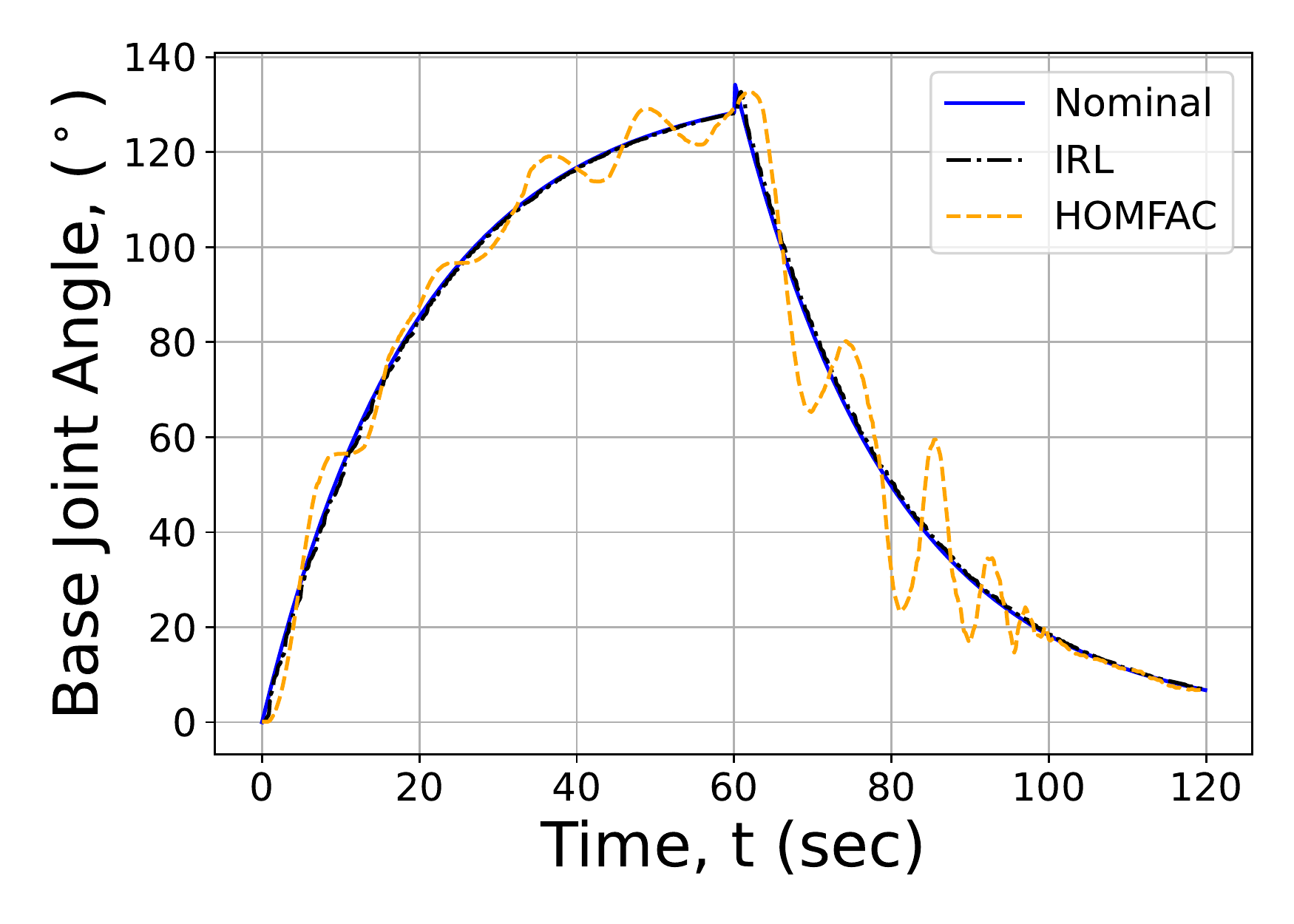}%
  }
     \hfill\mbox{}
  \caption{Base trajectories for the first three experiments.}
  \label{fig:actual_trajectories_j1}
\end{figure}

\begin{figure}[!ht]
  \centering
  \subcaptionbox{Experiment 1%
   \label{fig:actual_trajectories:j2_nm}}%
  {%
    \includegraphics[width=0.49\columnwidth]{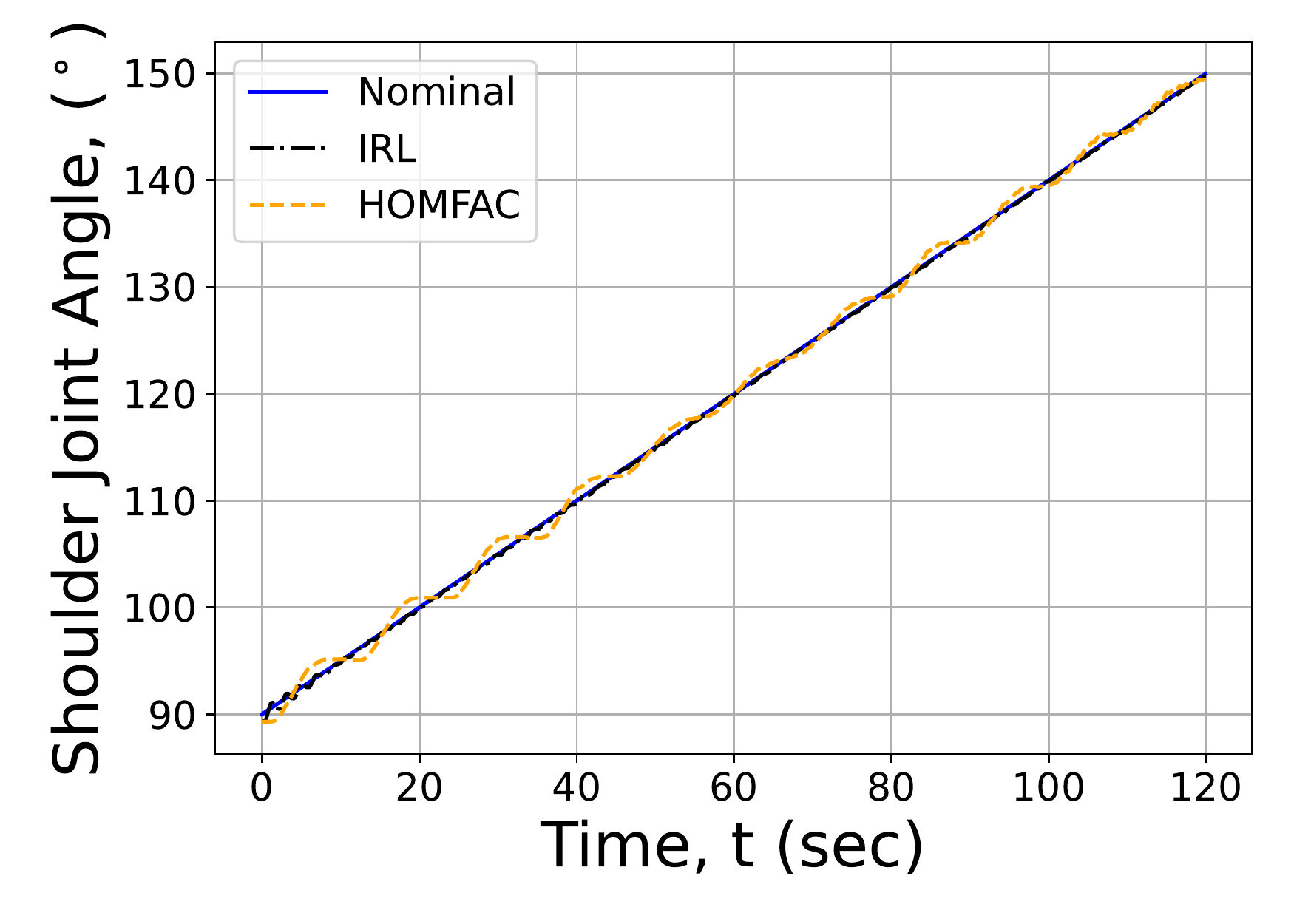}%
  } 
  \hfill%
  \subcaptionbox{Experiment 2%
   \label{fig:actual_trajectories:j2_3lb}}%
  {%
    \includegraphics[width=0.49\columnwidth]{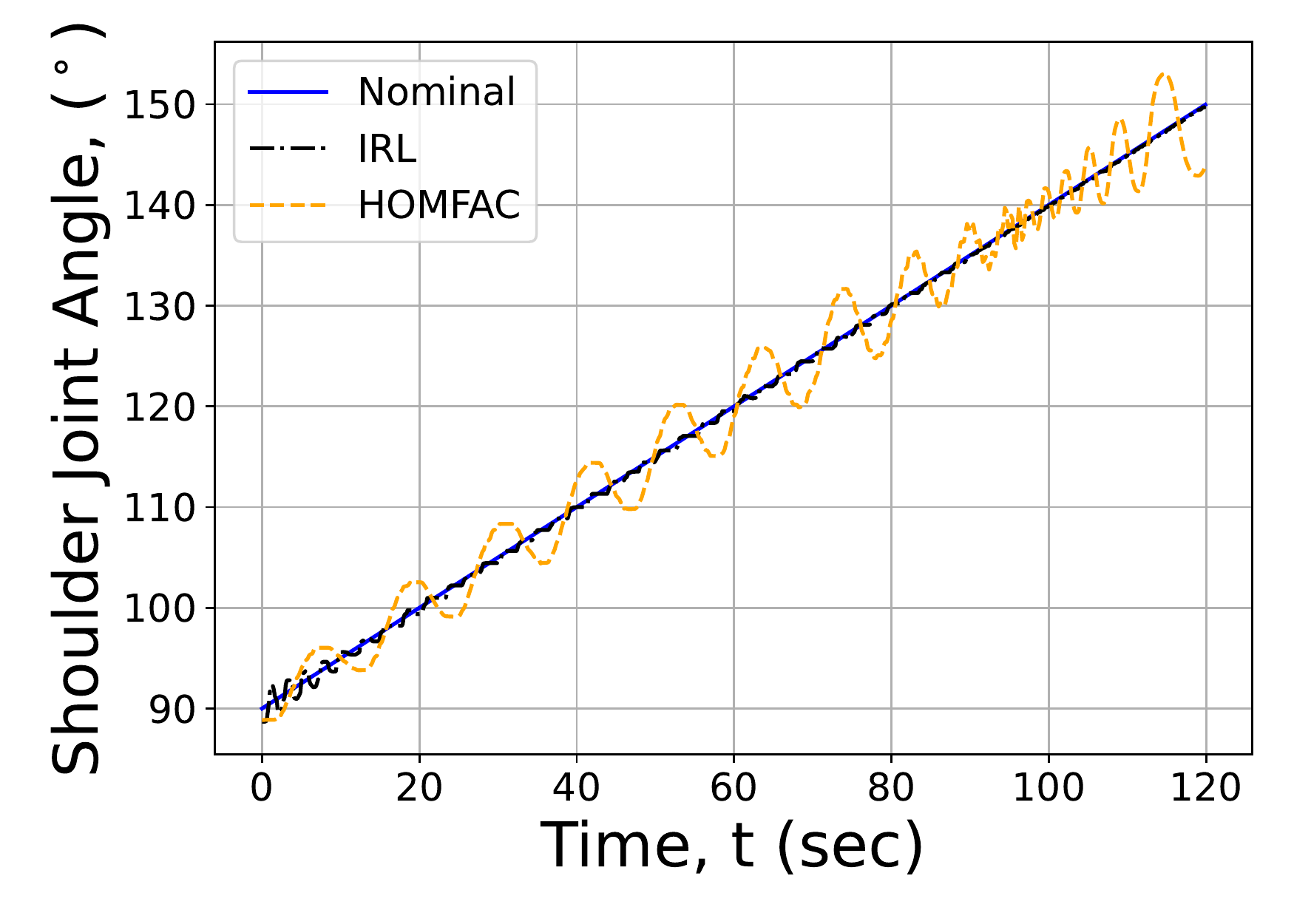}%
  }
  \\[2ex]
  \mbox{}\hfill
  \subcaptionbox{Experiment 3%
    \label{fig:actual_trajectories:j2_disturbed}}%
  {%
    \includegraphics[width=0.49\columnwidth]{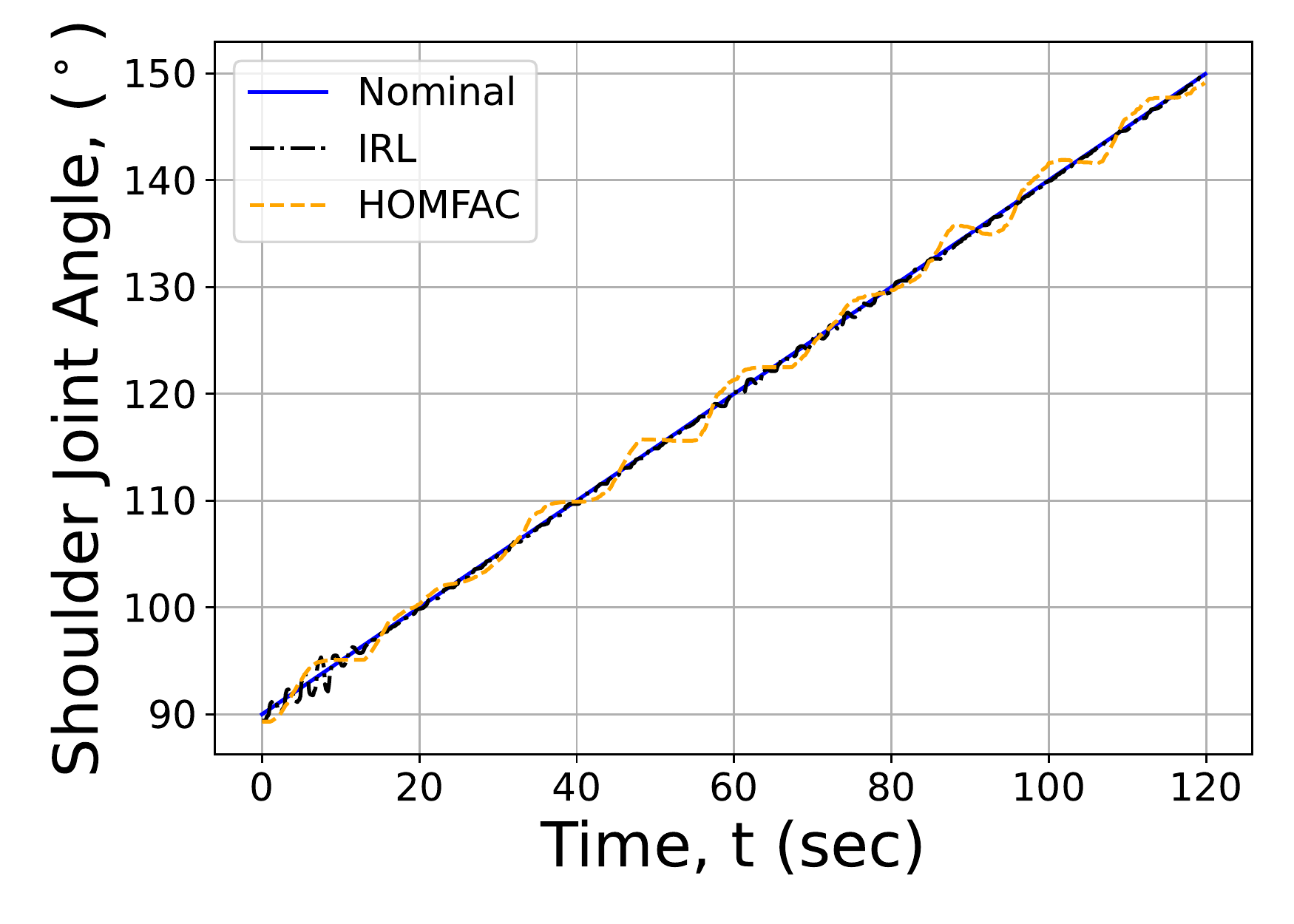}%
  }
     \hfill\mbox{}
  \caption{Shoulder trajectories for the first three experiments.}
  \label{fig:actual_trajectories_j2}
\end{figure}

\begin{figure}[!ht]
  \centering
  \subcaptionbox{Experiment 1%
   \label{fig:actual_trajectories:j3_nm}}%
  {%
    \includegraphics[width=0.49\columnwidth]{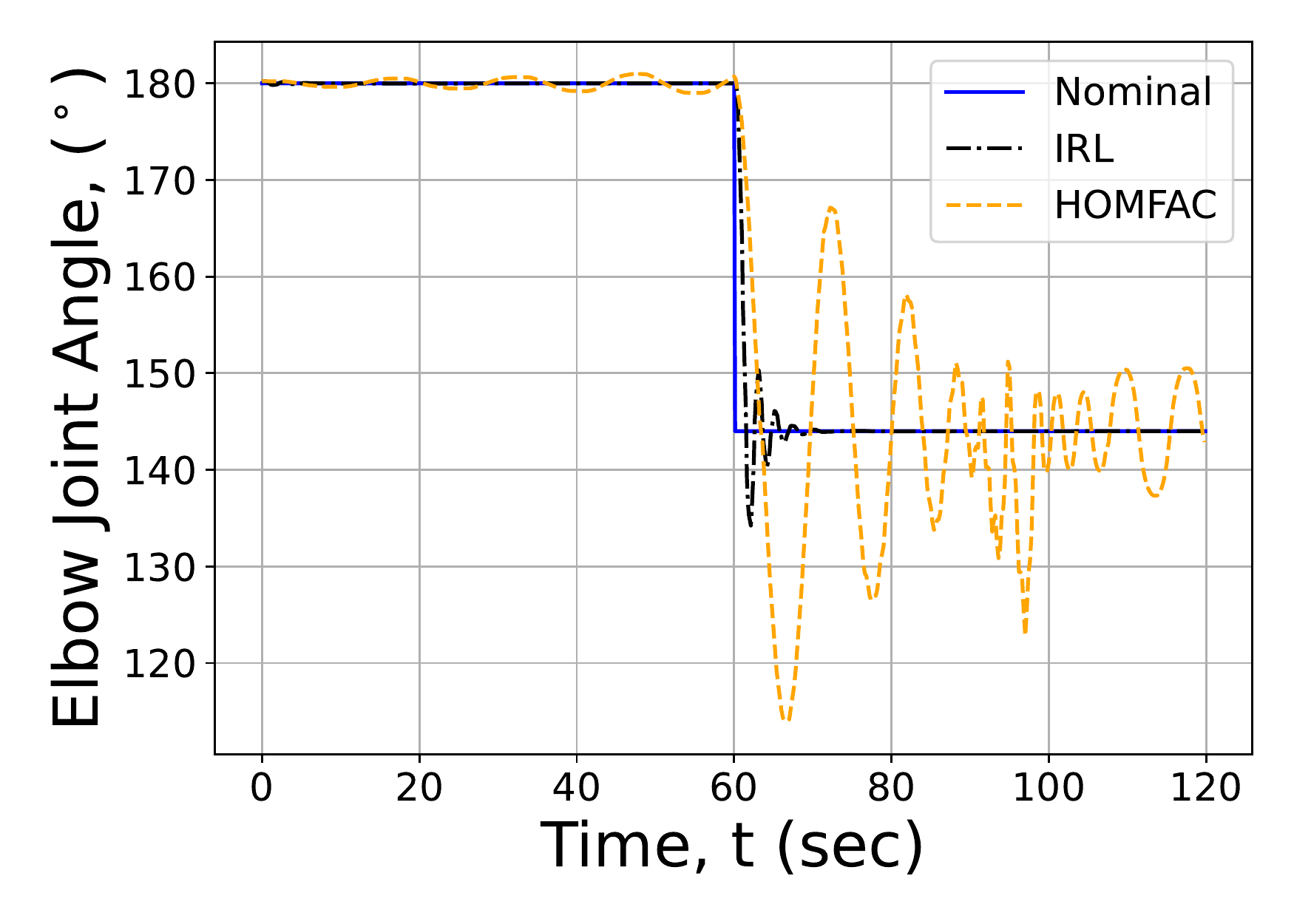}%
  } 
  \hfill%
  \subcaptionbox{Experiment 2%
   \label{fig:actual_trajectories:j3_3lb}}%
  {%
    \includegraphics[width=0.49\columnwidth]{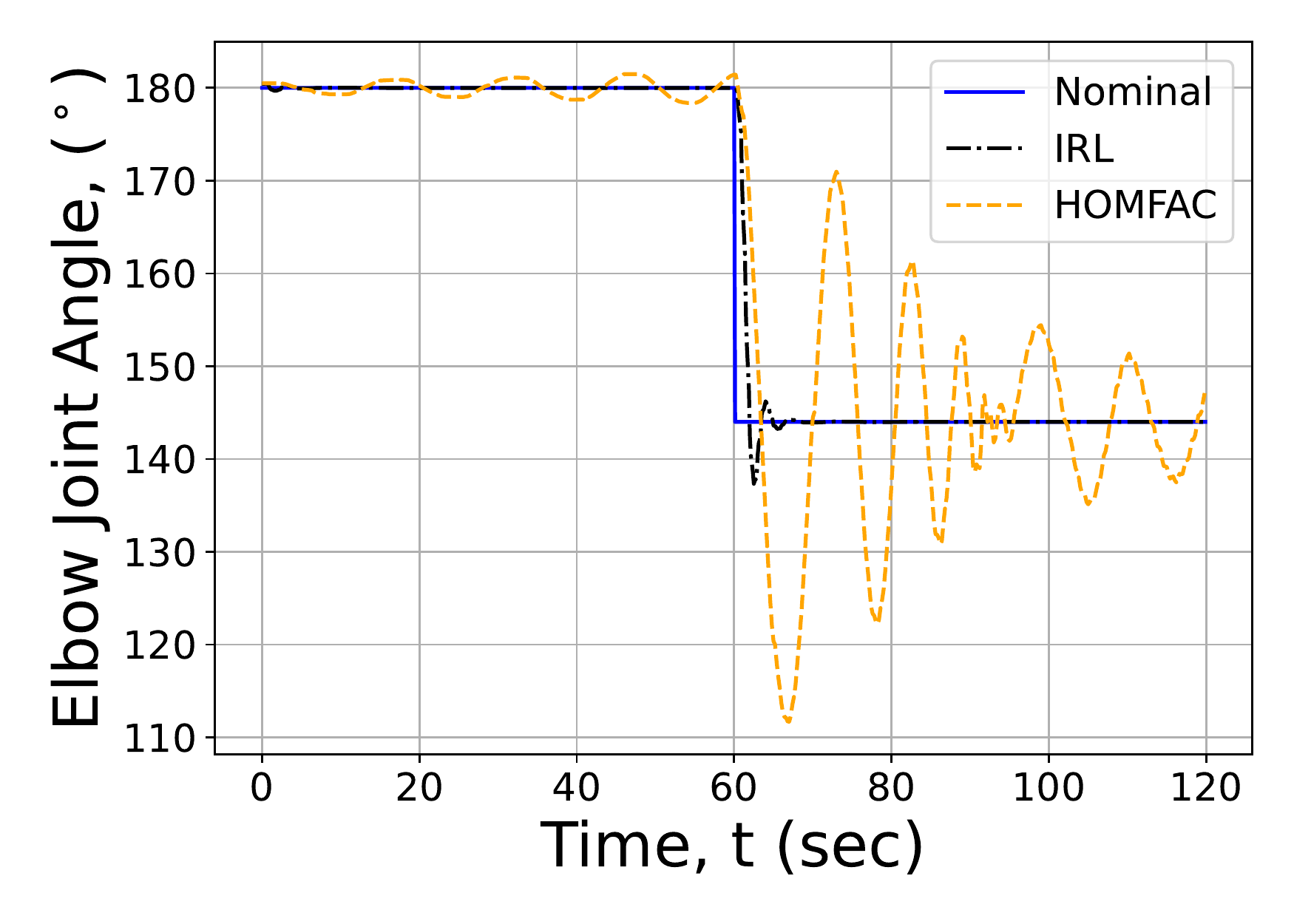}%
  }
  \\[2ex]
  \mbox{}\hfill
  \subcaptionbox{Experiment 3%
    \label{fig:actual_trajectories:j3_disturbed}}%
  {%
    \includegraphics[width=0.49\columnwidth]{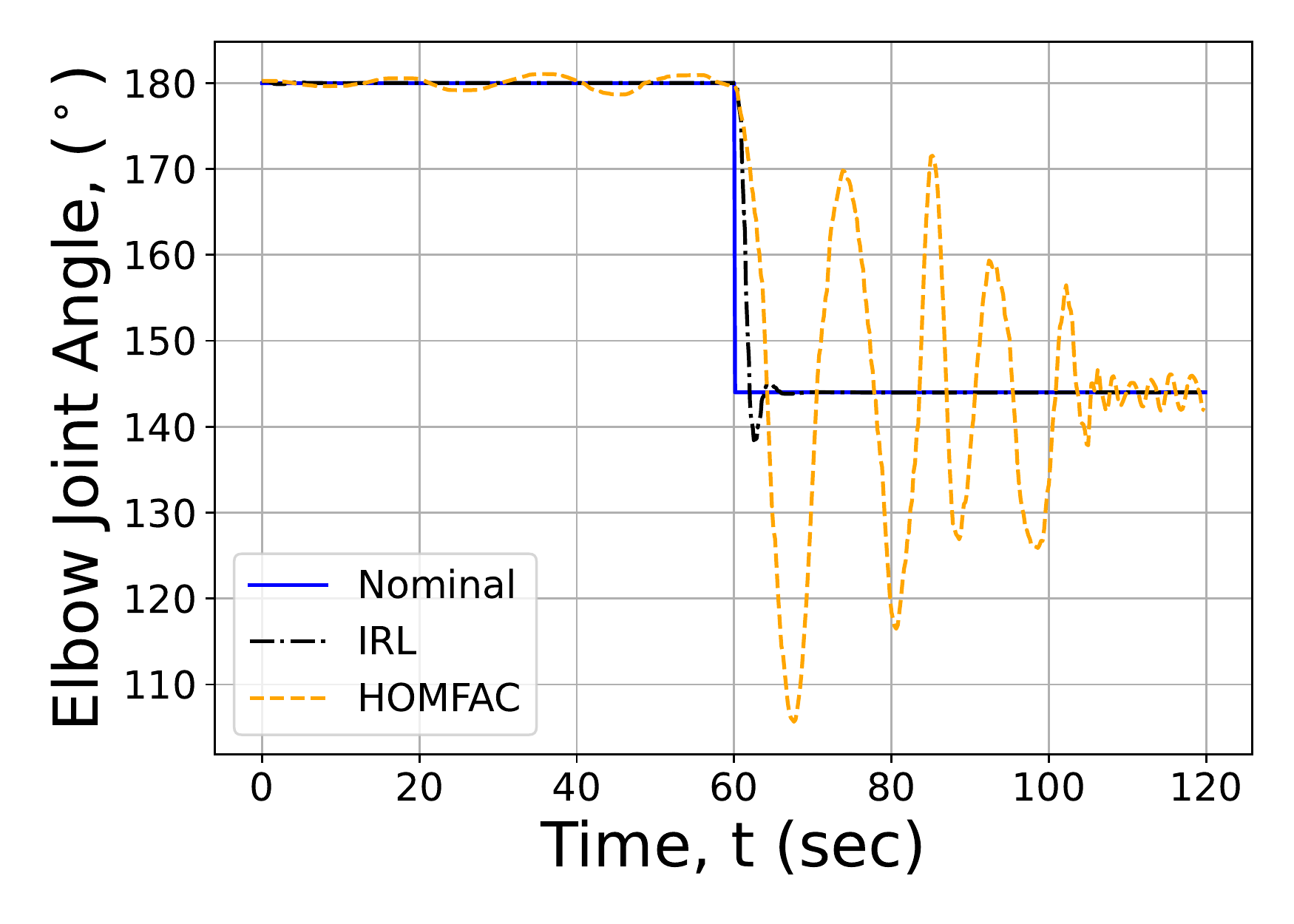}%
  }
     \hfill\mbox{}
  \caption{Elbow trajectories for the first three experiments.}
  \label{fig:actual_trajectories_j3}
\end{figure}

\begin{figure}[!ht]
  \centering
  \subcaptionbox{Experiment 1%
   \label{fig:actual_trajectories:j4_nm}}%
  {%
    \includegraphics[width=0.49\columnwidth]{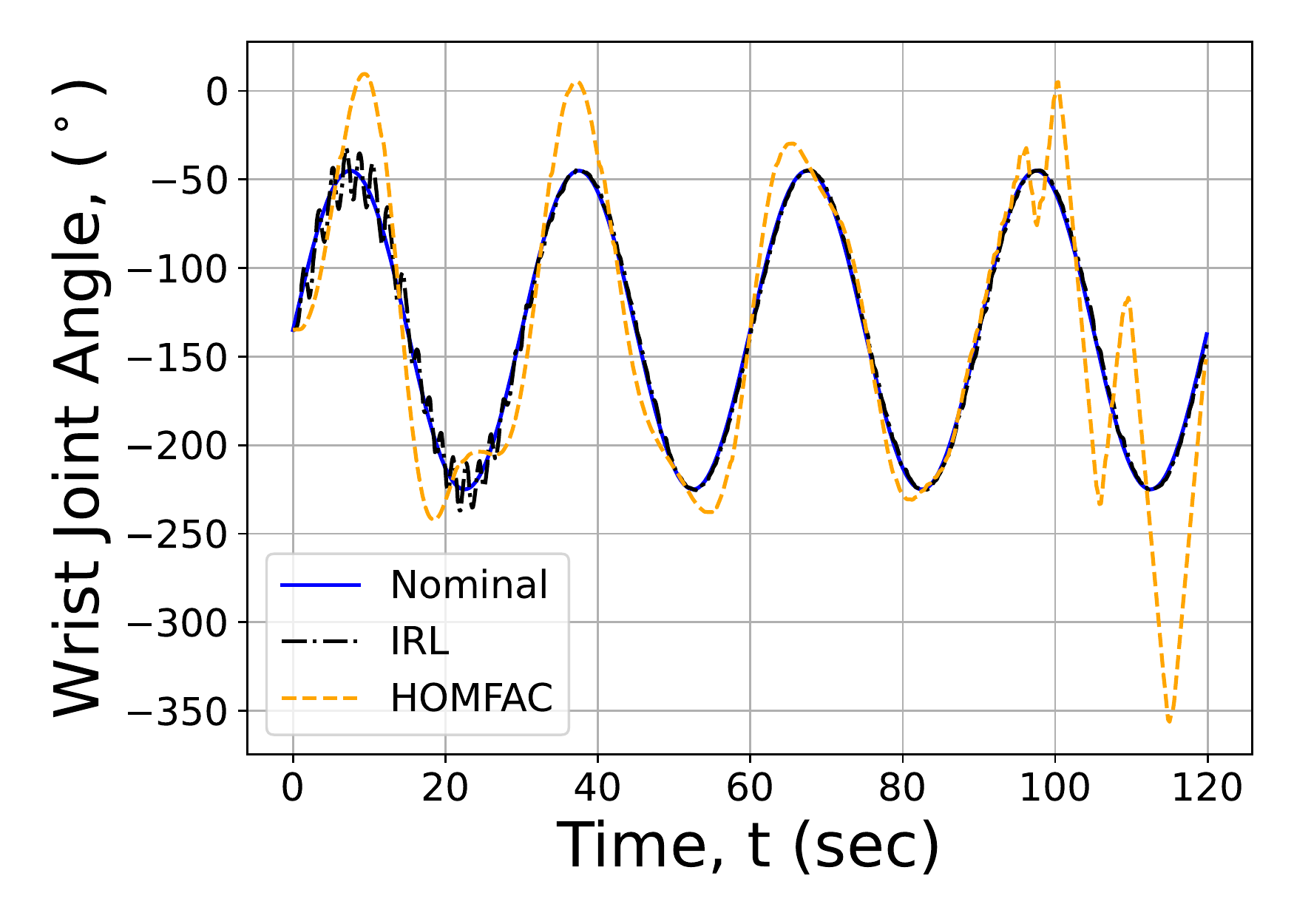}%
  } 
  \hfill%
  \subcaptionbox{Experiment 2%
   \label{fig:actual_trajectories:j4_3lb}}%
  {%
    \includegraphics[width=0.49\columnwidth]{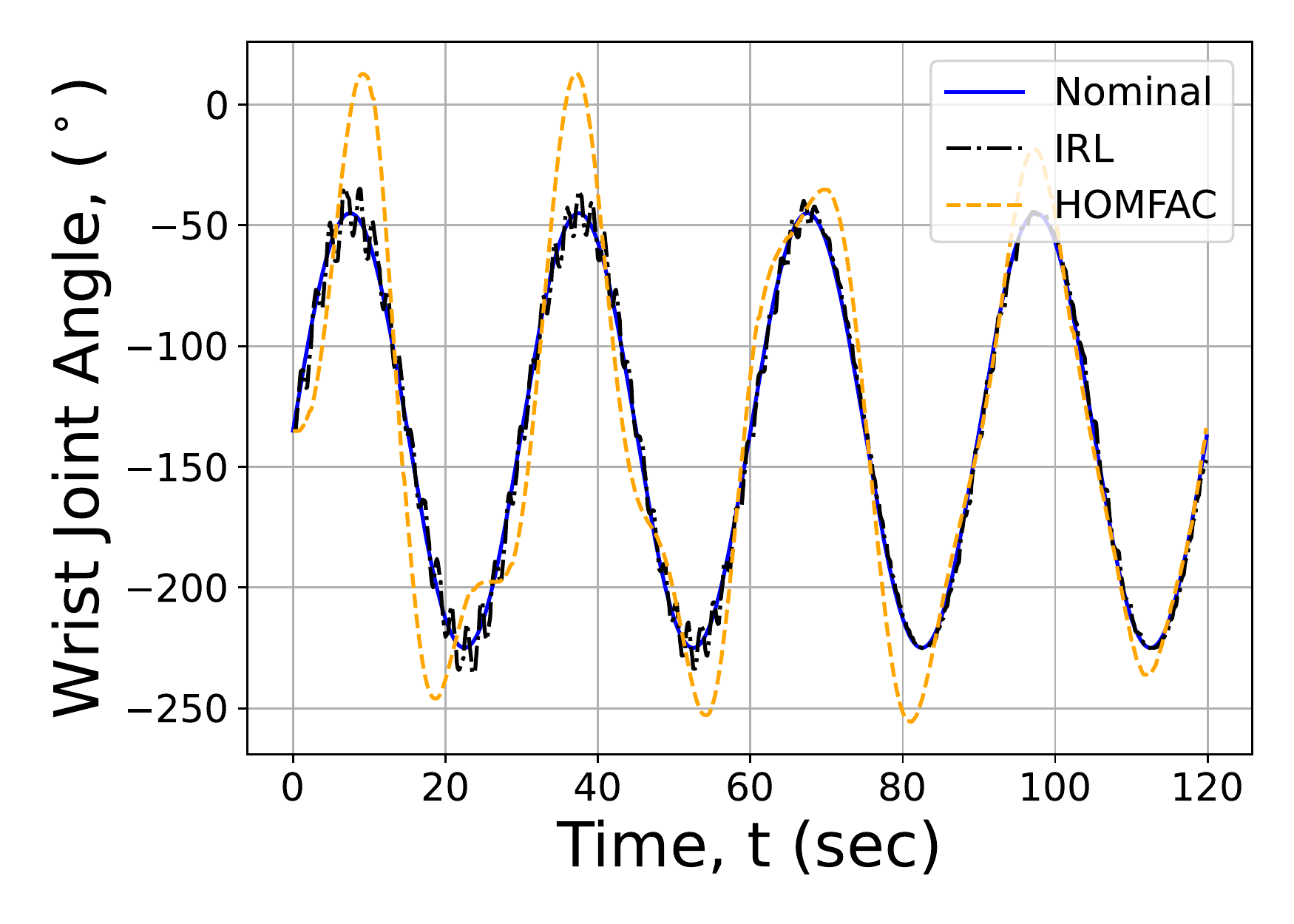}%
  }
  \\[2ex]
  \mbox{}\hfill
  \subcaptionbox{Experiment 3%
    \label{fig:actual_trajectories:j4_disturbed}}%
  {%
    \includegraphics[width=0.49\columnwidth]{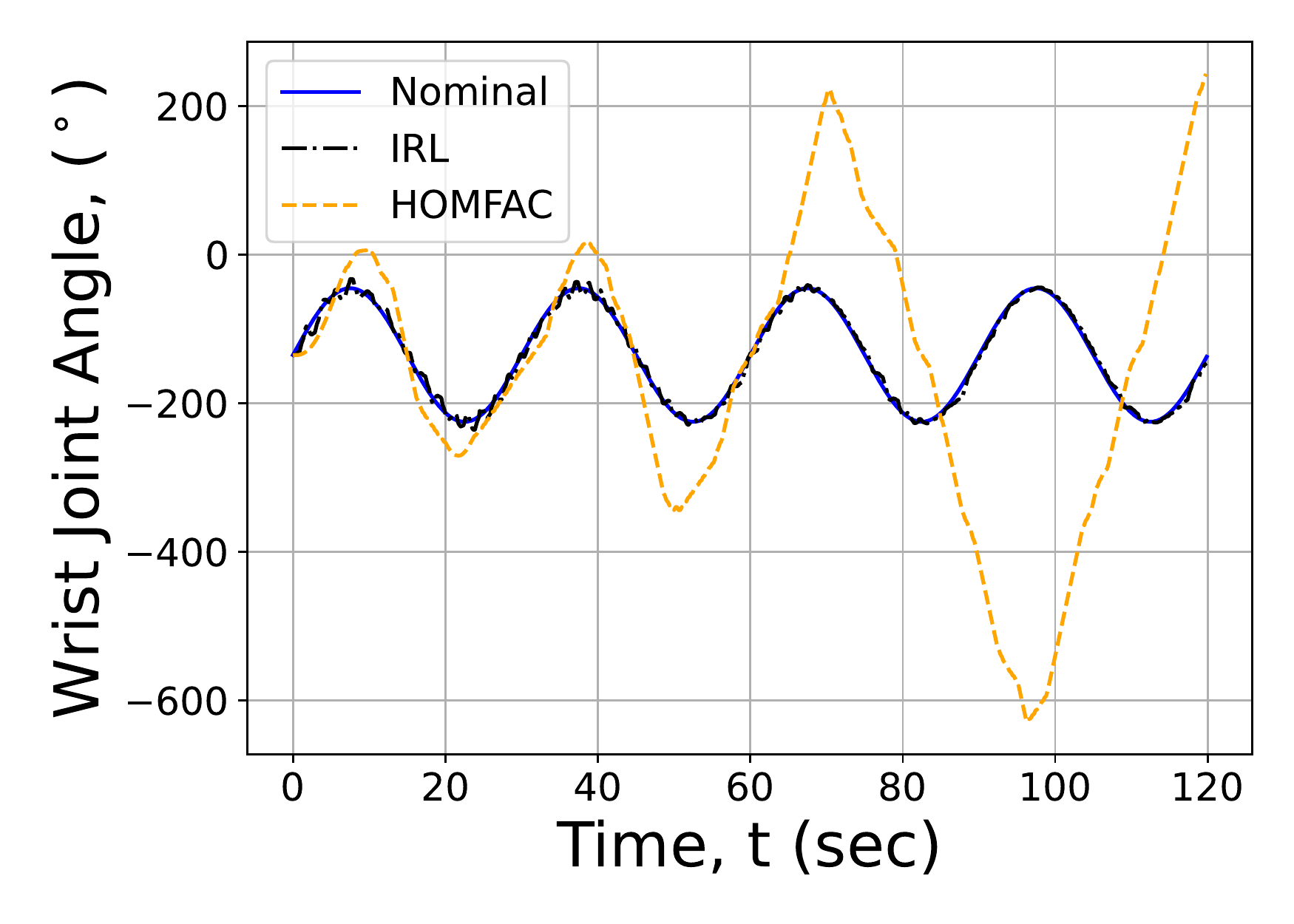}%
  }
     \hfill\mbox{}
  \caption{Wrist trajectories for the first three experiments.}
  \label{fig:actual_trajectories_j4}
\end{figure}

The joint trajectories achieved with the IRL and HOMFAC algorithms are logged at run-time and plotted against the nominal trajectories for comparison. The results for the first three experiments are illustrated in \Cref{fig:actual_trajectories_j1,fig:actual_trajectories_j2,fig:actual_trajectories_j3,fig:actual_trajectories_j4}.
It is observed that the IRL algorithm is able to converge rapidly to the reference signals, which is not always the case for the HOMFAC technique.
These figures clearly demonstrate the superiority of the IRL algorithm over the HOMFAC. This is more evident in the cases where there is a sudden or/and continuous changes in the reference trajectories (e.g., \Cref{fig:actual_trajectories_j3,fig:actual_trajectories_j4}). As a matter of fact, in some experiments, the HOMFAC exhibits excessive oscillations and even divergence, as in \Cref{fig:actual_trajectories_j3} and \Cref{fig:actual_trajectories:j4_disturbed}, respectively, while the IRL algorithm shows a smooth and rapid convergence. Quantitatively, \Cref{fig:actual_trajectories:j3_disturbed} shows a maximum elbow joint overshoot using the HOMFAC algorithm of approximately 28\% as opposed to approximately 4\% for the IRL.
For the purpose of readability and completeness, the adaptations of the actuator gains are presented in the Appendix.

\begin{figure}[!ht]
  \centering
  \subcaptionbox{Joint~1%
   \label{fig:actual_trajectories:j1_noise}}%
  {%
    \includegraphics[width=0.49\columnwidth]{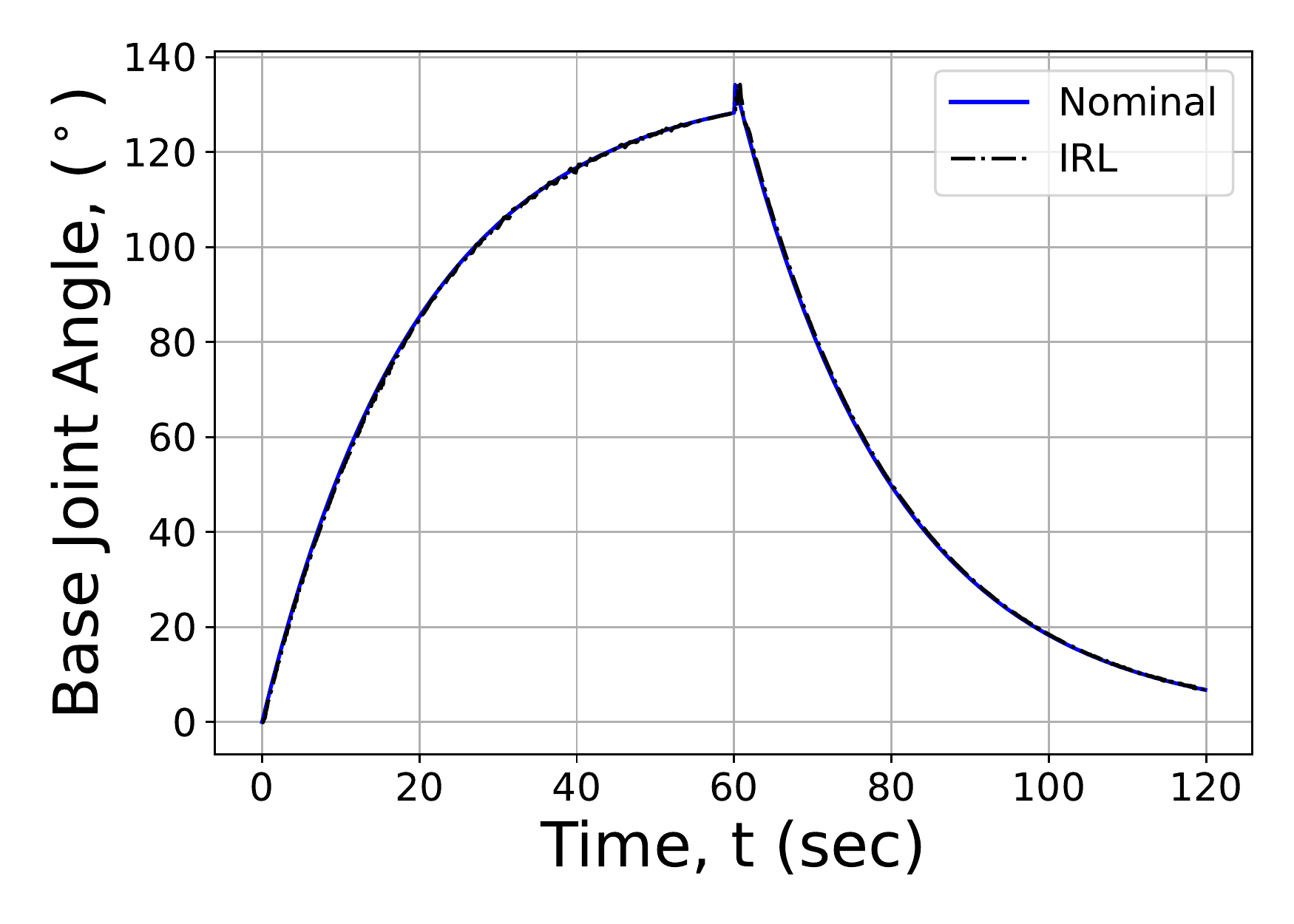}%
  } 
  \hfill%
  \subcaptionbox{Joint~2%
   \label{fig:actual_trajectories:j2_noise}}%
  {%
    \includegraphics[width=0.49\columnwidth]{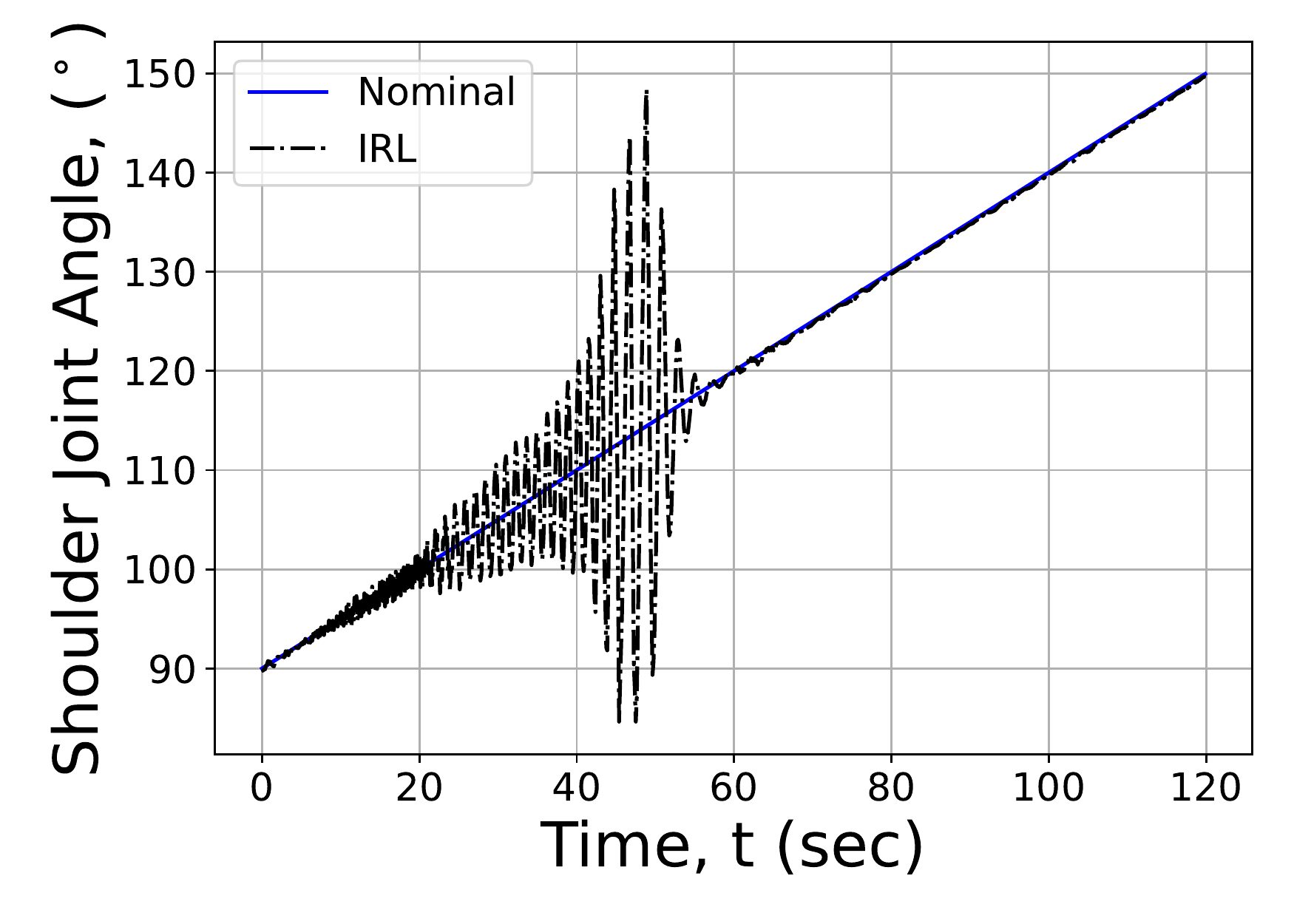}%
  }
  \\[2ex]
  \mbox{}\hfill
  \subcaptionbox{Joint~3%
    \label{fig:actual_trajectories:j3_noise}}%
  {%
    \includegraphics[width=0.49\columnwidth]{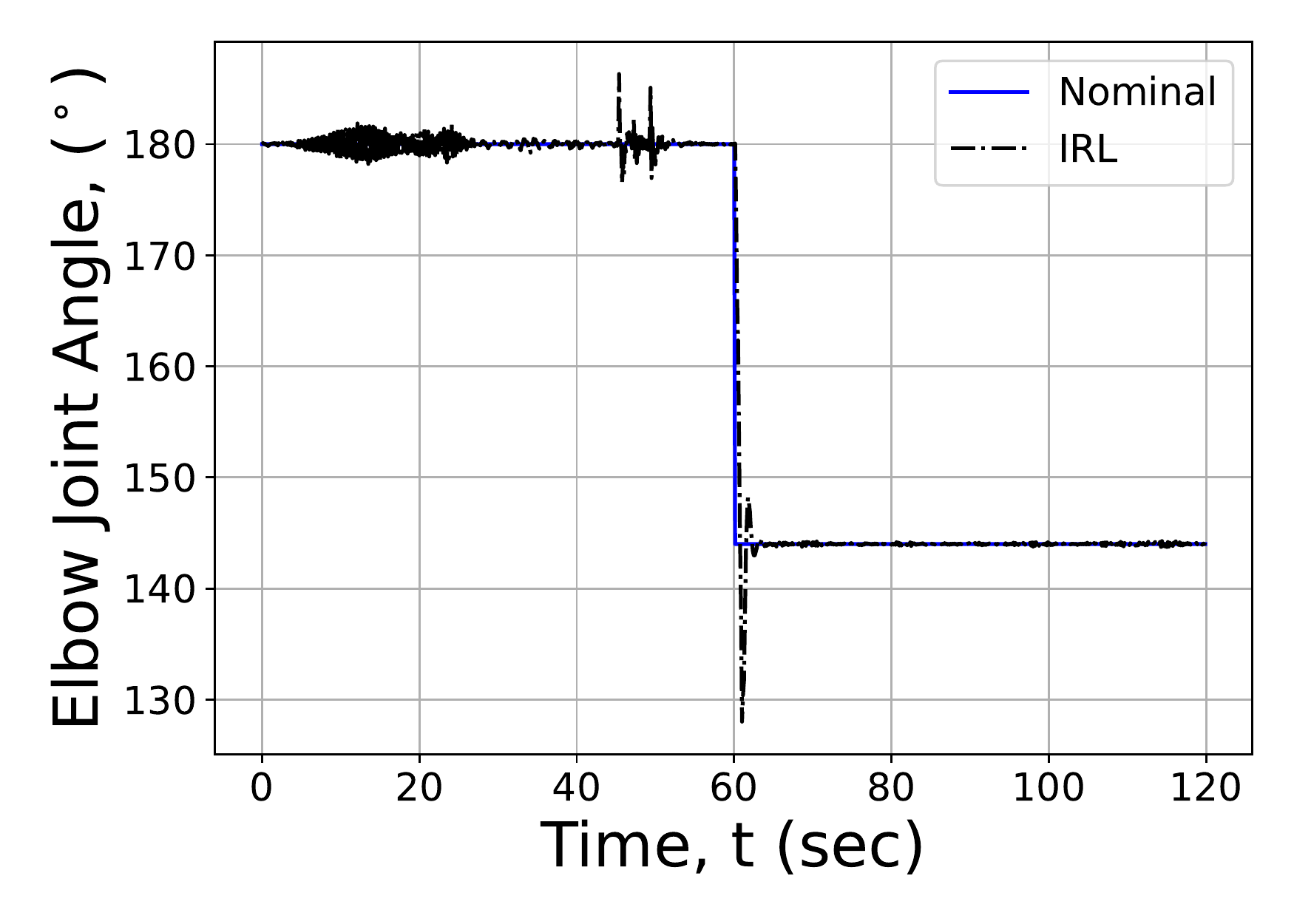}%
  }
     \hfill
  \subcaptionbox{Joint~4%
   \label{fig:actual_trajectories:j4_noise}}%
  {%
    \includegraphics[width=0.49\columnwidth]{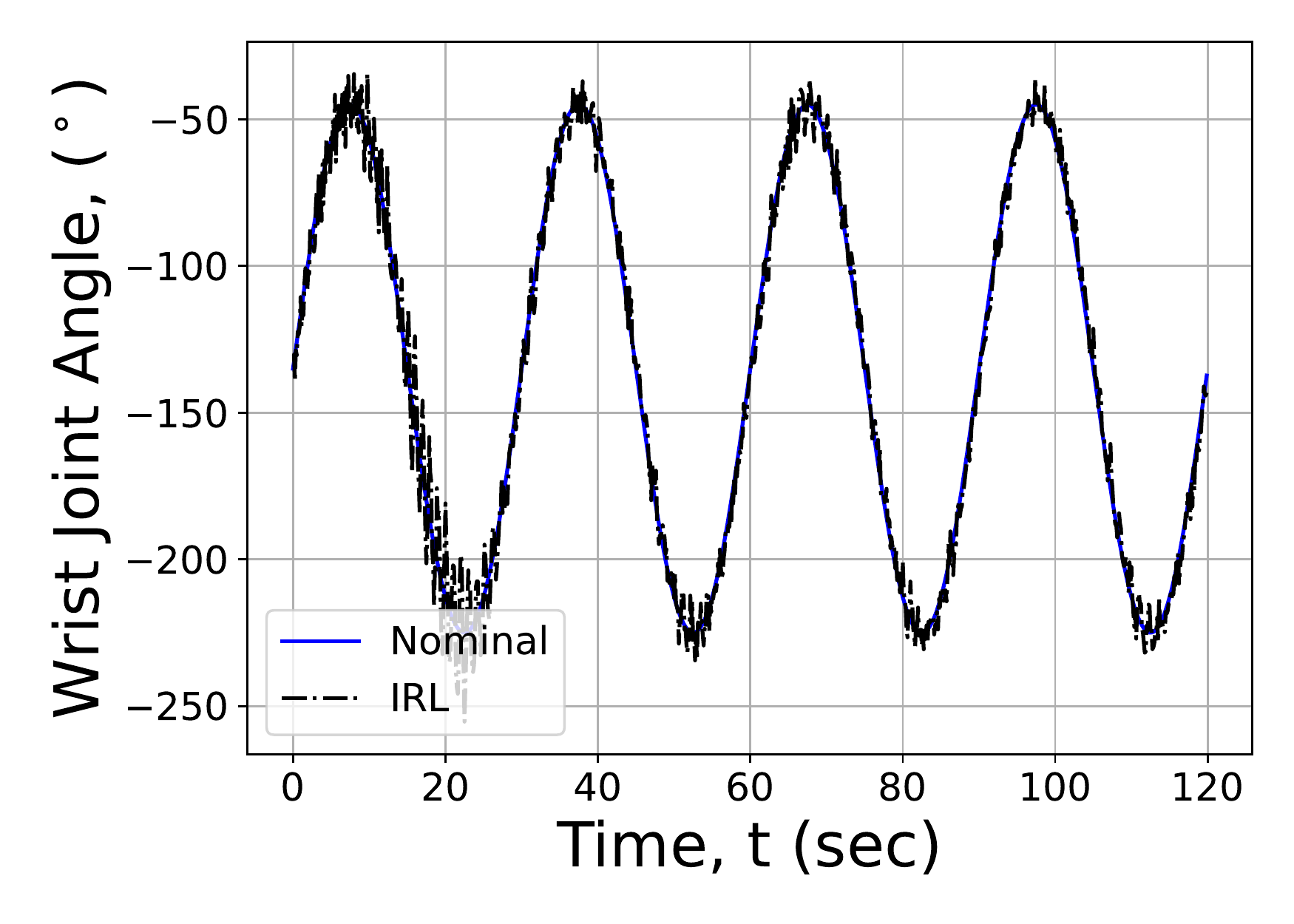}%
  }
  \\[2ex]
  \mbox{}\hfill
  \caption{Joint trajectories for experiment~4.}
  \label{fig:actual_trajectories:noise}
\end{figure}

\begin{figure}[!ht]
  \centering
  \subcaptionbox{Joint~3--Actor~1, $\omega$%
   \label{fig:actor_adaptation:j3_a1_noise}}%
  {%
    \includegraphics[width=0.49\columnwidth]{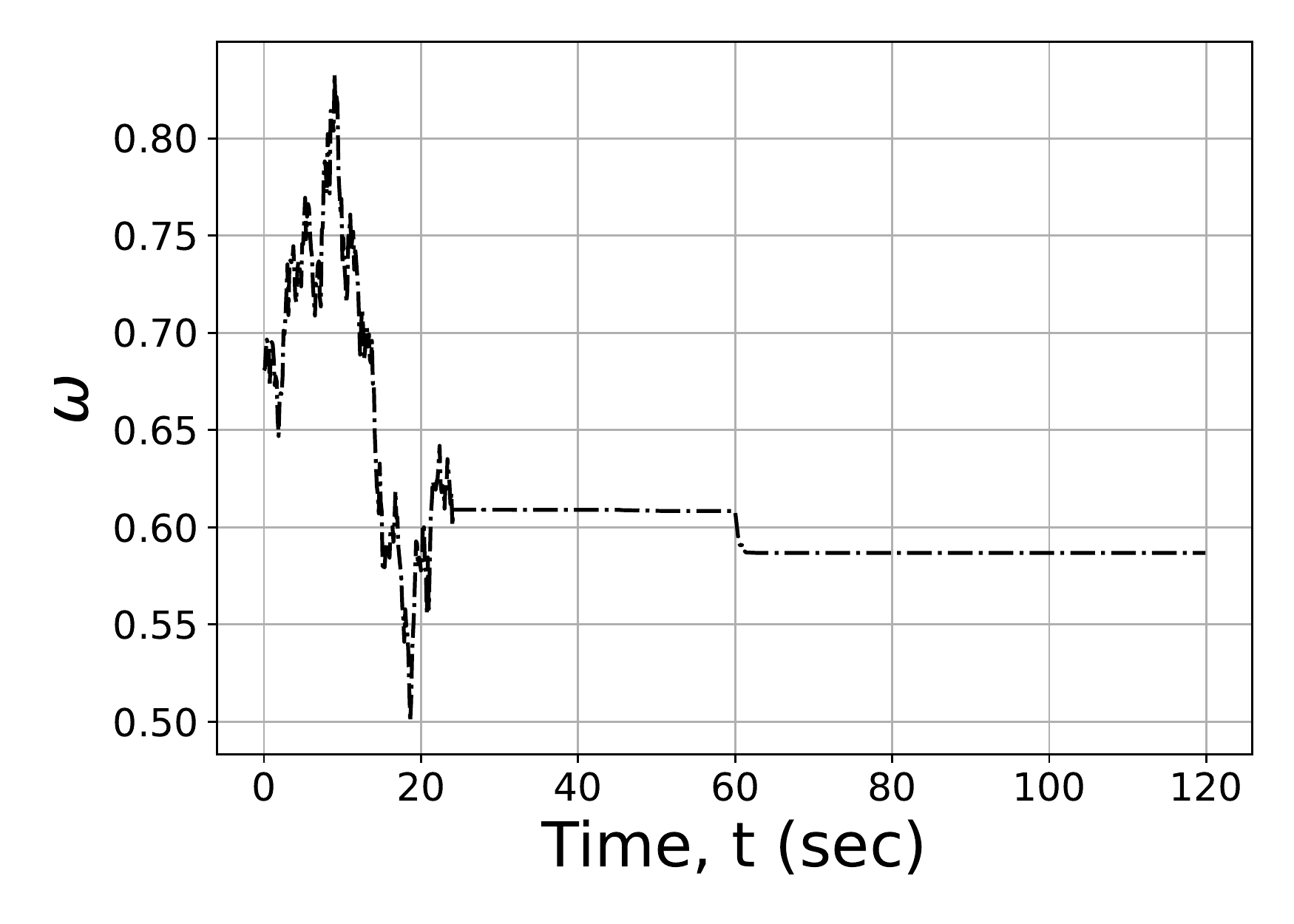}%
  } 
  \hfill%
  \subcaptionbox{Joint~3--Actor~2, $\omega_{\nu}$%
   \label{fig:actor_adaptation:j3_a2_noise}}%
  {%
    \includegraphics[width=0.49\columnwidth]{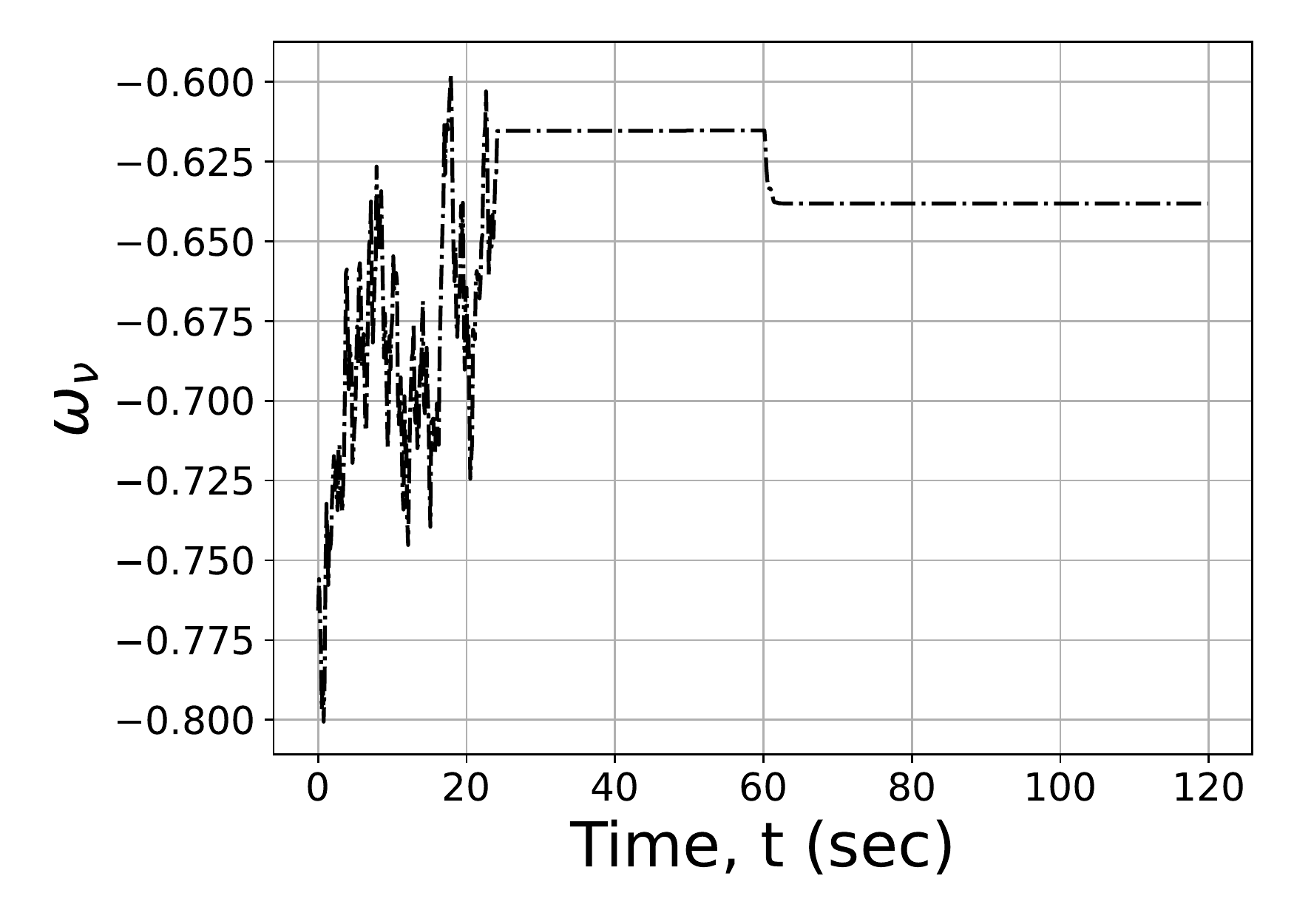}%
  }
  \\[2ex]
  \mbox{}\hfill
  \subcaptionbox{Joint~3--Actor~3, $\omega_{2\nu}$%
   \label{fig:actor_adaptation:j3_a3_noise}}%
  {%
    \includegraphics[width=0.49\columnwidth]{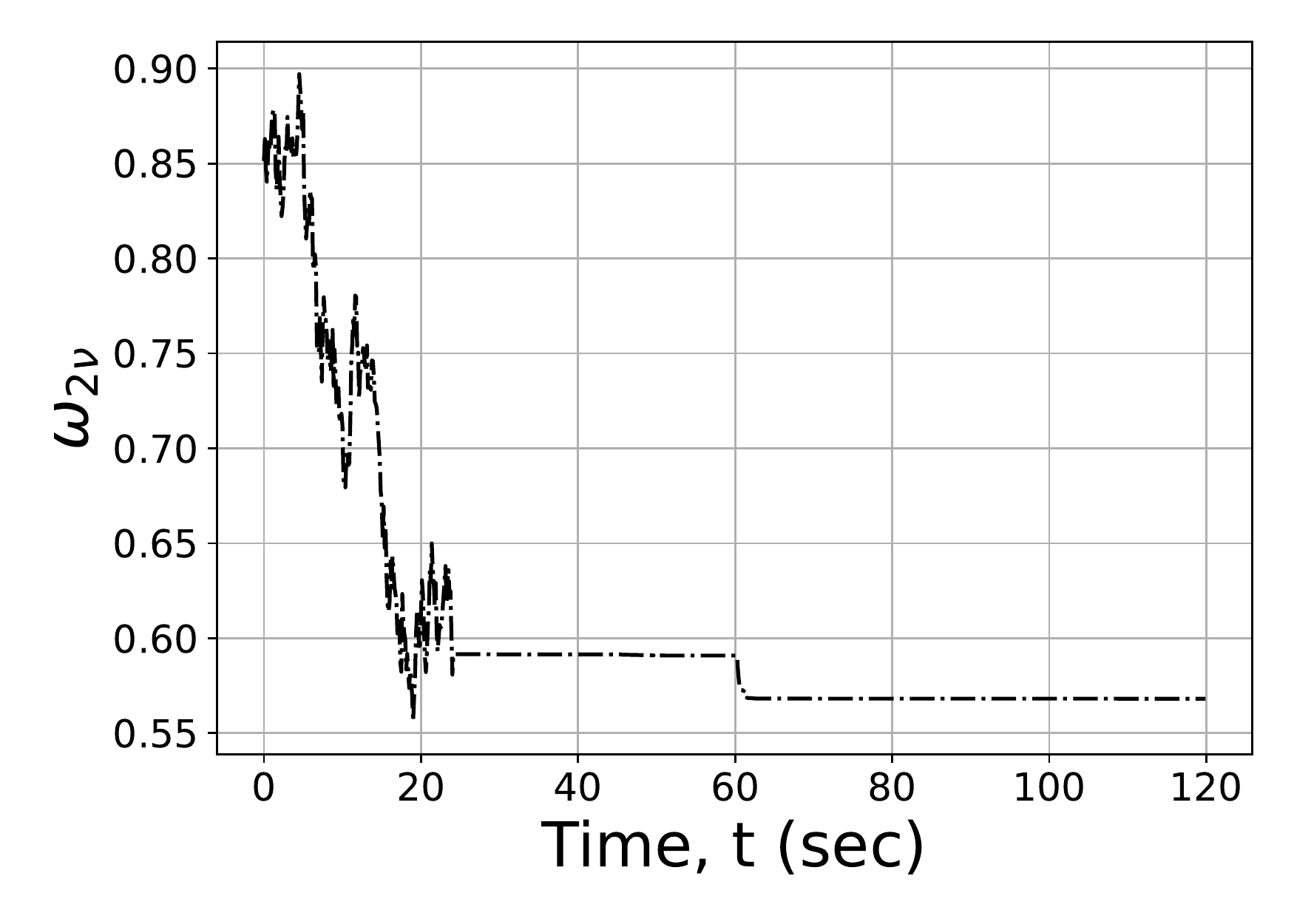}%
  }
     \hfill\mbox{}
  \caption{Adaptation of the elbow actor gains for experiment~4.}  
  \label{fig:actor_adaptation_j3_noise}
\end{figure}

The tracking performance of the IRL algorithm for experiment~4 is shown in \Cref{fig:actual_trajectories:noise}. The adaptations of the elbow actor gains are depicted in \Cref{fig:actor_adaptation_j3_noise}. For completeness, the gain figures for the rest of the joints can be found in {Appendix A}. Despite the extra noise injected in the first \SI{24}{\second} of the experiment, the IRL algorithm maintained a trajectory profile which oscillates closely around the reference-signal during this period of time before it rapidly converges after that. This is particularly clear for joints~3 and~4 (the oscillations in joint~1 are too small to be noticed). The significant oscillations in the shoulder joint between the \SI{20}{\second} and \SI{60}{\second} point marks are due to the dependence of motion between joints~2 and~3. The axes of rotation for these two joints are parallel to one another, implying that any jerk in joint~2 influences joint~3. This is clearly observed by considering the response of joints~2 and~3 between the \SI{45}{\second} and \SI{50}{\second} point marks. During this interval, an increase in oscillation is observed in both joints before the behavior is rectified by the IRL algorithm and the trajectory error is successfully reduced preceding the step change in joint~3 at the \SI{60}{\second} mark. The actor's effort in counter-acting the noise is manifested in the dynamic behavior of its gains during the first \SI{24}{\second}. They smoothly converge soon after that.
For readability, the results of the experiments with longer disturbance periods are included in {Appendix B}.

\begin{figure}[!ht]
	\centering	\includegraphics[width=0.75\columnwidth]{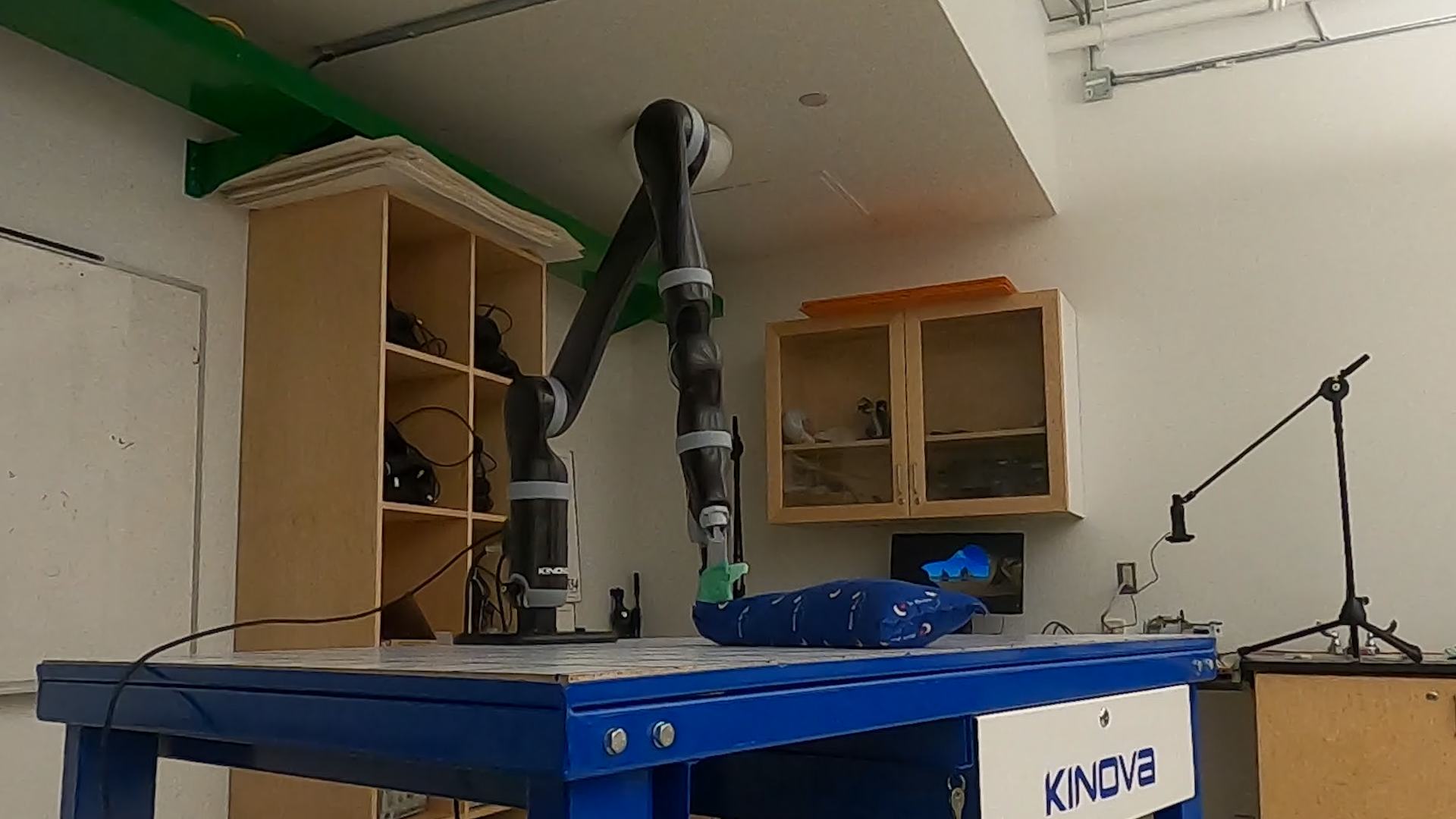}
	\caption{Robot initial position for experiment~5.}
	\label{fig:time_vary_setup}
\end{figure}

\begin{figure}[!ht]
	\centering	\includegraphics[width=0.8\columnwidth]{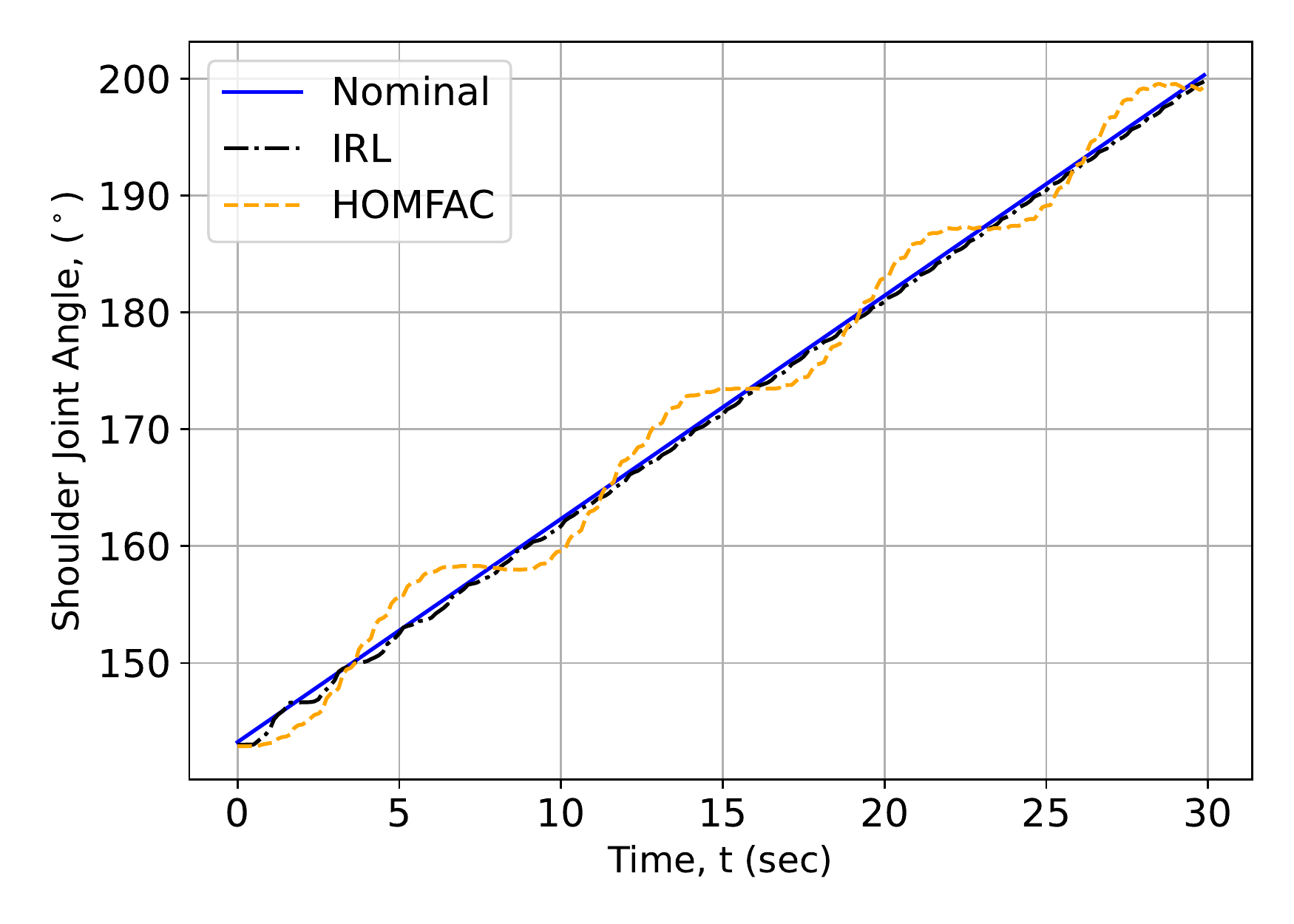}
	\caption{Shoulder trajectories for experiment~5.}
	\label{fig:time_vary_traj}
\end{figure}

\begin{figure}[!ht]
  \centering
  \subcaptionbox{Joint~2--Actor~1, $\omega$%
   \label{fig:actor_adaptation:j2_1_time_vary}}%
  {%
    \includegraphics[width=0.49\columnwidth]{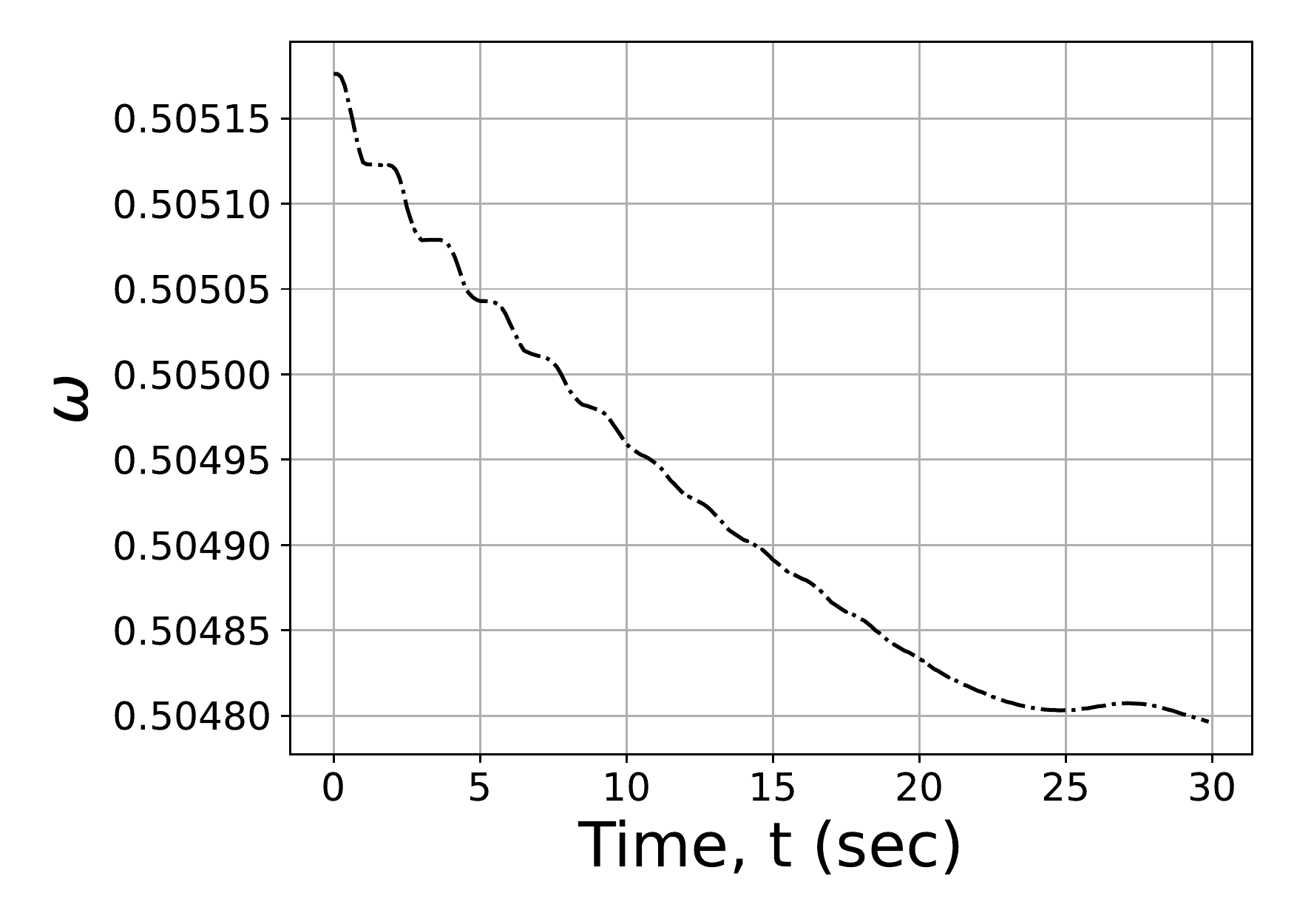}%
  } 
  \hfill%
  \subcaptionbox{Joint~2--Actor~2, $\omega_{\nu}$%
   \label{fig:actor_adaptation:j2_2_time_vary}}%
  {%
    \includegraphics[width=0.49\columnwidth]{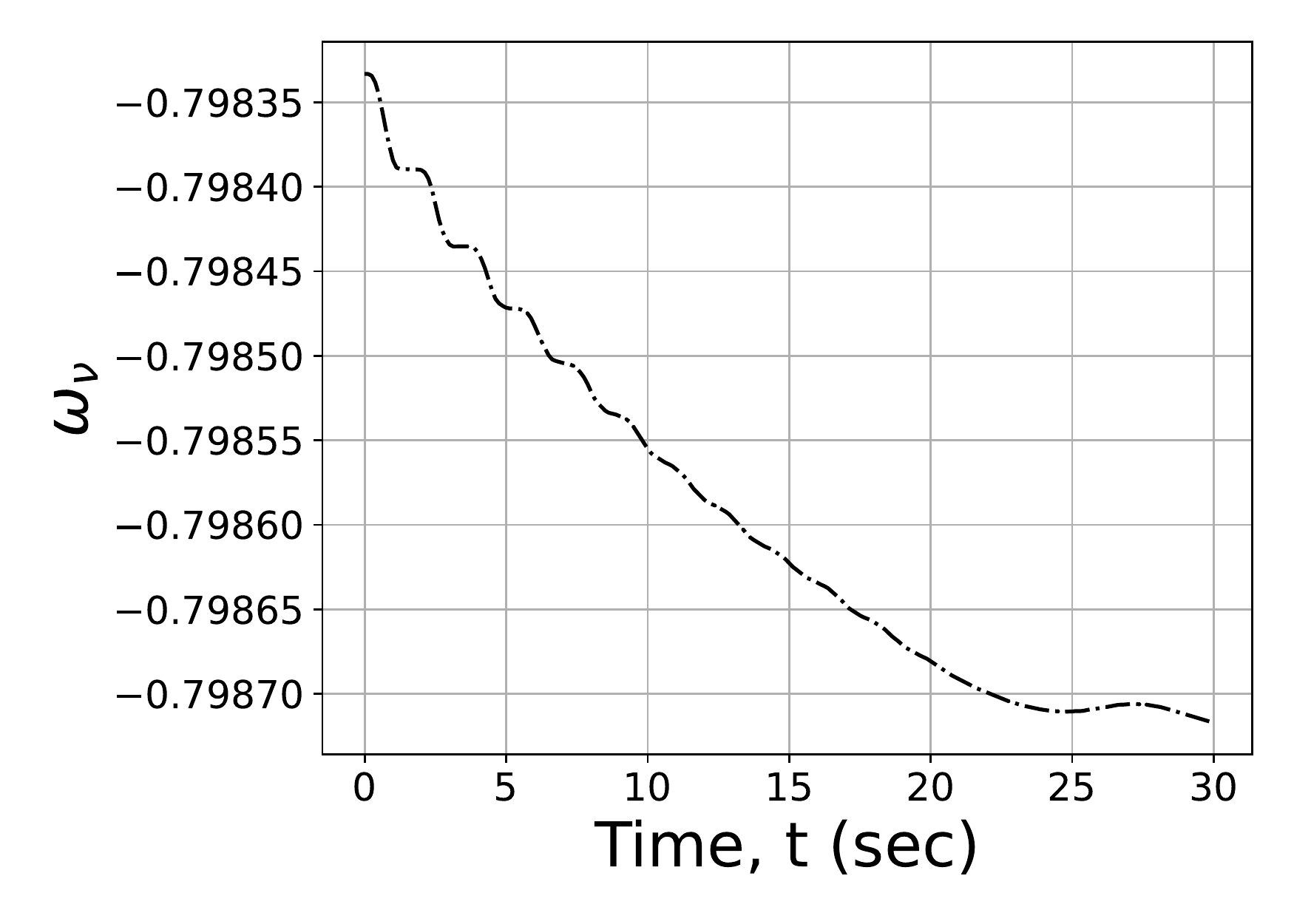}%
  }
  \\[2ex]
  \mbox{}\hfill
  \subcaptionbox{Joint~2--Actor~3, $\omega_{2\nu}$%
   \label{fig:actor_adaptation:j2_3_time_vary}}%
  {%
    \includegraphics[width=0.49\columnwidth]{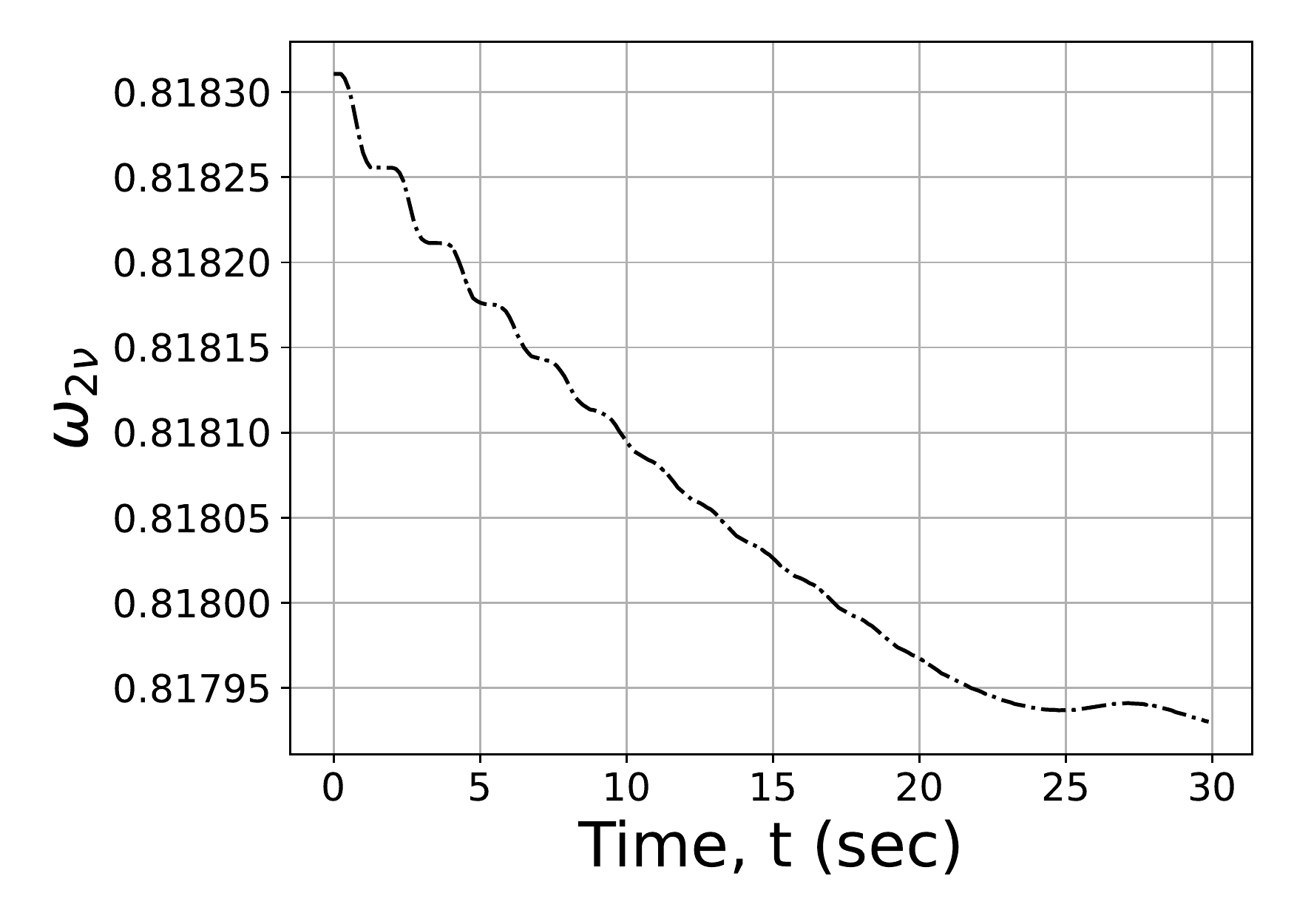}%
  }
     \hfill\mbox{}
  \caption{Adaptation of the shoulder actor gains for experiment~5.}  
  \label{fig:actor_adaptation_j2_time_vary}
\end{figure}

The robot's initial configuration and the results for experiment~5 are visualized in \Cref{fig:time_vary_setup,fig:time_vary_traj}, respectively. These results are consistent with those observed earlier in that, the HOMFAC algorithm seems to ill-cope with continuously changing dynamics (due to the time-varying payload in this case). Nonetheless, the IRL method easily converges despite a few minor initial oscillation cycles. As expected, the learning process of the IRL algorithm remains active while the payload is varying and stabilizes right after it reaches a constant value. This is shown in the variation of the actor gains plotted in \Cref{fig:actor_adaptation_j2_time_vary}.

\section{Conclusion}
\label{sec:Con}

The paper discusses the need for reliable real-time processing schemes of sensor readings using robust machine learning algorithms. To that end, an integral reinforcement learning approach is developed for the control of a class of nonlinear systems. This is accomplished in real-time without prior knowledge of the robot dynamics or explicitly incorporating the desired trajectory in the adapted strategies. The solution is realized using a value iteration process. The convergence and stability characteristics of the adaptive learning solution are formally analyzed. The proposed technique is demonstrated on a 6-DoF Kinova robotic arm and is compared with another model-free controller (HOMFAC). Results show the superiority of the former approach under various dynamics and reference signals, including high-frequency measurement noise.

\clearpage
\appendix

\section{Actors Adaptations for Experiments~1-3}
\label{Appendix A}

\begin{figure}[!htb]
  \centering
  \subcaptionbox{Joint~1--Actor~1, $\omega$%
   \label{fig:actor_adaptation:j1_a1}}%
  {%
    \includegraphics[width=0.49\columnwidth]{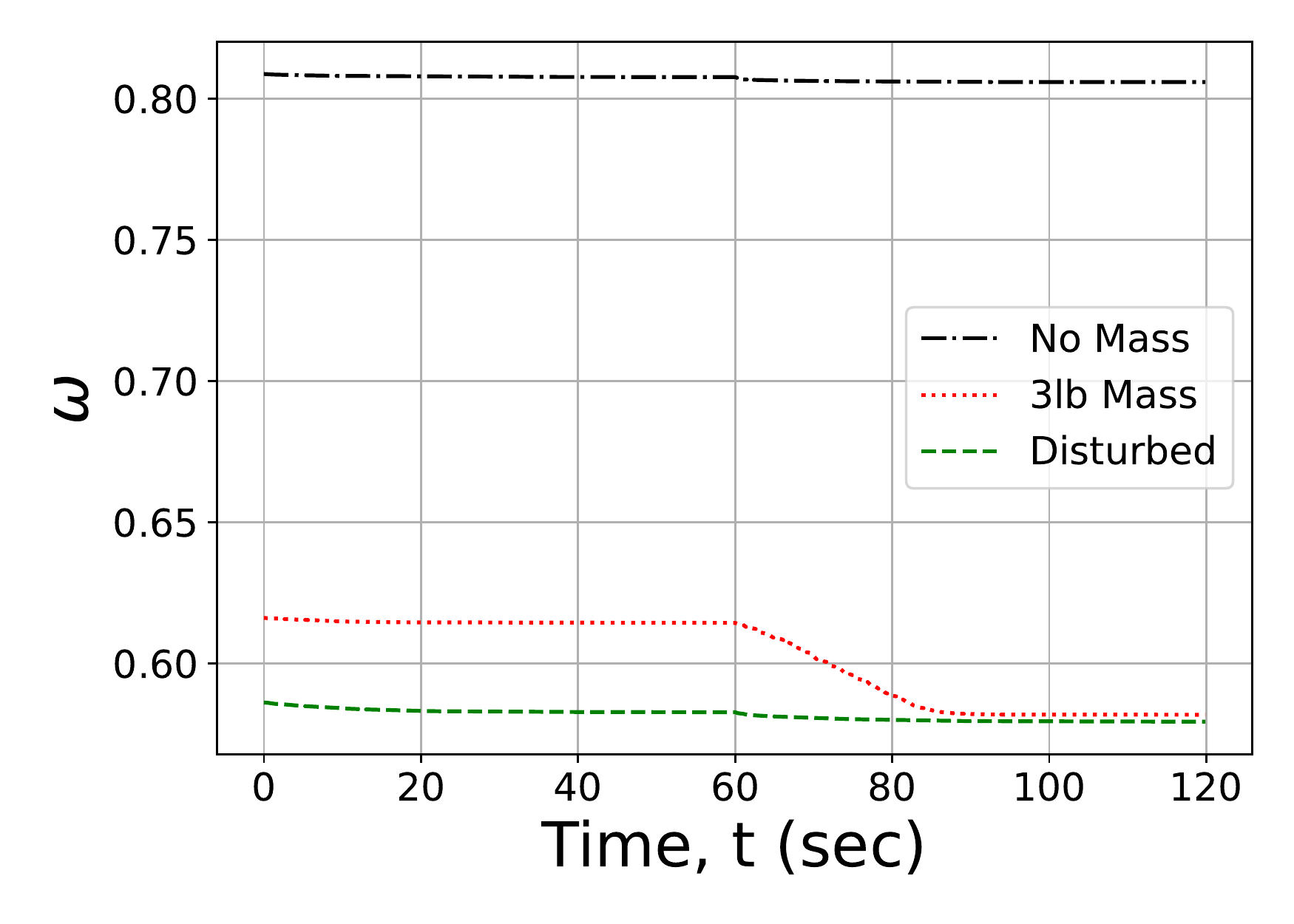}%
  } 
  \hfill%
  \subcaptionbox{Joint~1--Actor~2, $\omega_{\nu}$%
   \label{fig:actor_adaptation:j1_a2}}%
  {%
    \includegraphics[width=0.49\columnwidth]{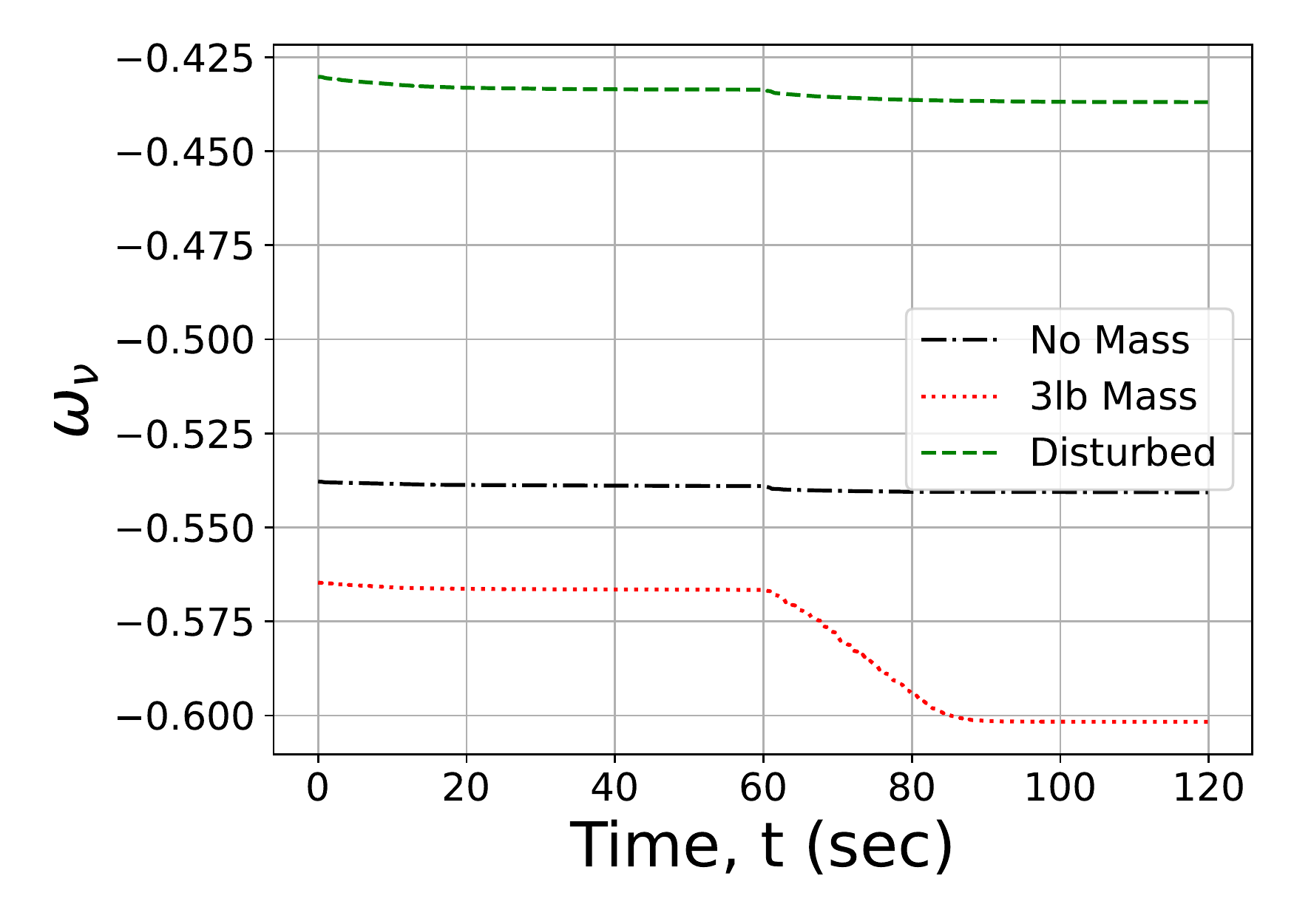}%
  }
  \\[2ex]
  \mbox{}\hfill
  \subcaptionbox{Joint~1--Actor~3, $\omega_{2\nu}$%
   \label{fig:actor_adaptation:j1_a3}}%
  {%
    \includegraphics[width=0.49\columnwidth]{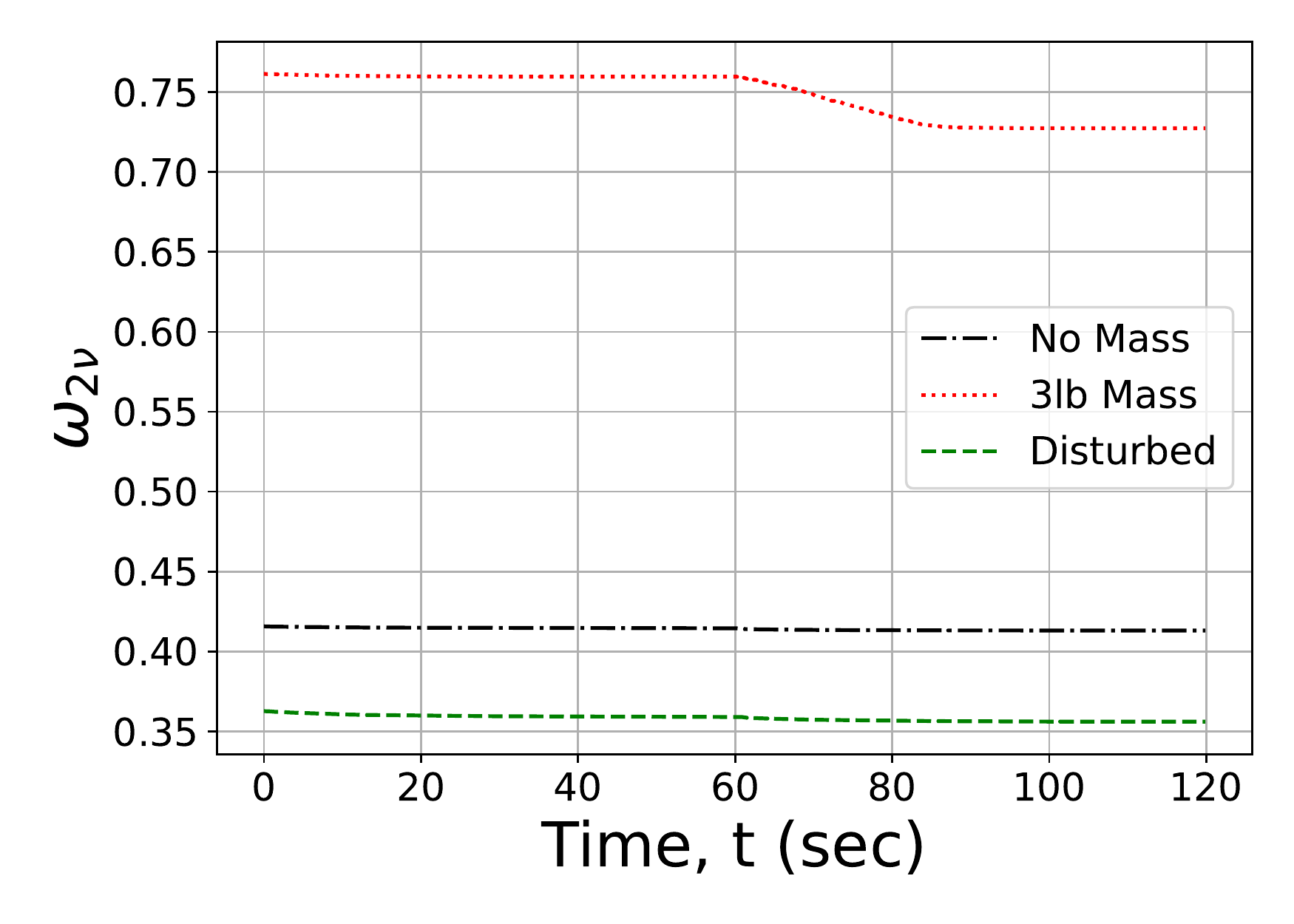}%
  }
     \hfill\mbox{}
  \caption{Adaptation of the base actor gains for the first three experiments.}
  \label{fig:actor_adaptation_j1}
\end{figure}

\begin{figure}[!htb]
  \centering
  \subcaptionbox{Joint~2--Actor~1, $\omega$%
   \label{fig:actor_adaptation:j2_a1}}%
  {%
    \includegraphics[width=0.49\columnwidth]{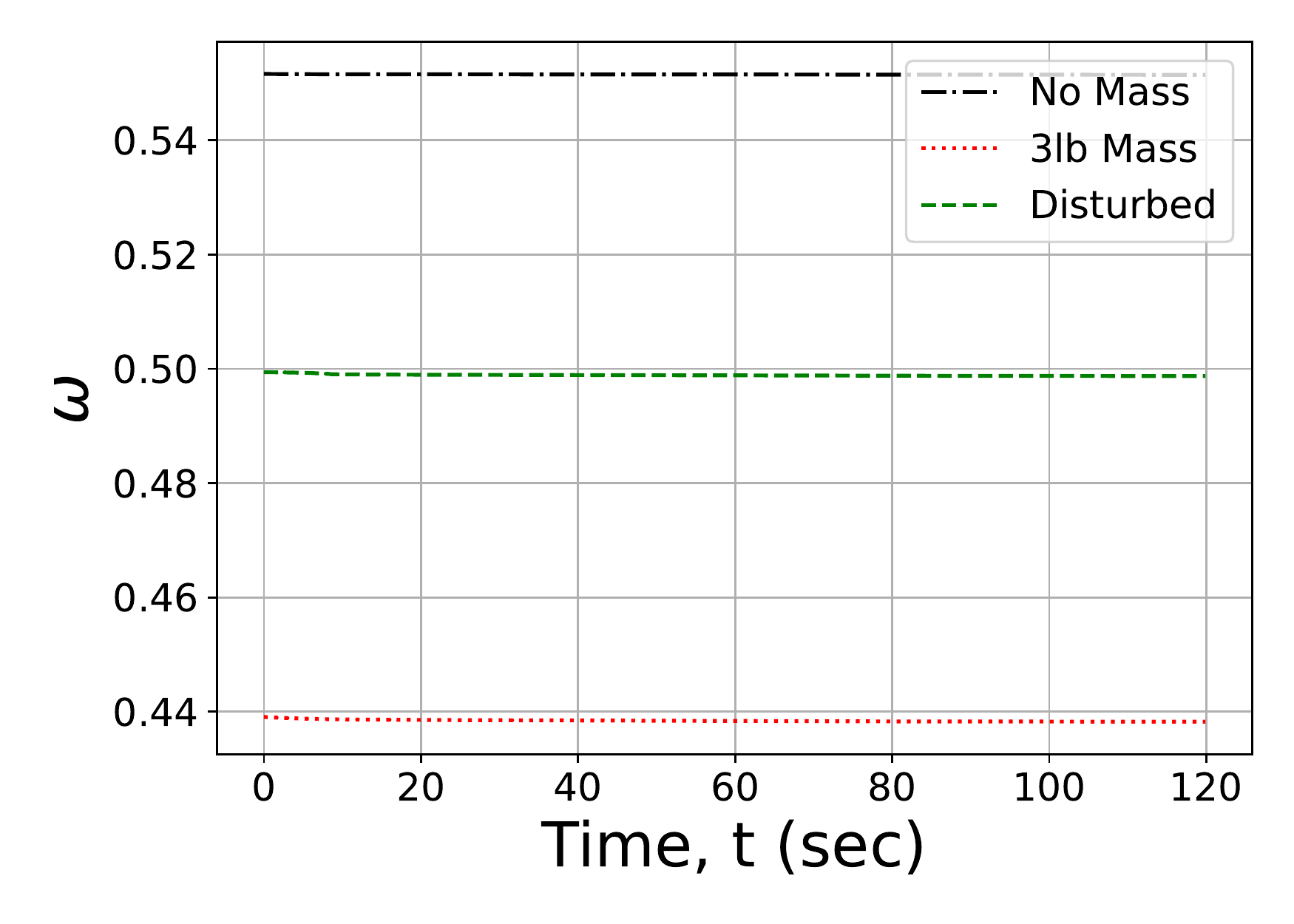}%
  } 
  \hfill%
  \subcaptionbox{Joint~2--Actor~2, $\omega_{\nu}$%
   \label{fig:actor_adaptation:j2_a2}}%
  {%
    \includegraphics[width=0.49\columnwidth]{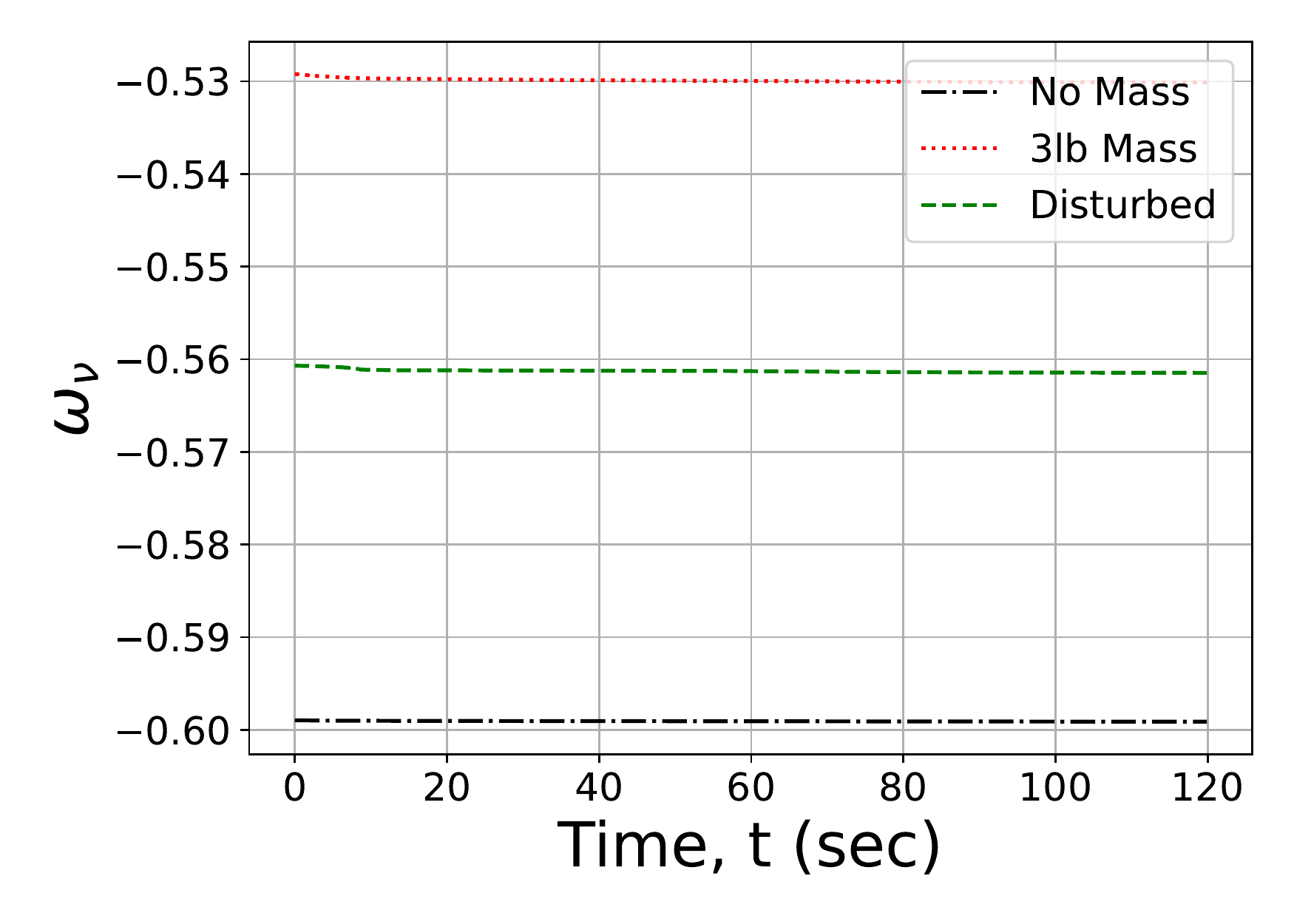}%
  }
  \\[2ex]
  \mbox{}\hfill
  \subcaptionbox{Joint~2--Actor~3, $\omega_{2\nu}$%
   \label{fig:actor_adaptation:j2_a3}}%
  {%
    \includegraphics[width=0.49\columnwidth]{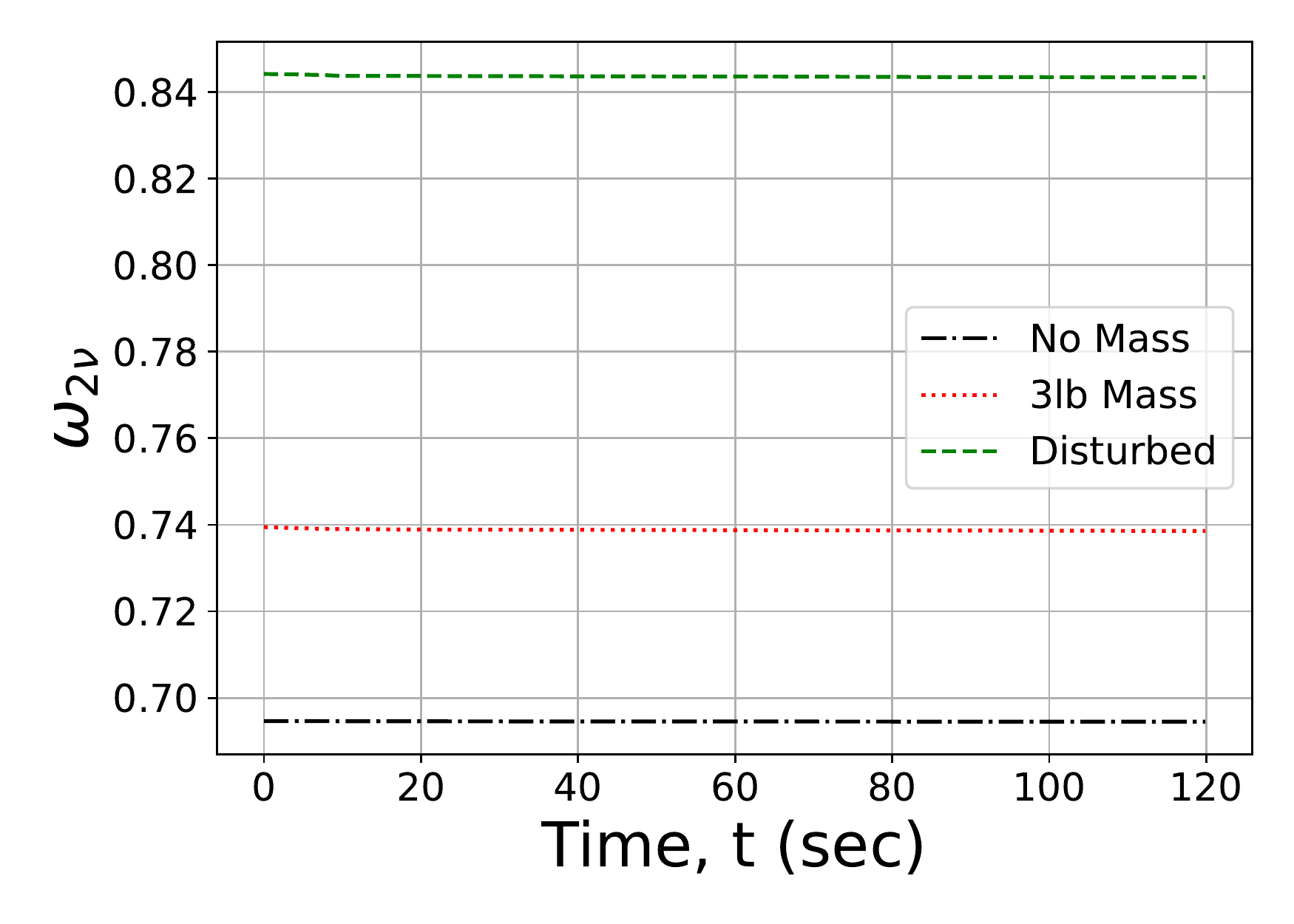}%
  }
     \hfill\mbox{}
  \caption{Adaptation of the shoulder actor gains for the first three experiments.}
  \label{fig:actor_adaptation_j2}
\end{figure}

\begin{figure}[!htb]
  \centering
  \subcaptionbox{Joint~3--Actor~1, $\omega$%
   \label{fig:actor_adaptation:j3_a1}}%
  {%
    \includegraphics[width=0.49\columnwidth]{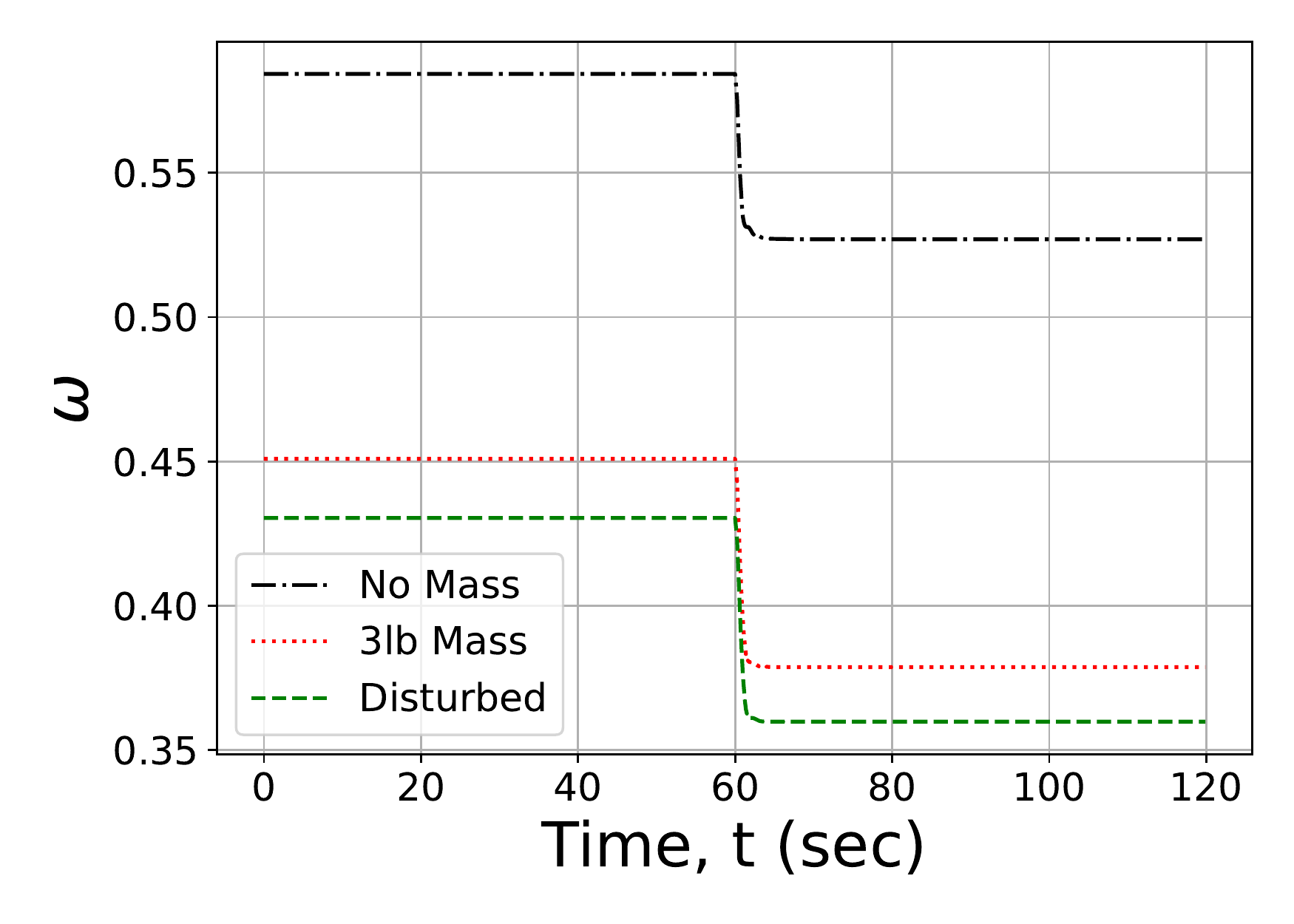}%
  } 
  \hfill%
  \subcaptionbox{Joint~3--Actor~2, $\omega_{\nu}$%
   \label{fig:actor_adaptation:j3_a2}}%
  {%
    \includegraphics[width=0.49\columnwidth]{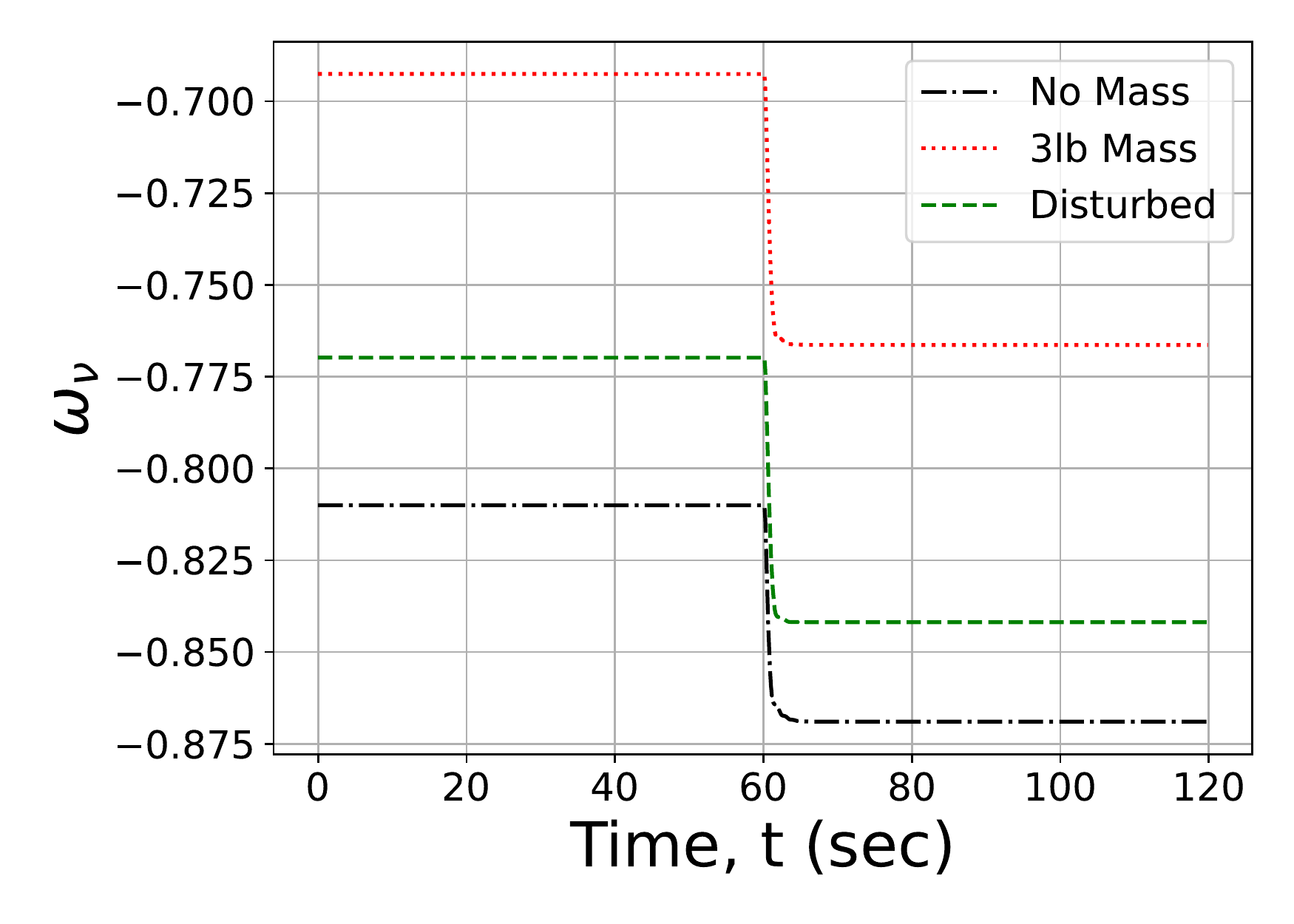}%
  }
  \\[2ex]
  \mbox{}\hfill
  \subcaptionbox{Joint~3--Actor~3, $\omega_{2\nu}$%
   \label{fig:actor_adaptation:j3_a3}}%
  {%
    \includegraphics[width=0.49\columnwidth]{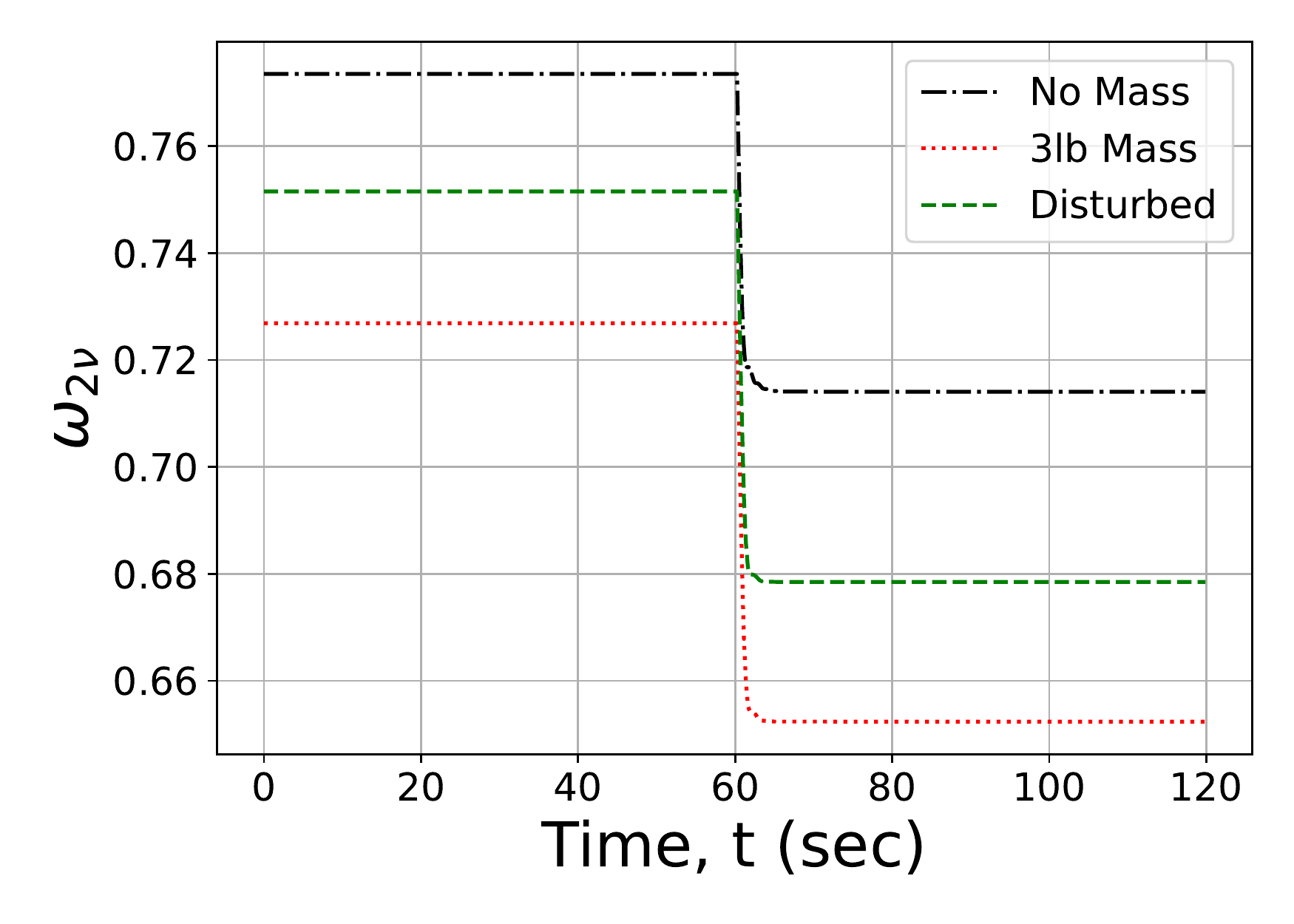}%
  }
     \hfill\mbox{}     
  \caption{Adaptation of the elbow actor gains for the first three experiments.}
  \label{fig:actor_adaptation_j3}
\end{figure}

\begin{figure}[!htb]
  \centering
  \subcaptionbox{Joint~4--Actor~1, $\omega$%
   \label{fig:actor_adaptation:j4_a1}}%
  {%
    \includegraphics[width=0.49\columnwidth]{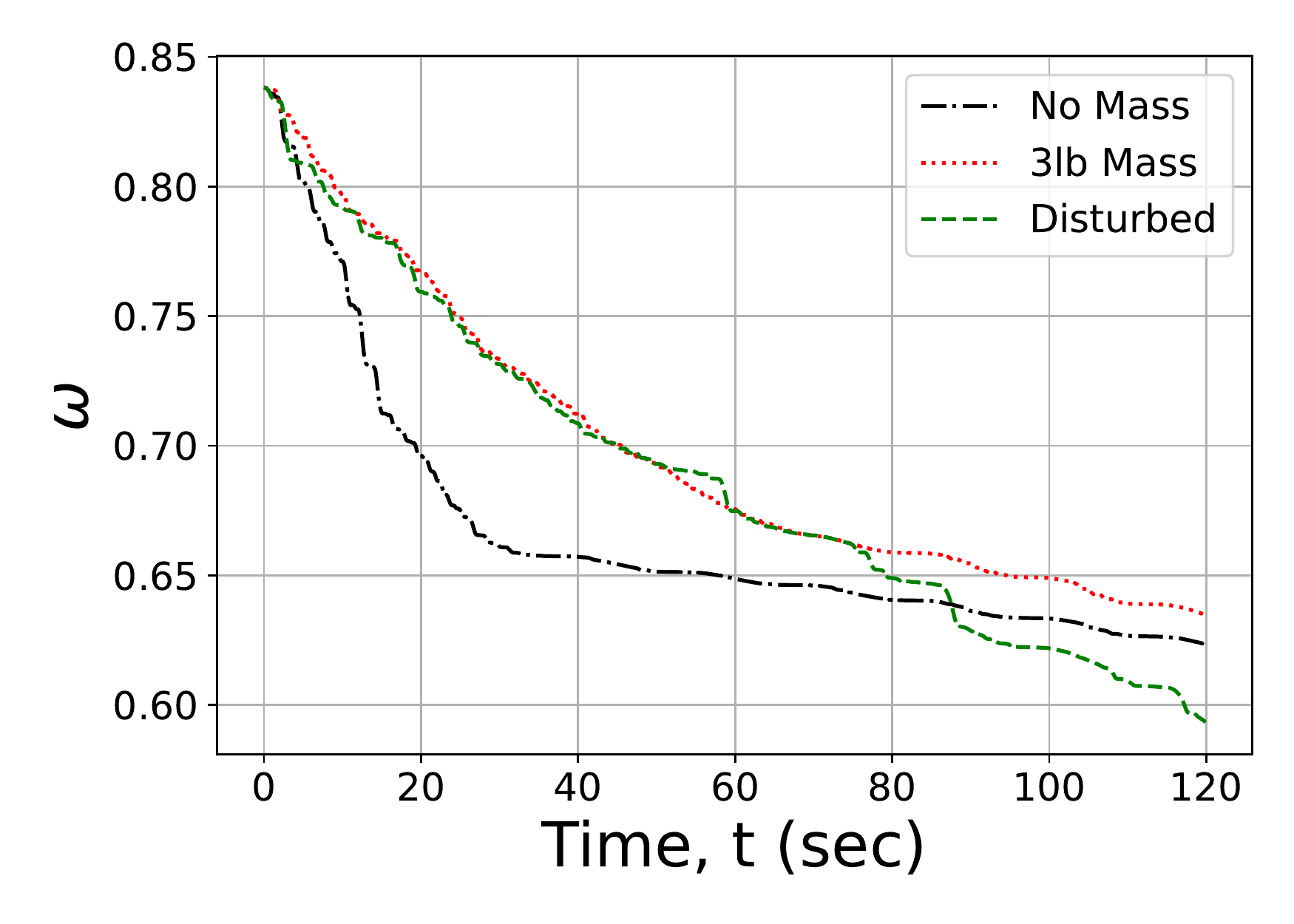}%
  } 
  \hfill%
  \subcaptionbox{Joint~4--Actor~2, $\omega_{\nu}$%
   \label{fig:actor_adaptation:j4_a2}}%
  {%
    \includegraphics[width=0.49\columnwidth]{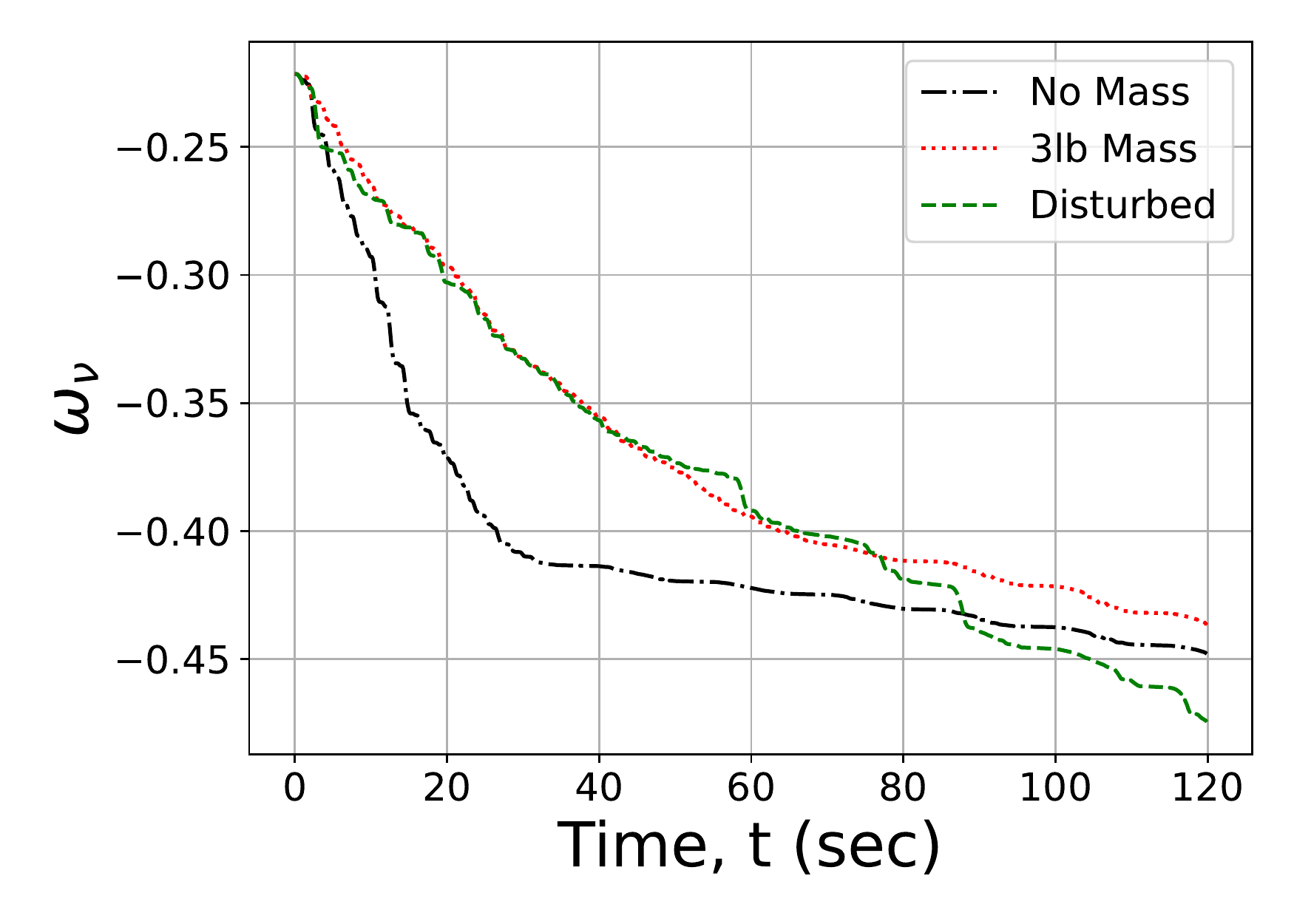}%
  }
  \\[2ex]
  \mbox{}\hfill
  \subcaptionbox{Joint~4--Actor~3, $\omega_{2\nu}$%
   \label{fig:actor_adaptation:j4_a3}}%
  {%
    \includegraphics[width=0.49\columnwidth]{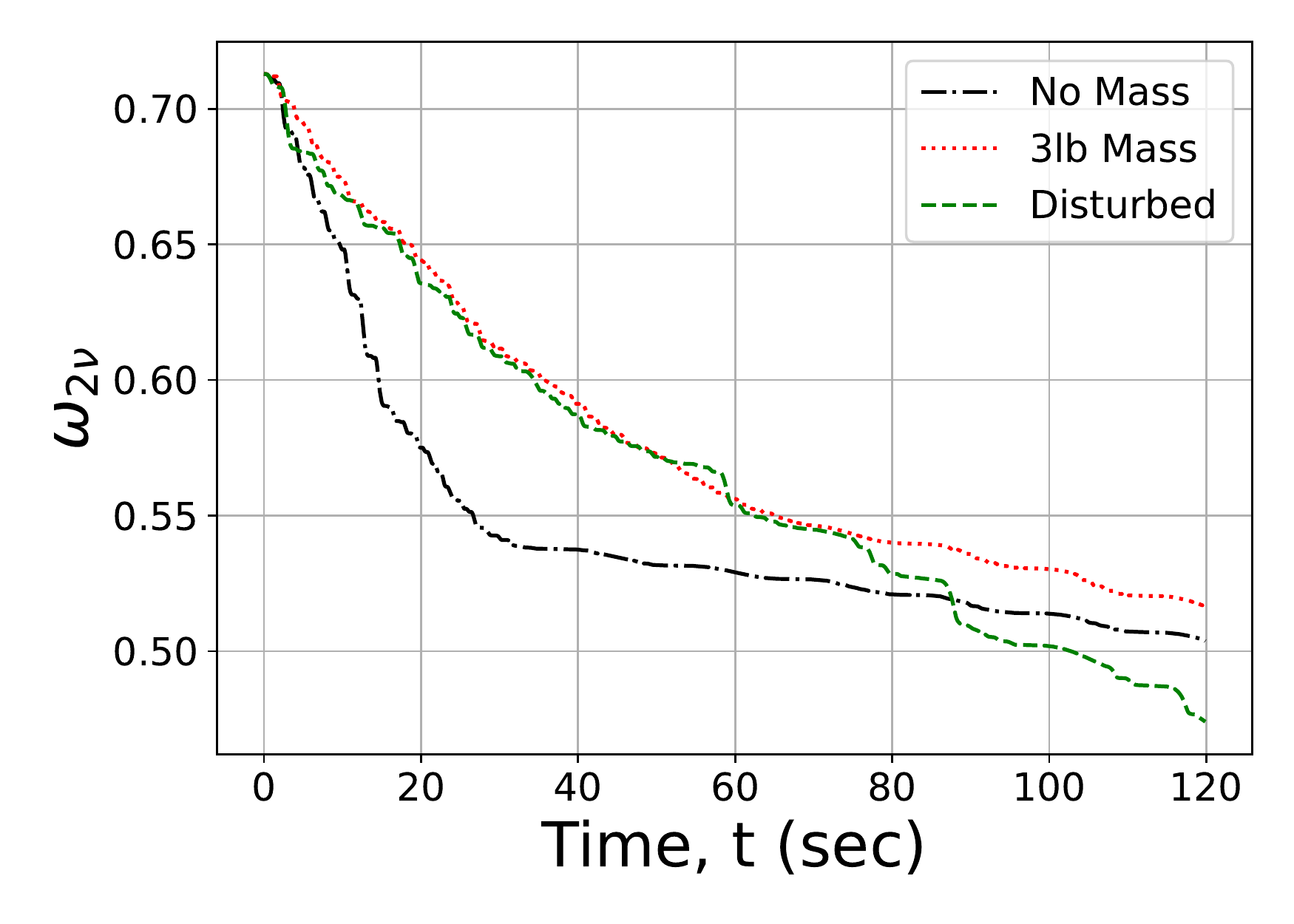}%
  }
     \hfill\mbox{}
  \caption{Adaptation of the wrist actor gains for the first three experiments.}
  \label{fig:actor_adaptation_j4}
\end{figure}

\mbox{}
\clearpage

\section{Supplementary Data for Noise Experiments}
\label{Appendix B}

\begin{figure}[!htb]
  \centering
  \subcaptionbox{Joint~1%
   \label{fig:actual_trajectories:40j1_noise}}%
  {%
    \includegraphics[width=0.49\columnwidth]{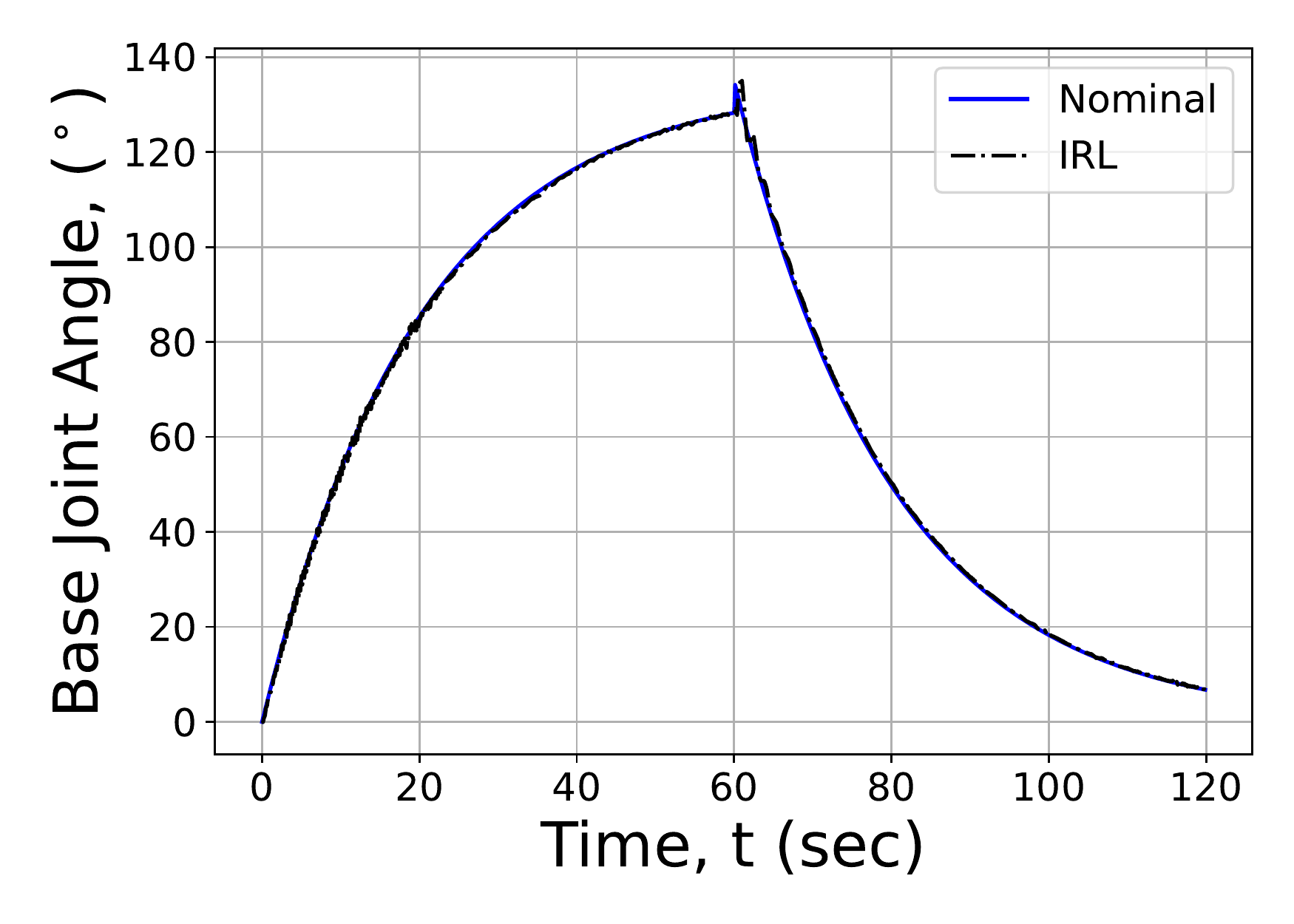}%
  } 
  \hfill%
  \subcaptionbox{Joint~2%
   \label{fig:actual_trajectories:40j2_noise}}%
  {%
    \includegraphics[width=0.49\columnwidth]{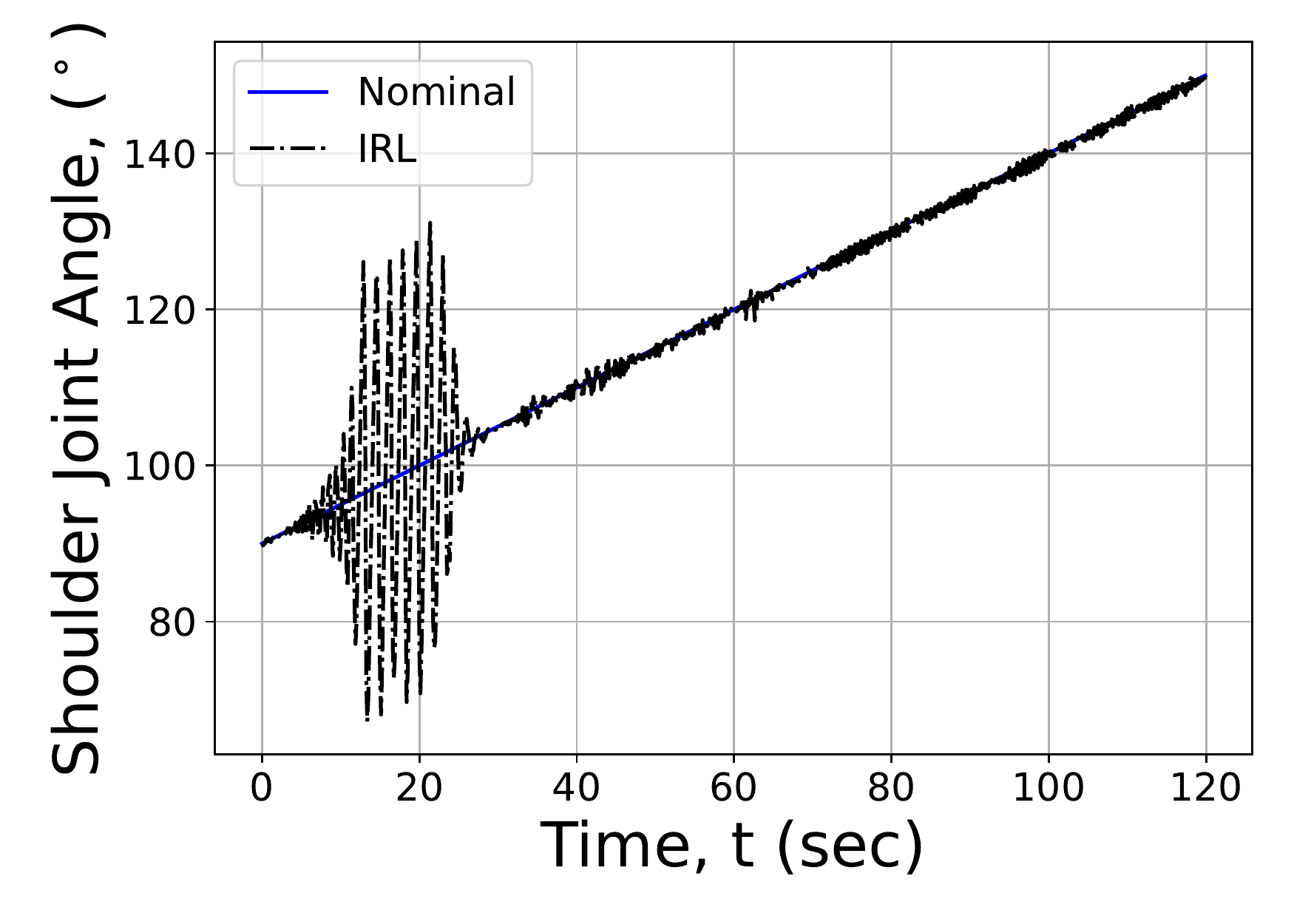}%
  }
  \\[2ex]
  \mbox{}\hfill
  \subcaptionbox{Joint~3%
    \label{fig:actual_trajectories:40j3_noise}}%
  {%
    \includegraphics[width=0.49\columnwidth]{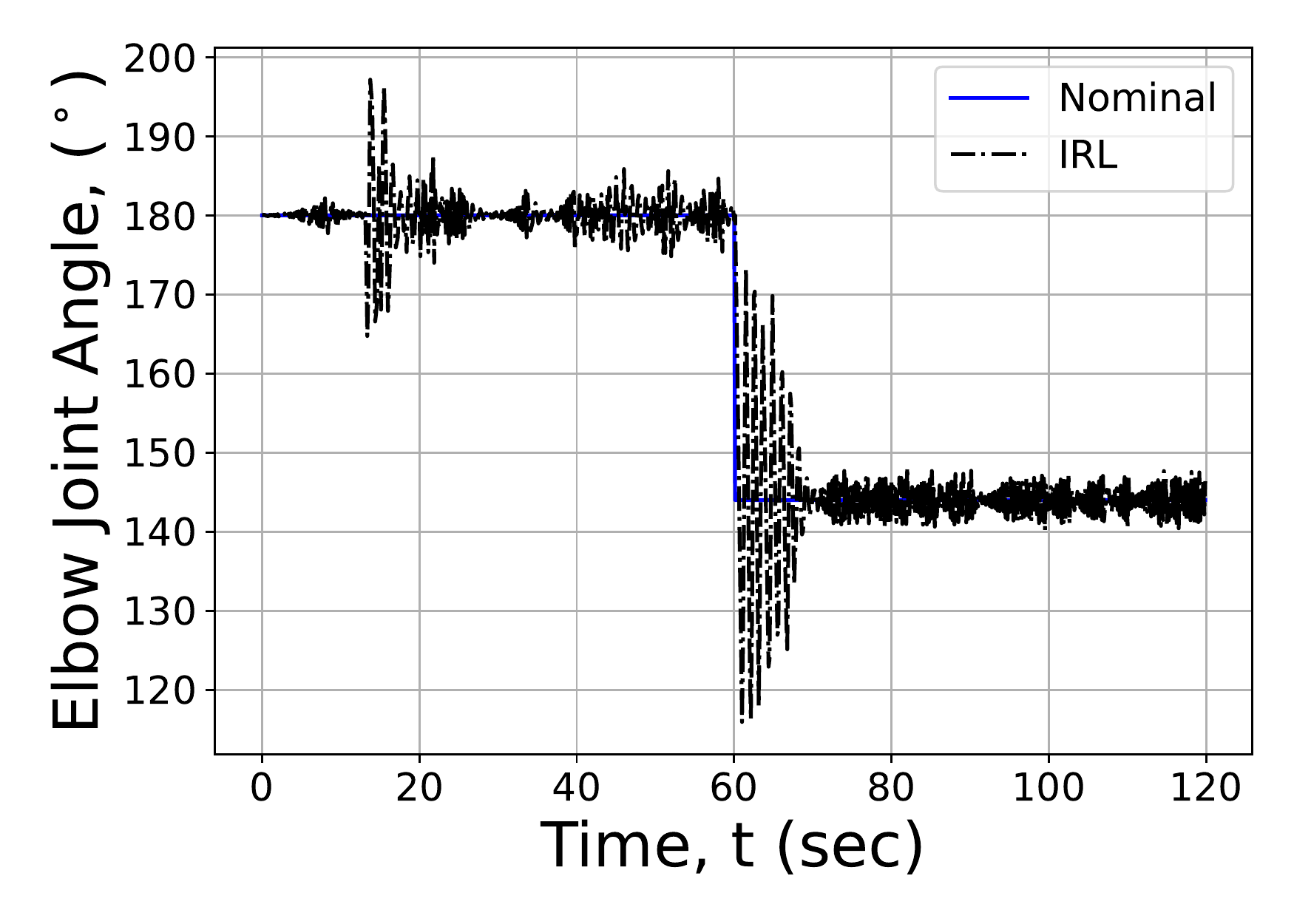}%
  }
     \hfill
  \subcaptionbox{Joint~4%
   \label{fig:actual_trajectories:j4_noise}}%
  {%
    \includegraphics[width=0.49\columnwidth]{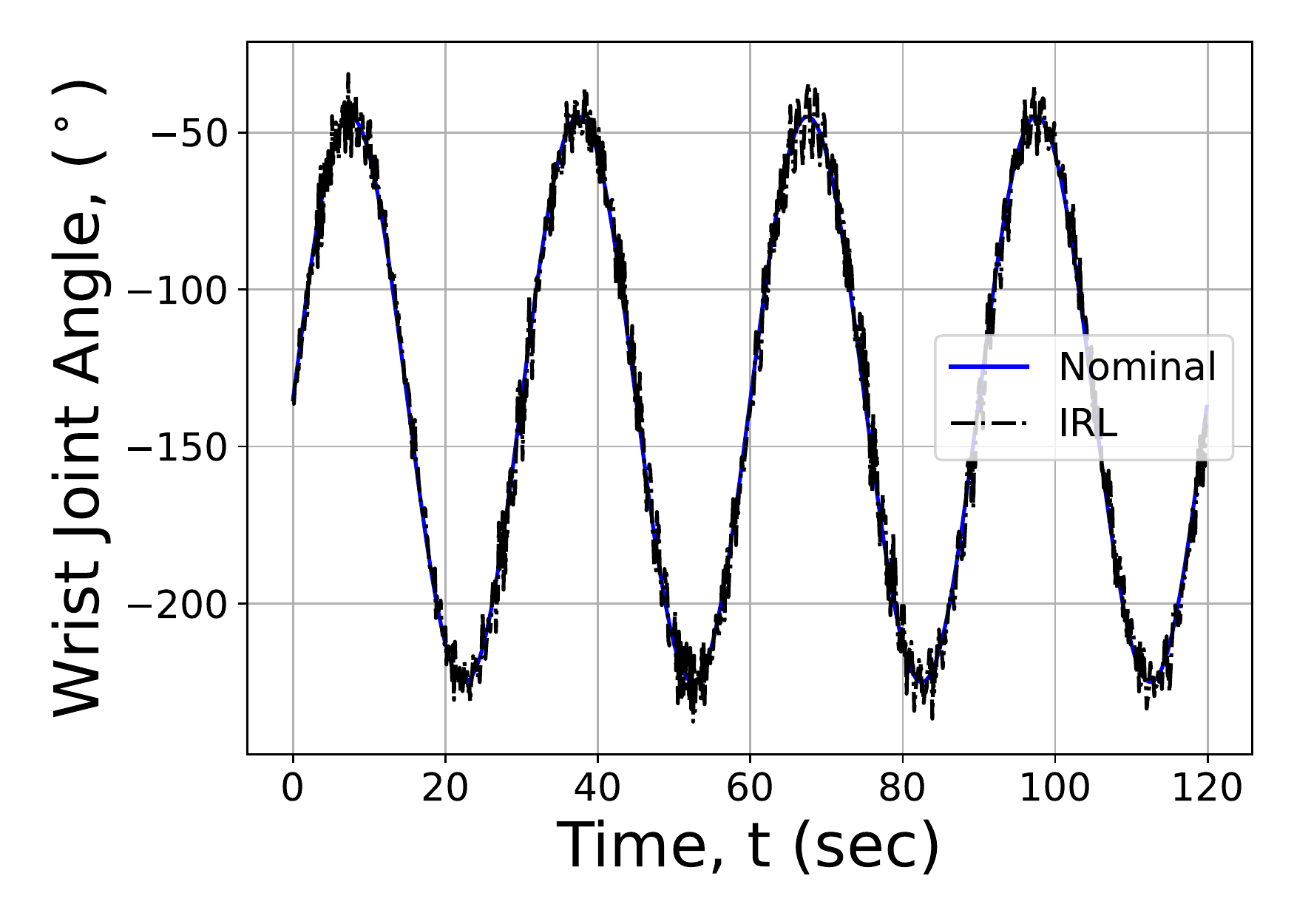}%
  }
  \\[2ex]
  \mbox{}\hfill
  \caption{Joint trajectories for experiment~4 with noise added to actor weights over the first 40\% of the experiment.}
  \label{fig:actual_trajectories:40noise}
\end{figure}

\begin{figure}[!htb]
  \centering
  \subcaptionbox{Joint~1%
   \label{fig:actual_trajectories:60j1_noise}}%
  {%
    \includegraphics[width=0.49\columnwidth]{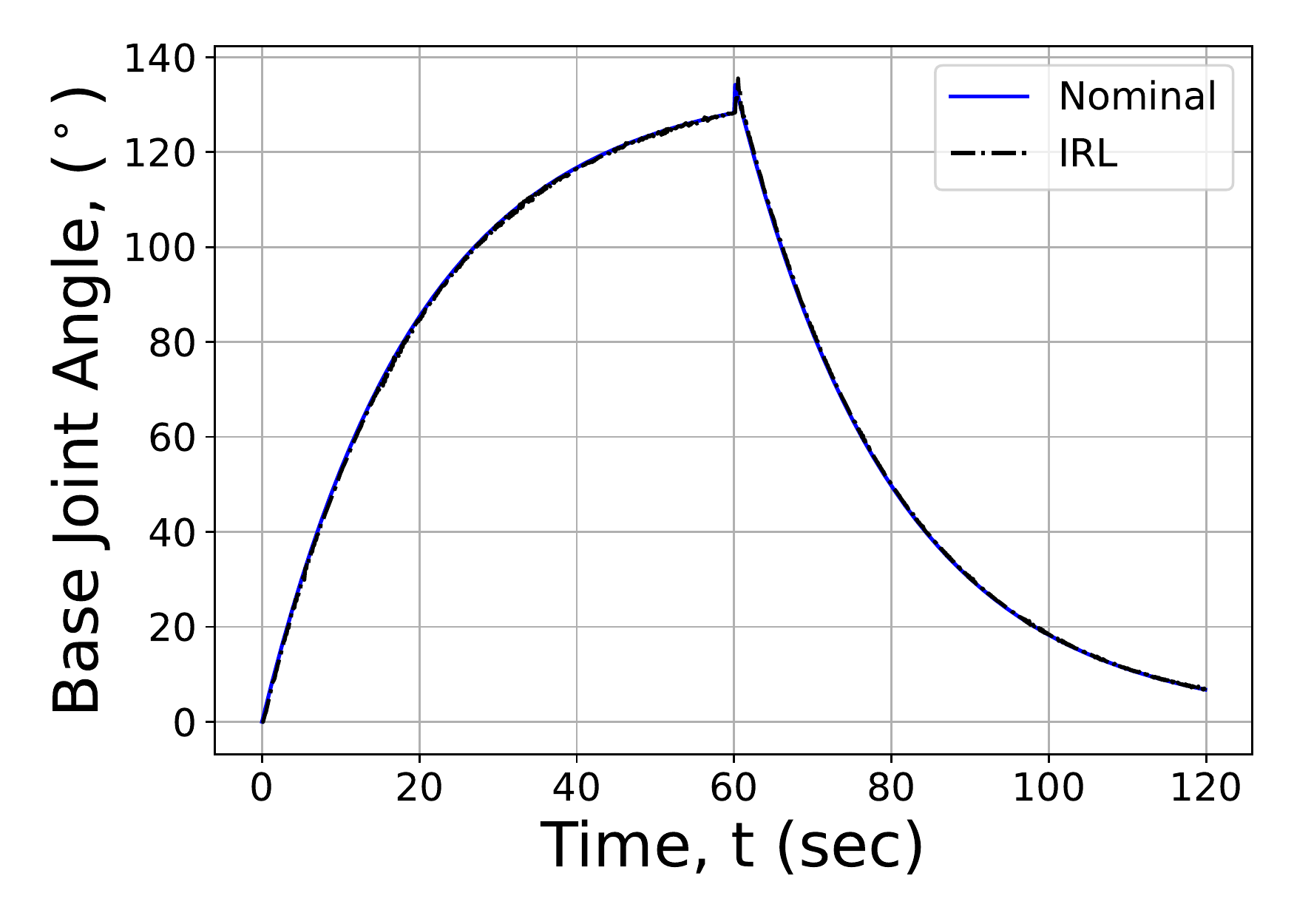}%
  } 
  \hfill%
  \subcaptionbox{Joint~2%
   \label{fig:actual_trajectories:60j2_noise}}%
  {%
    \includegraphics[width=0.49\columnwidth]{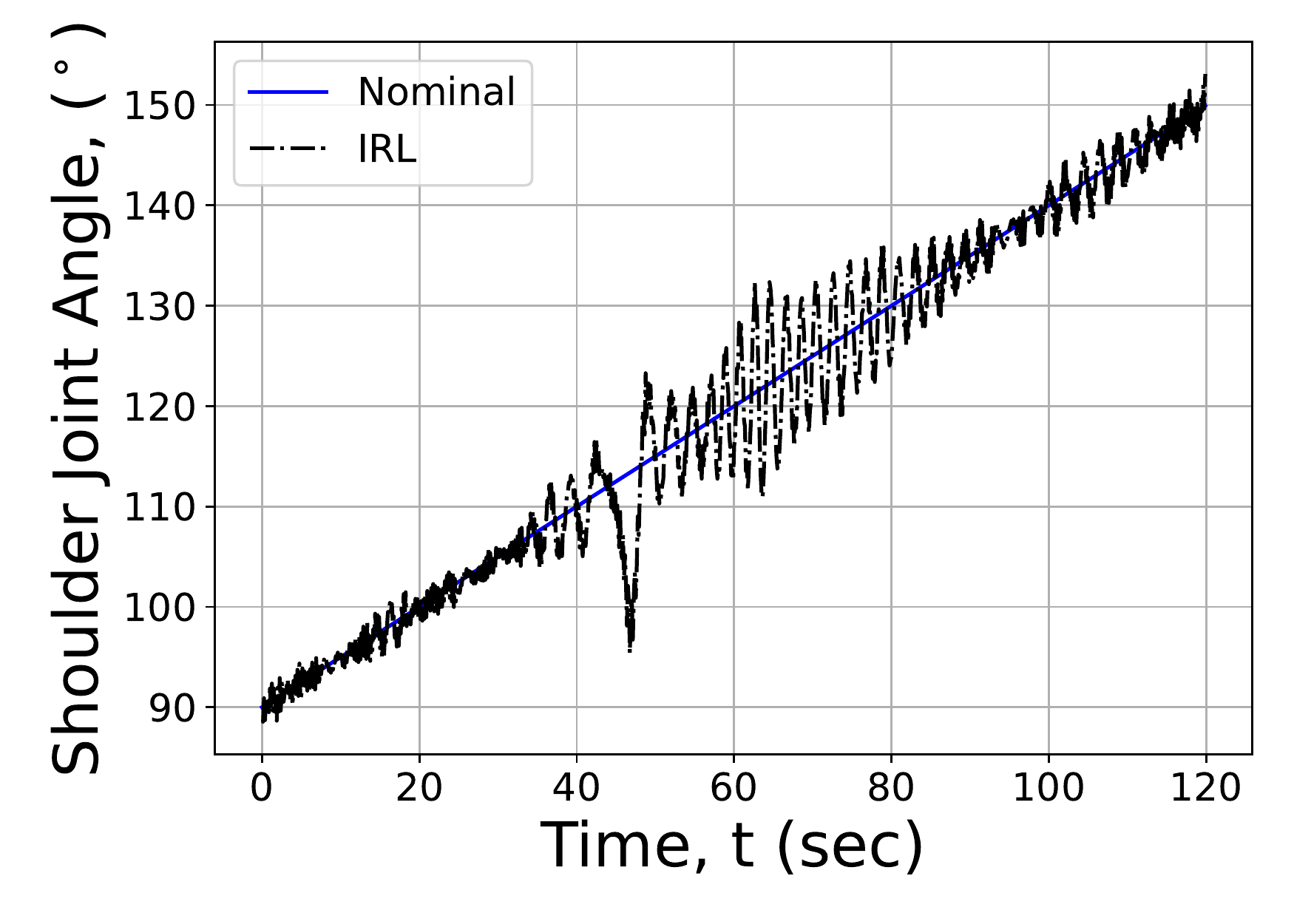}%
  }
  \\[2ex]
  \mbox{}\hfill
  \subcaptionbox{Joint~3%
    \label{fig:actual_trajectories:60j3_noise}}%
  {%
    \includegraphics[width=0.49\columnwidth]{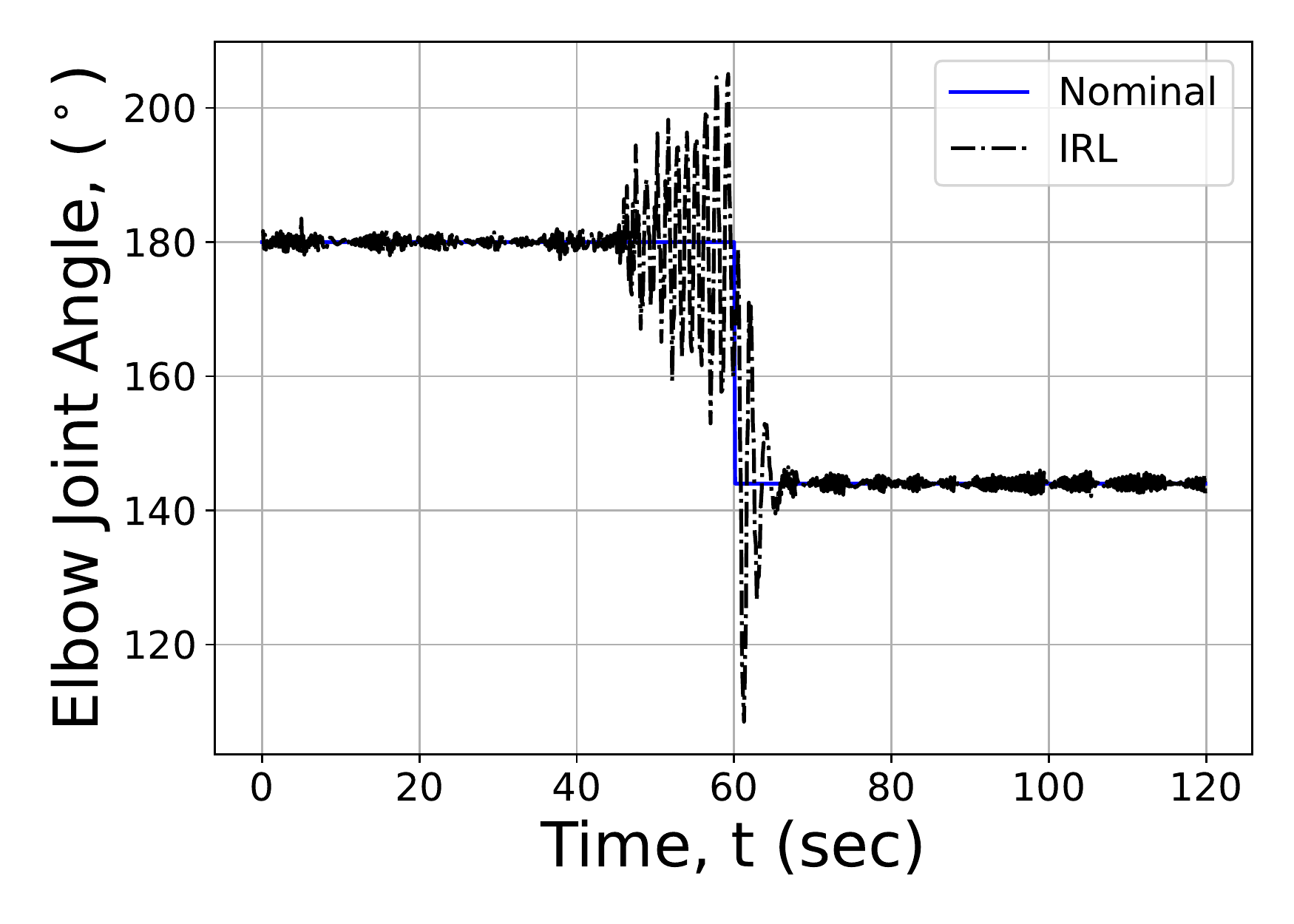}%
  }
     \hfill
  \subcaptionbox{Joint~4%
   \label{fig:actual_trajectories:60j4_noise}}%
  {%
    \includegraphics[width=0.49\columnwidth]{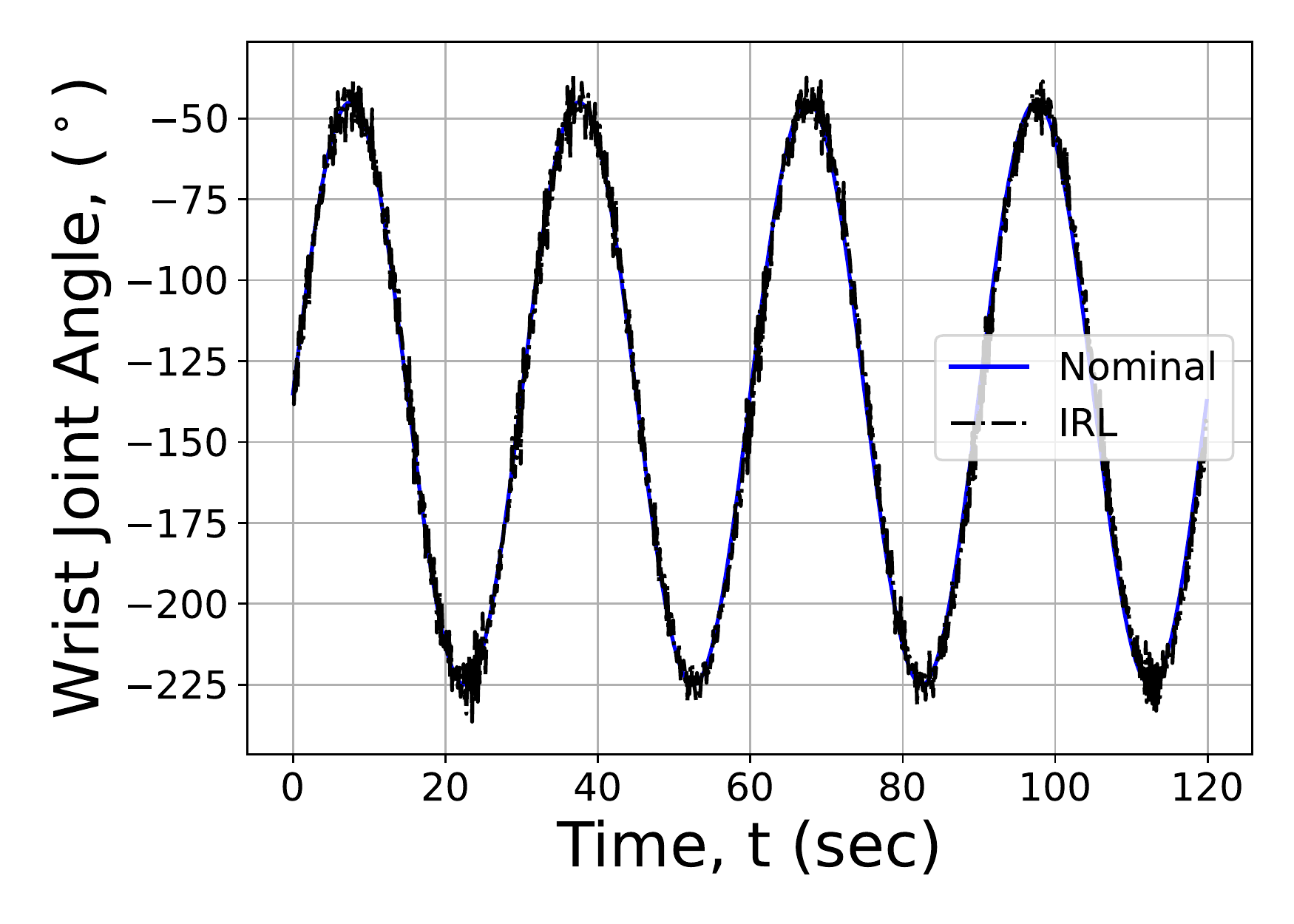}%
  }
  \\[2ex]
  \mbox{}\hfill
  \caption{Joint trajectories for experiment~4 with noise added to actor weights over the first 60\% of the experiment.}
  \label{fig:actual_trajectories:60noise}
\end{figure}

\begin{figure}[!htb]
  \centering
  \subcaptionbox{Joint~1%
   \label{fig:actual_trajectories:fullj1_noise}}%
 {%
   \includegraphics[width=0.49\columnwidth]{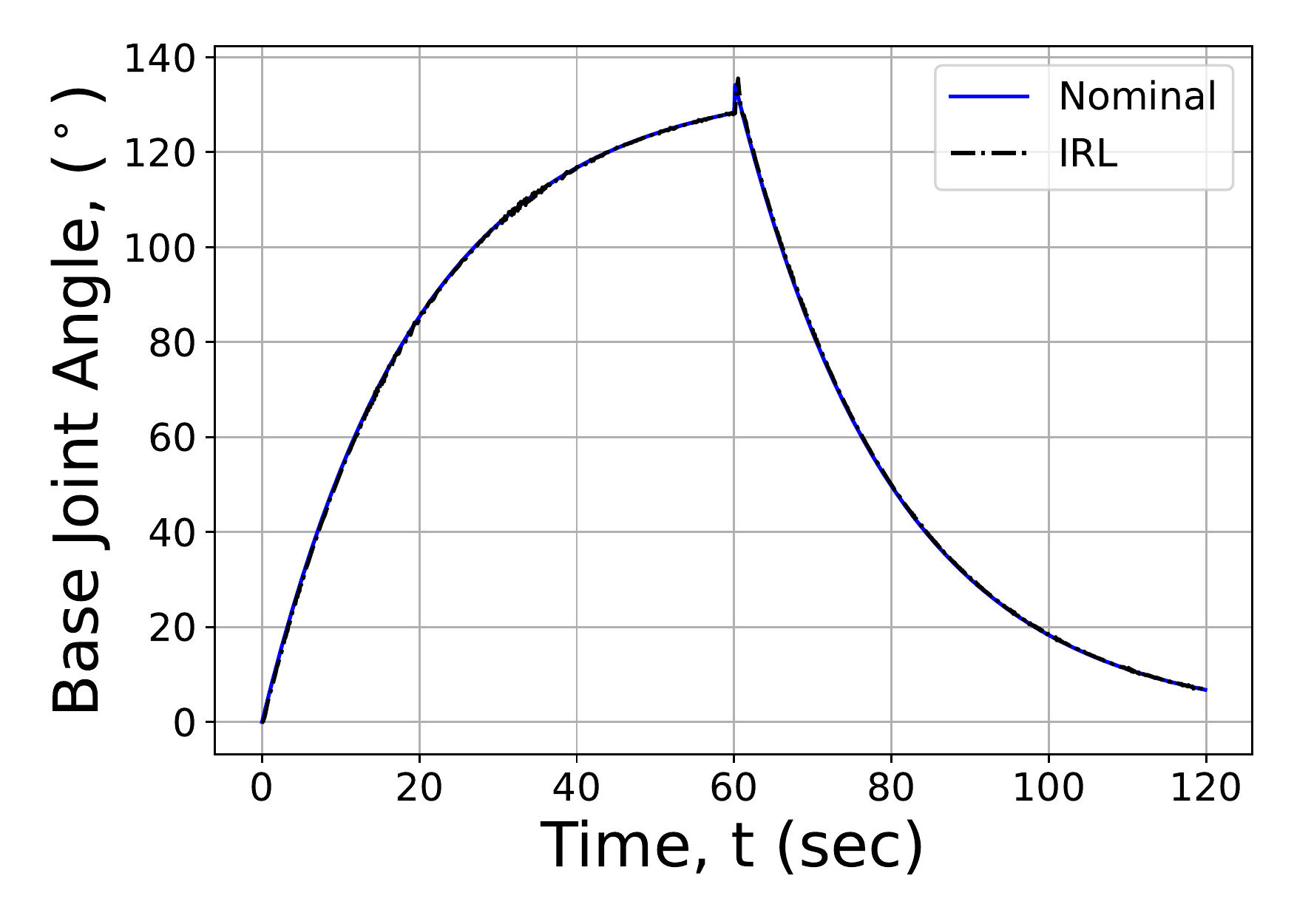}%
  } 
  \hfill%
  \subcaptionbox{Joint~2%
   \label{fig:actual_trajectories:fullj2_noise}}%
  {%
    \includegraphics[width=0.49\columnwidth]{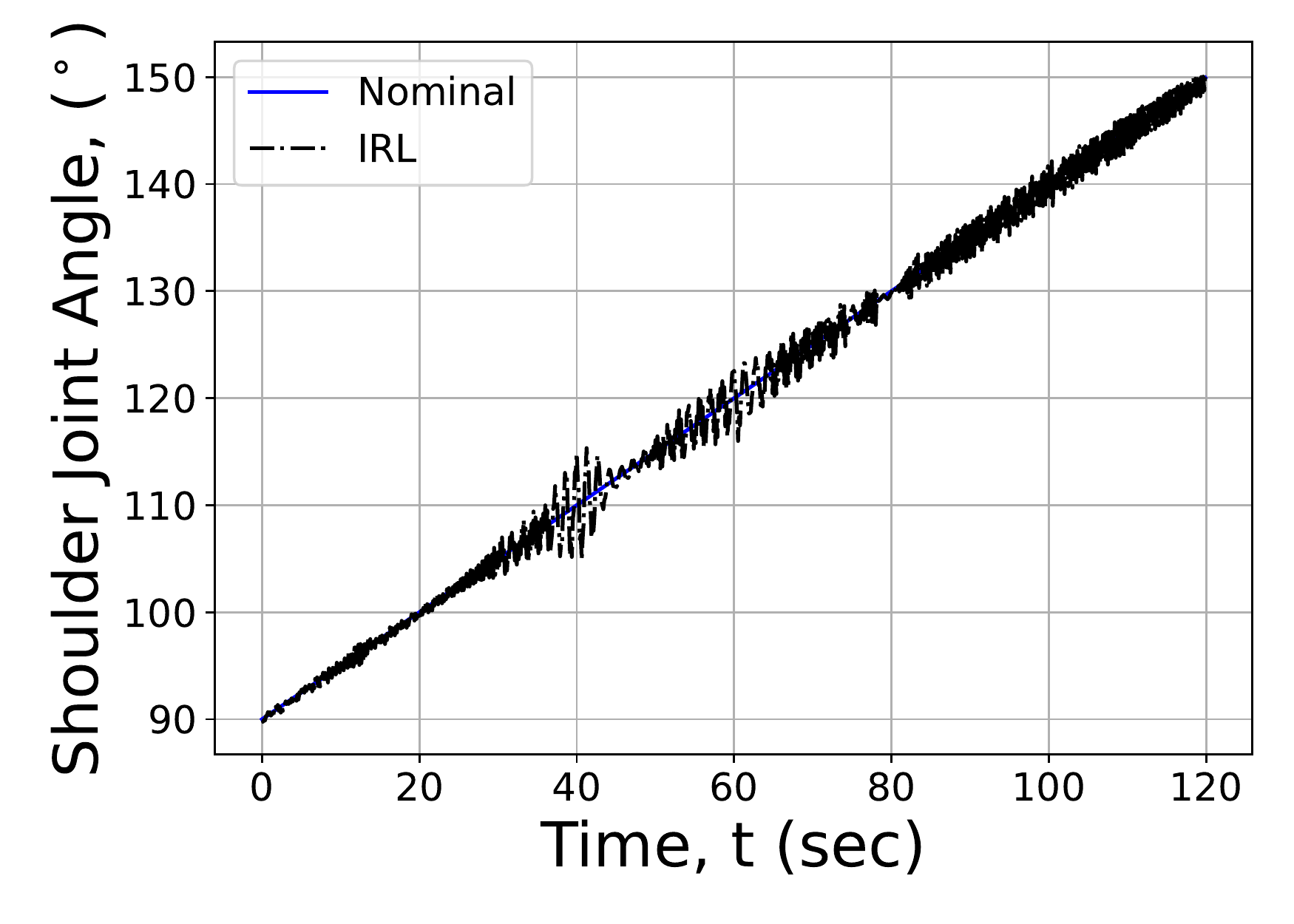}%
  }
  \\[2ex]
  \mbox{}\hfill
  \subcaptionbox{Joint~3%
    \label{fig:actual_trajectories:fullj3_noise}}%
  {%
    \includegraphics[width=0.49\columnwidth]{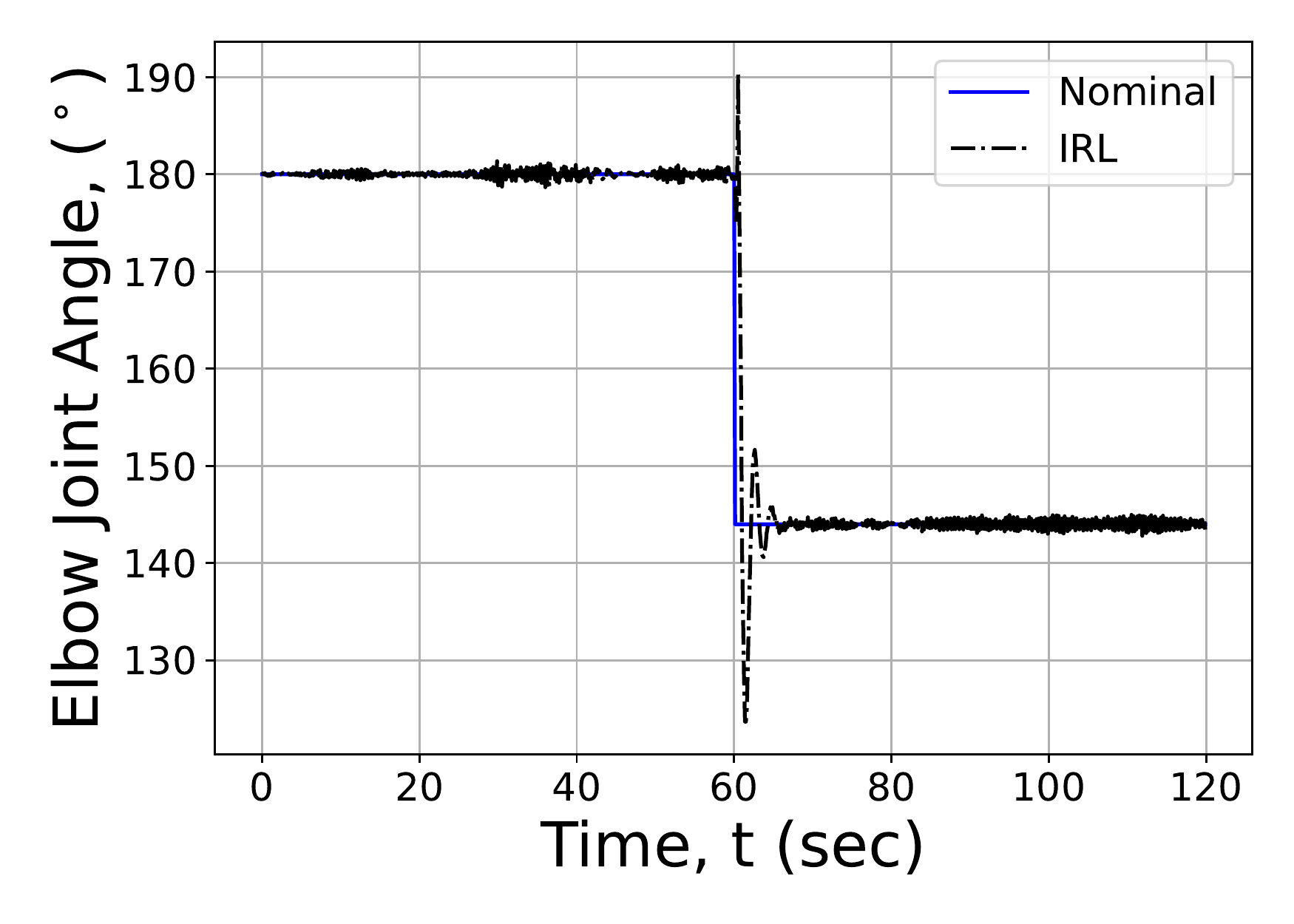}%
  }
     \hfill
  \subcaptionbox{Joint~4%
   \label{fig:actual_trajectories:j4_noise}}%
  {%
    \includegraphics[width=0.49\columnwidth]{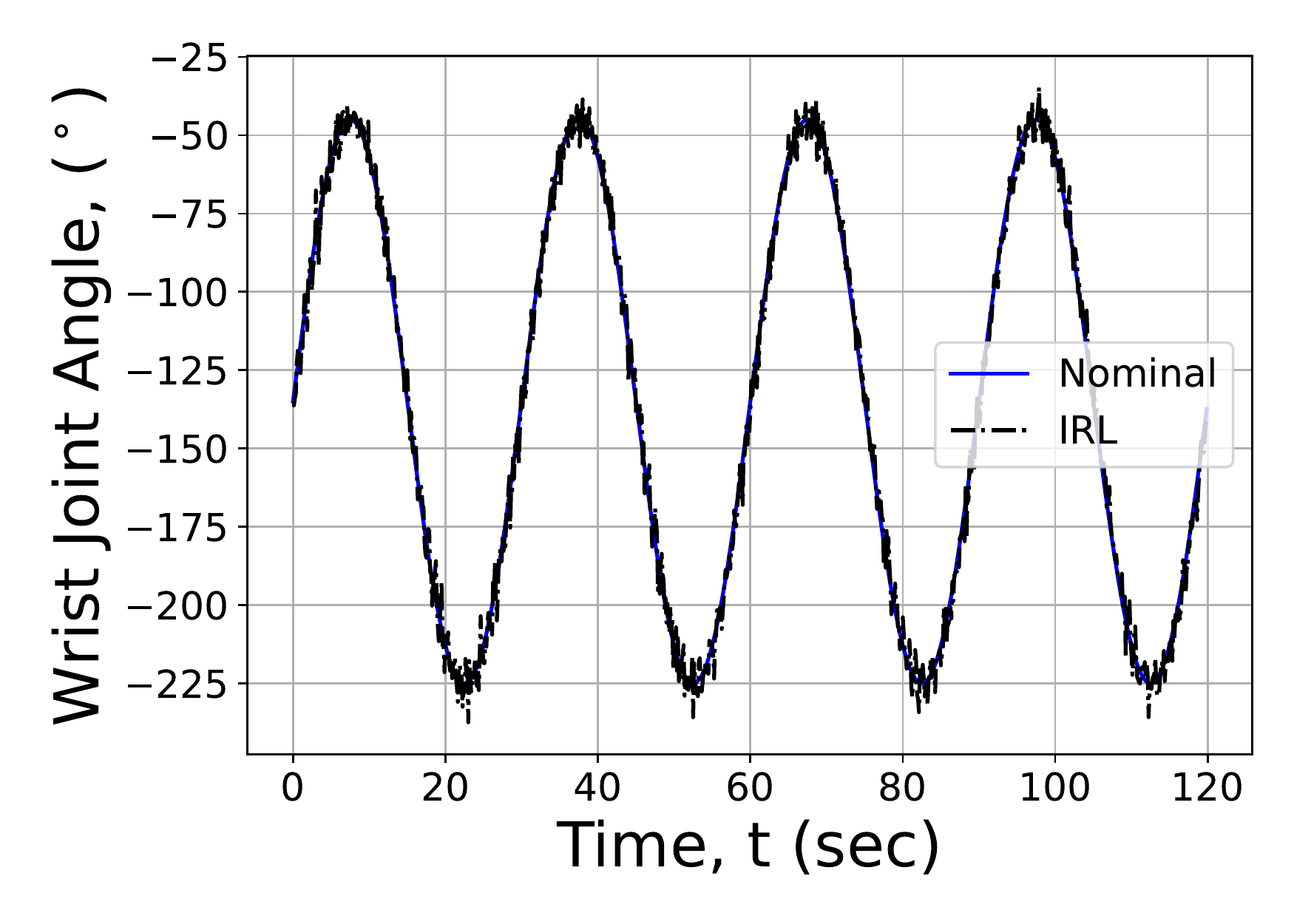}%
  }
  \\[2ex]
  \mbox{}\hfill
  \caption{Joint trajectories for experiment~4 with noise added to actor weights over the experiment.}
  \label{fig:actual_trajectories:full_noise}
\end{figure}

\begin{figure}[!htb]
  \centering
  \subcaptionbox{Joint~1--Actor~1, $\omega$%
   \label{fig:actor_adaptation:j1_a1_noise}}%
  {%
    \includegraphics[width=0.49\columnwidth]{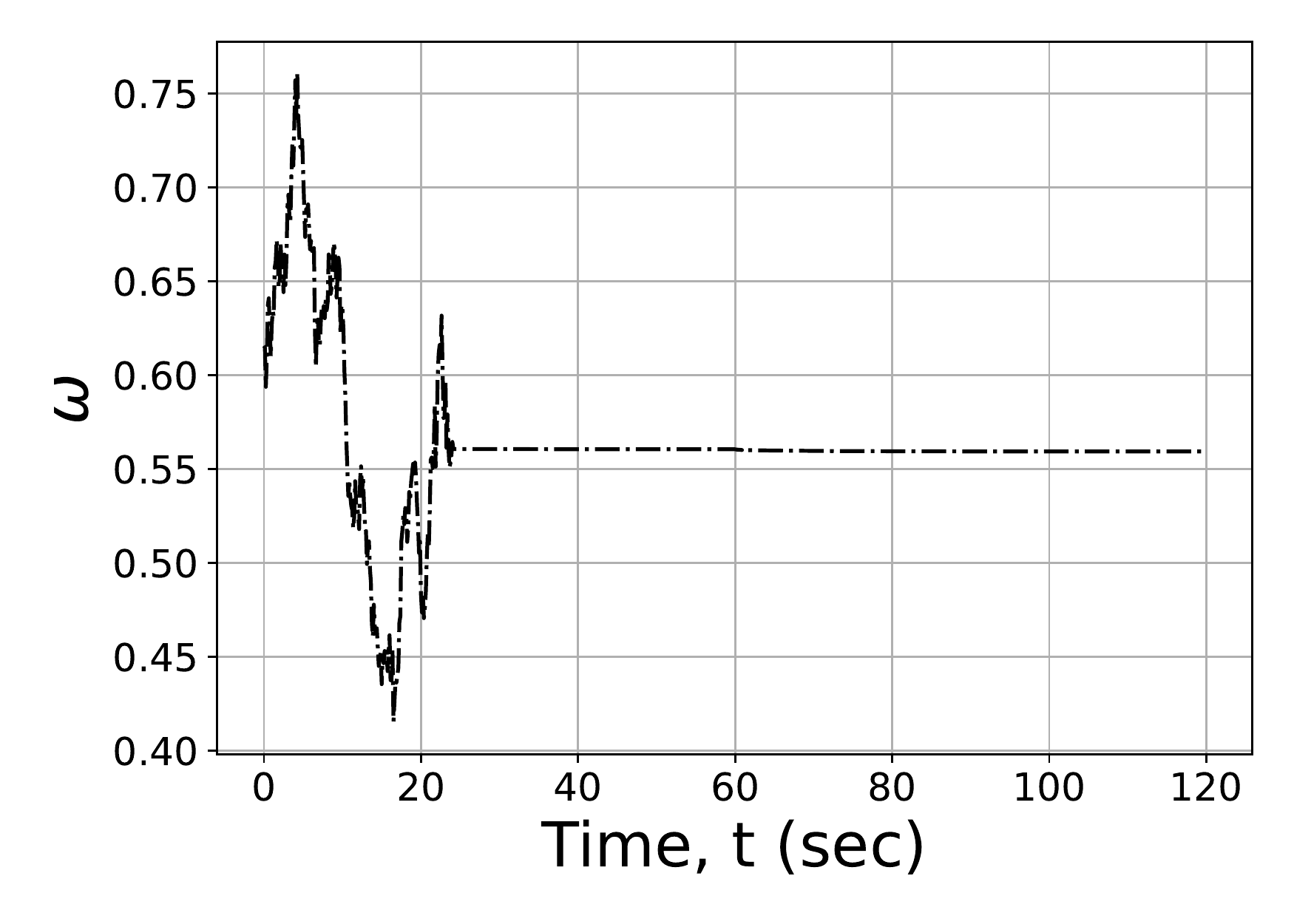}%
  } 
  \hfill%
  \subcaptionbox{Joint~1--Actor~2, $\omega_{\nu}$%
   \label{fig:actor_adaptation:j1_a2_noise}}%
  {%
    \includegraphics[width=0.49\columnwidth]{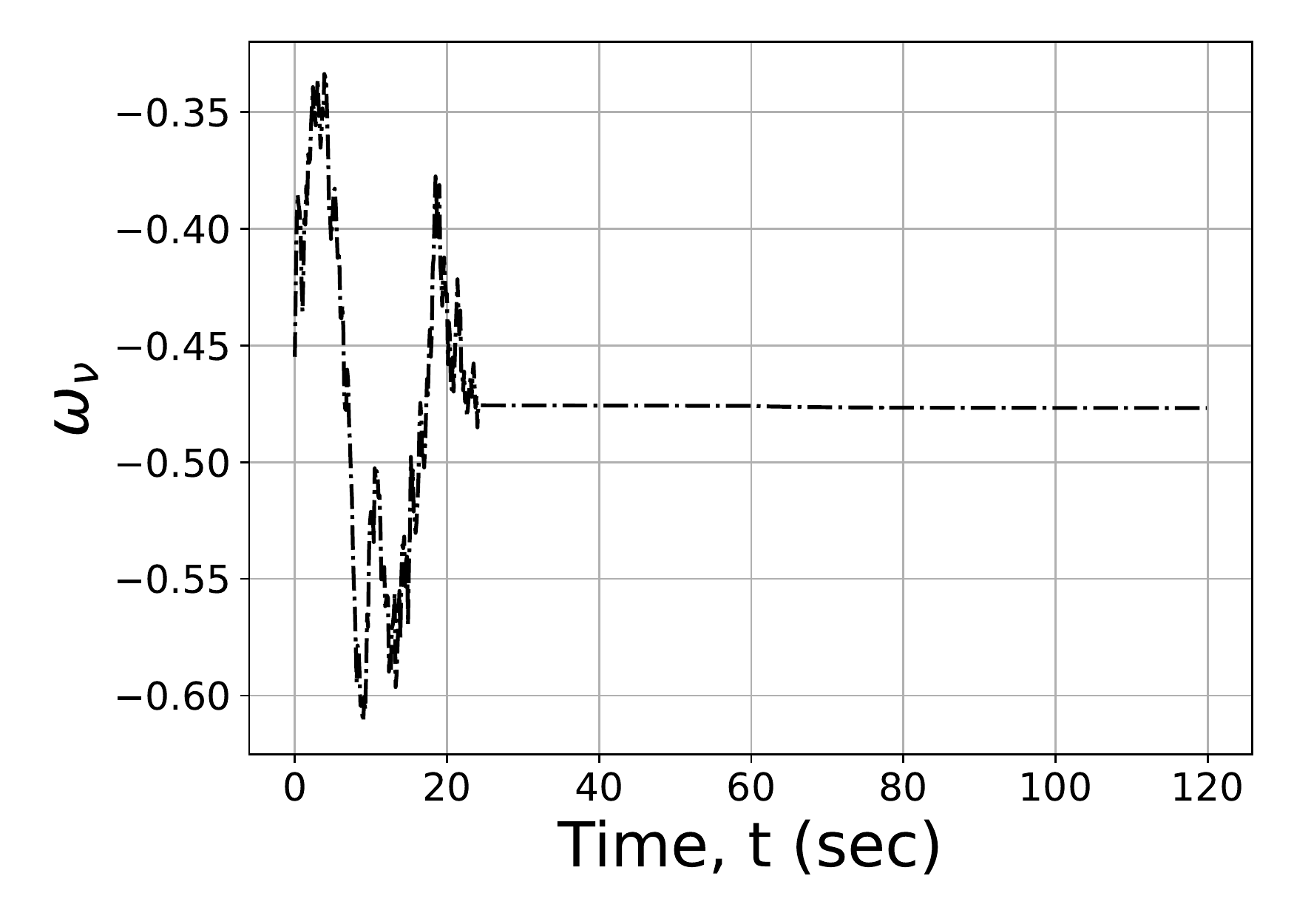}%
  }
  \\[2ex]
  \mbox{}\hfill
  \subcaptionbox{Joint~1--Actor~3, $\omega_{2\nu}$%
   \label{fig:actor_adaptation:j1_a3_noise}}%
  {%
    \includegraphics[width=0.49\columnwidth]{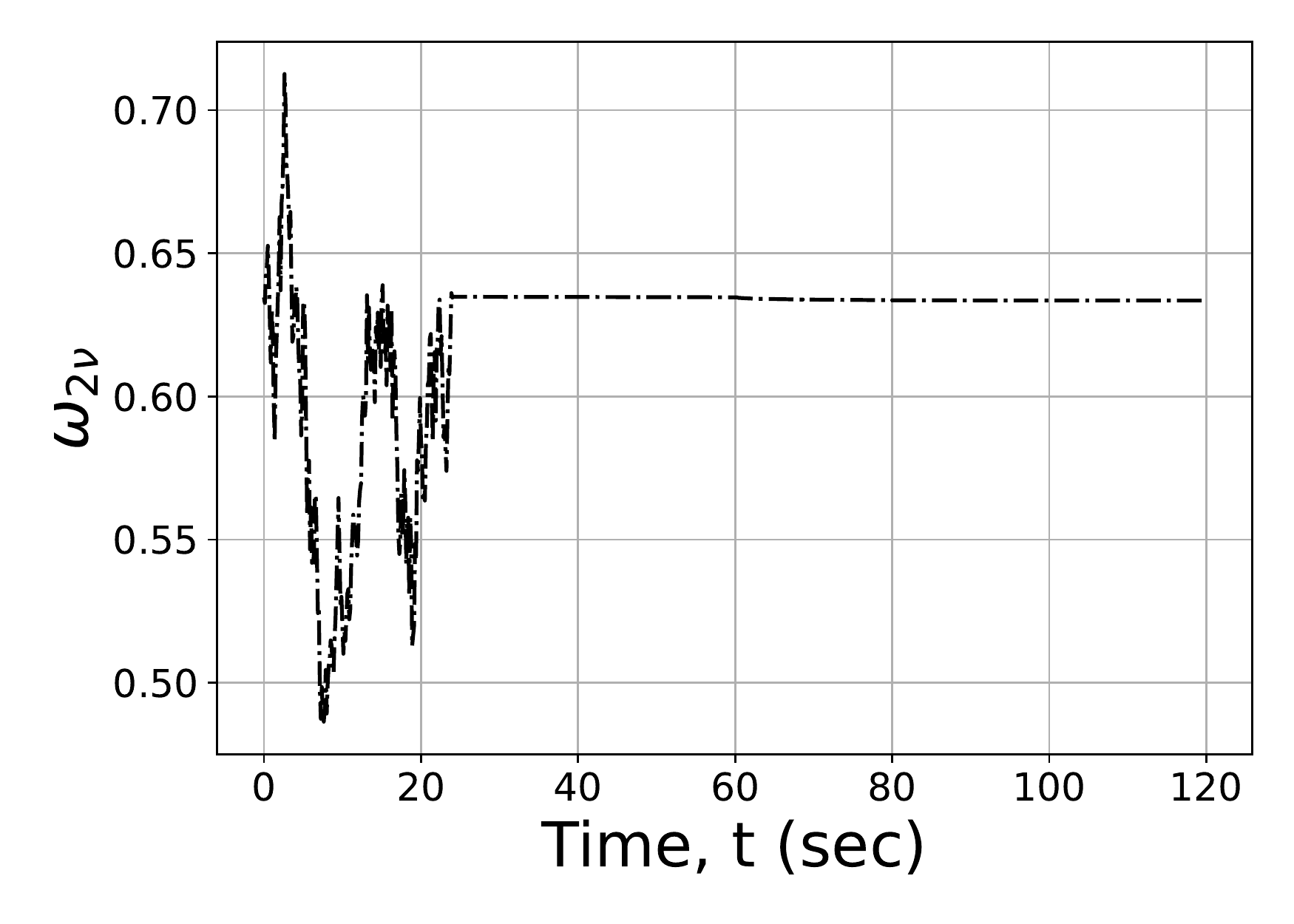}%
  }
     \hfill\mbox{}
  \caption{Adaptation of the base actor gains for experiment~4..}
  \label{fig:actor_adaptation_j1_noise}
\end{figure}

\begin{figure}[!htb]
  \centering
  \subcaptionbox{Joint~2--Actor~1, $\omega$%
   \label{fig:actor_adaptation:j2_a1_noise}}%
  {%
    \includegraphics[width=0.49\columnwidth]{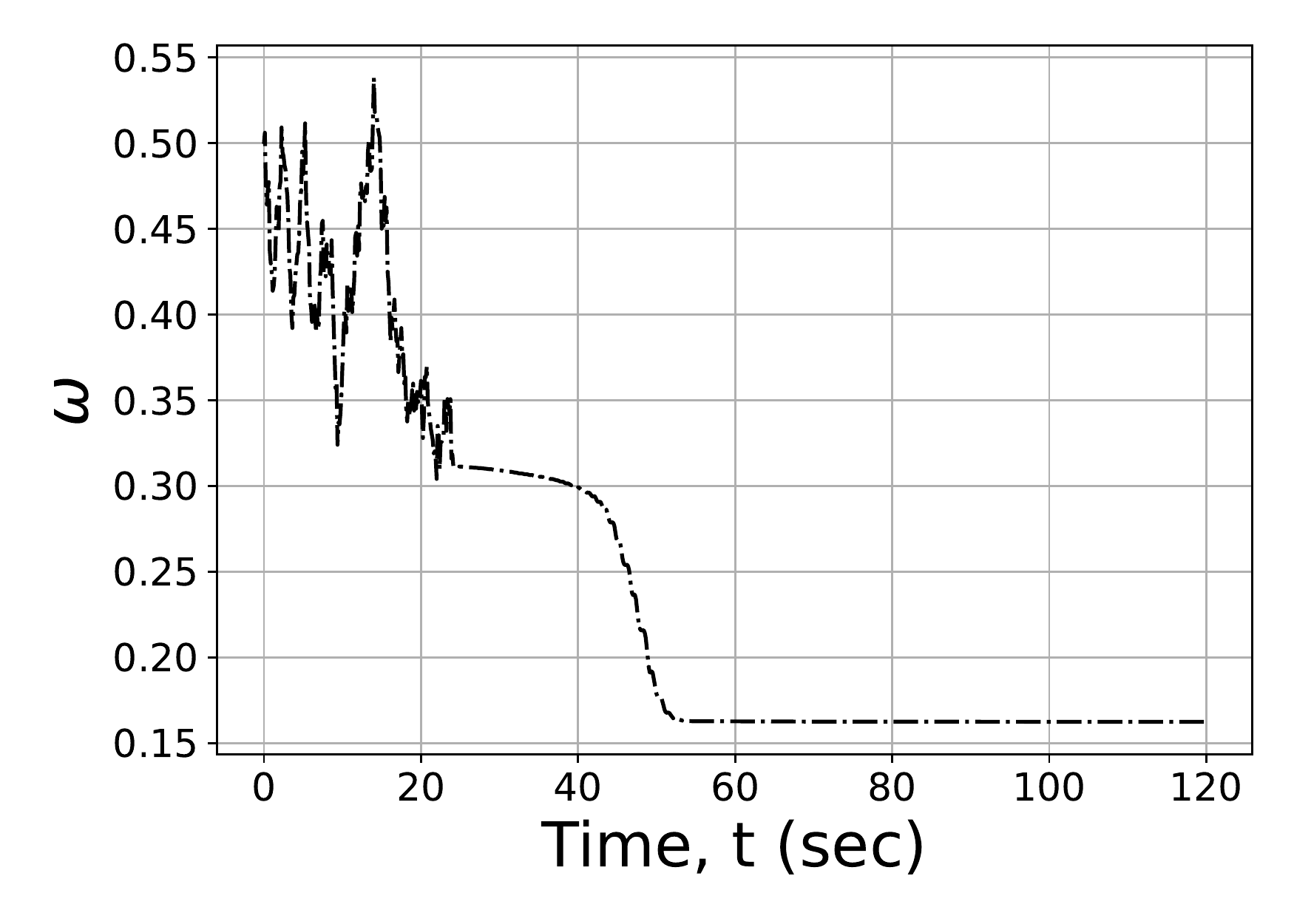}%
  } 
  \hfill%
  \subcaptionbox{Joint~2--Actor~2, $\omega_{\nu}$%
   \label{fig:actor_adaptation:j2_a2_noise}}%
  {%
    \includegraphics[width=0.49\columnwidth]{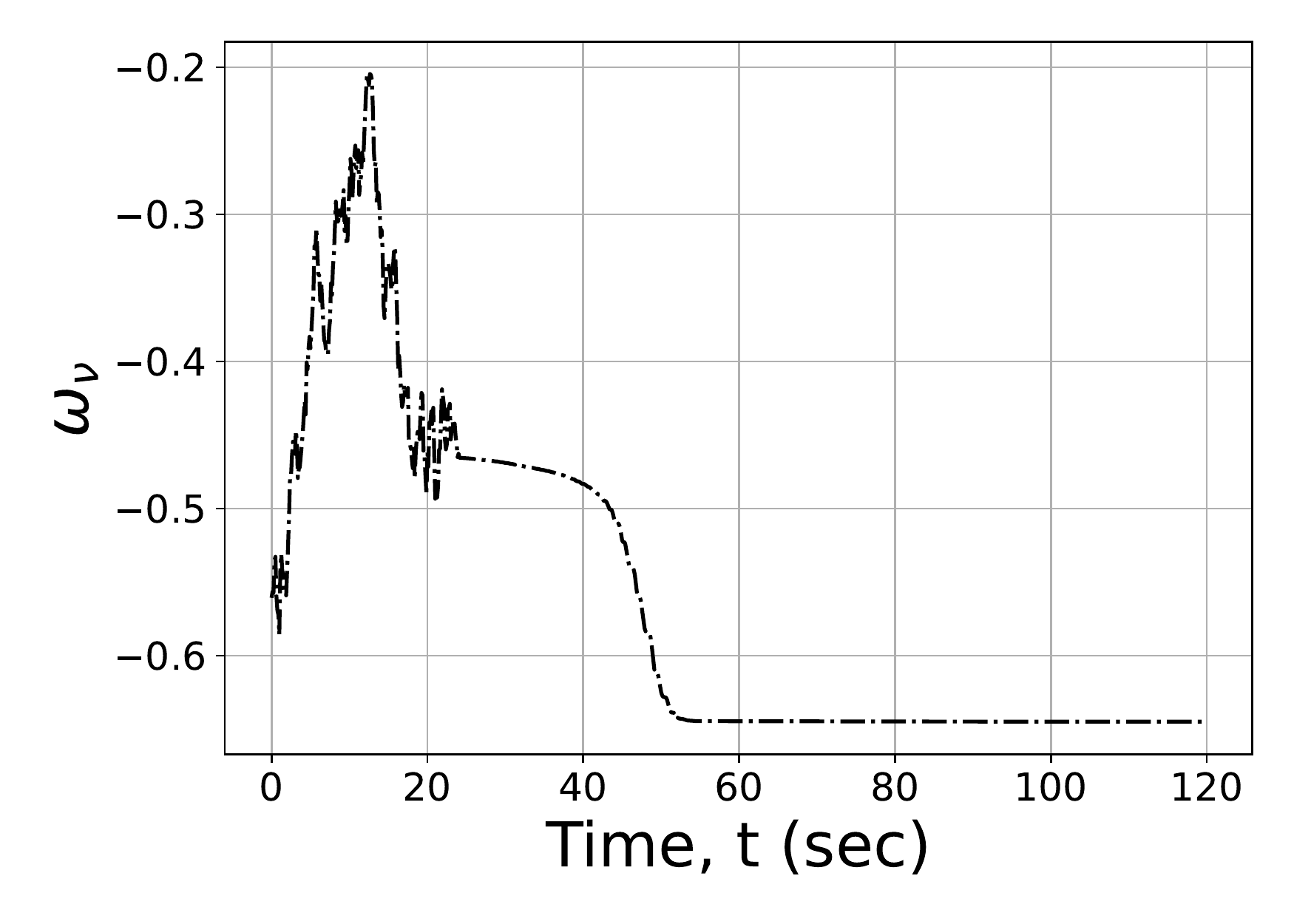}%
  }
  \\[2ex]
  \mbox{}\hfill
  \subcaptionbox{Joint~2--Actor~3, $\omega_{2\nu}$%
   \label{fig:actor_adaptation:j2_a3_noise}}%
  {%
    \includegraphics[width=0.49\columnwidth]{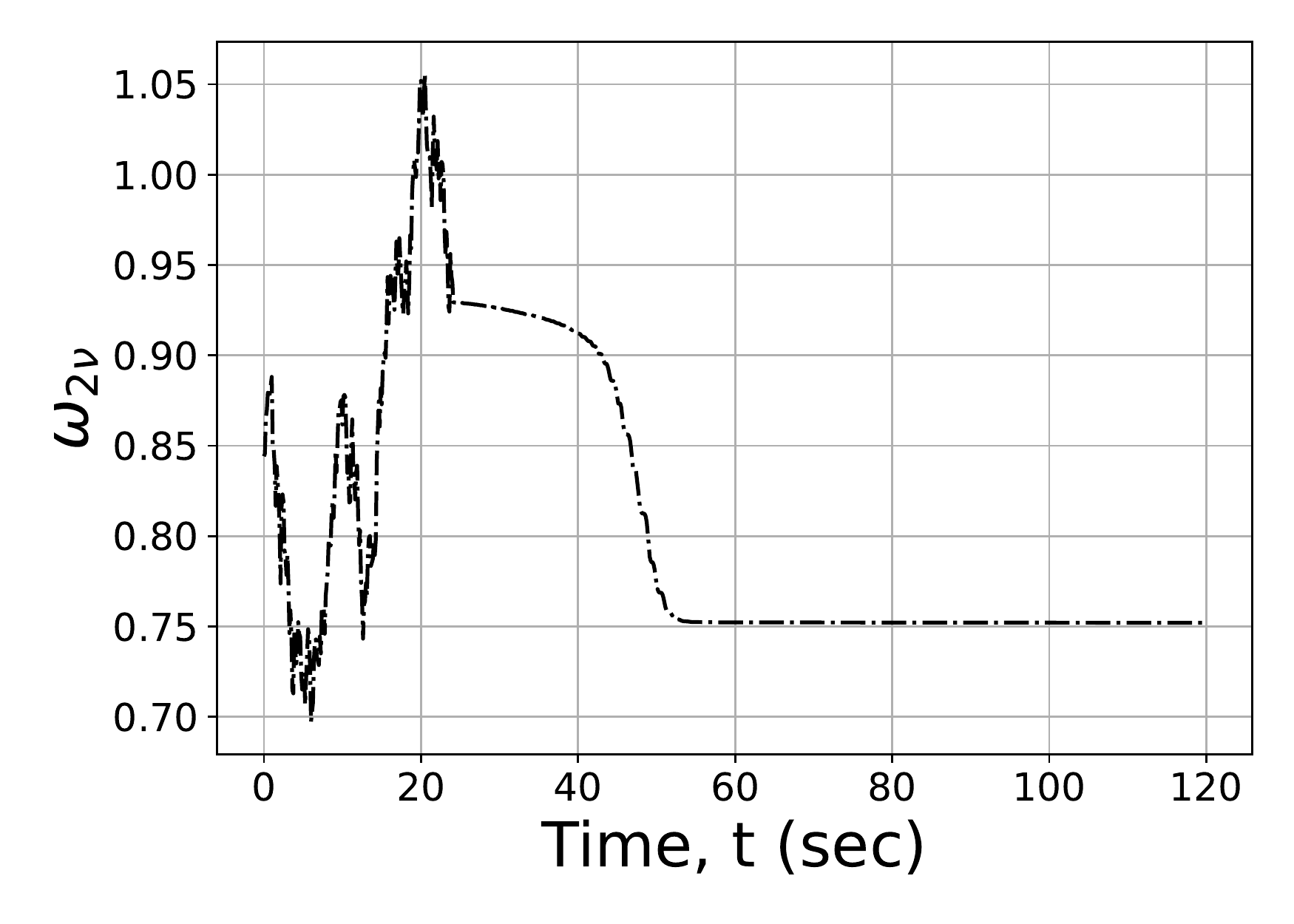}%
  }
     \hfill\mbox{}
  \caption{Adaptation of the shoulder actor gains for experiment~4.}
  \label{fig:actor_adaptation_j2_noise}
\end{figure}

\begin{figure}[!htb]
  \centering
  \subcaptionbox{Joint~4--Actor~1, $\omega$%
   \label{fig:actor_adaptation:j4_a1_noise}}%
  {%
    \includegraphics[width=0.49\columnwidth]{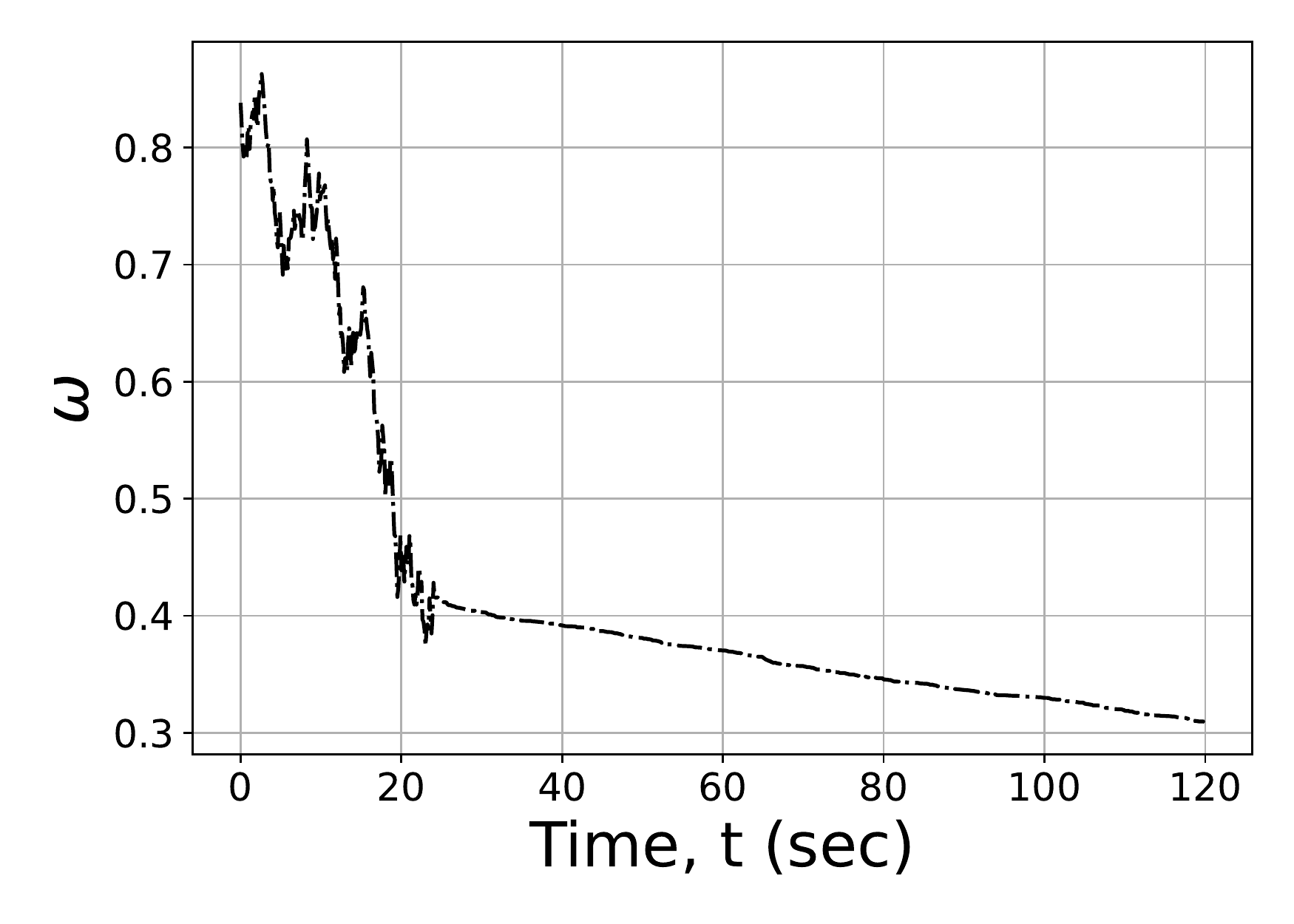}%
  } 
  \hfill%
  \subcaptionbox{Joint~4--Actor~2, $\omega_{\nu}$%
   \label{fig:actor_adaptation:j4_a2_noise}}%
  {%
    \includegraphics[width=0.49\columnwidth]{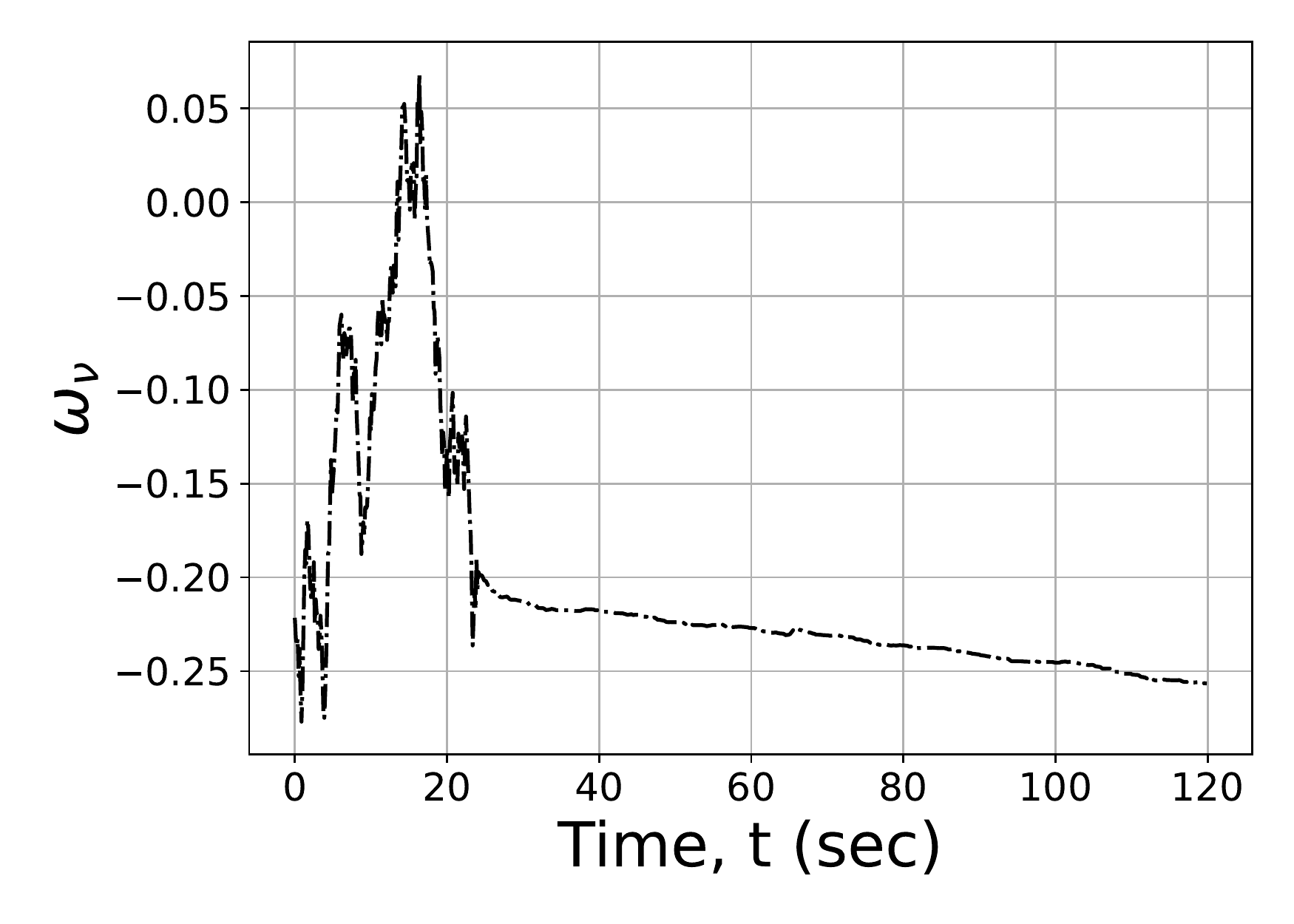}%
  }
  \\[2ex]
  \mbox{}\hfill
  \subcaptionbox{Joint~4--Actor~3, $\omega_{2\nu}$%
   \label{fig:actor_adaptation:j4_a3_noise}}%
  {%
    \includegraphics[width=0.49\columnwidth]{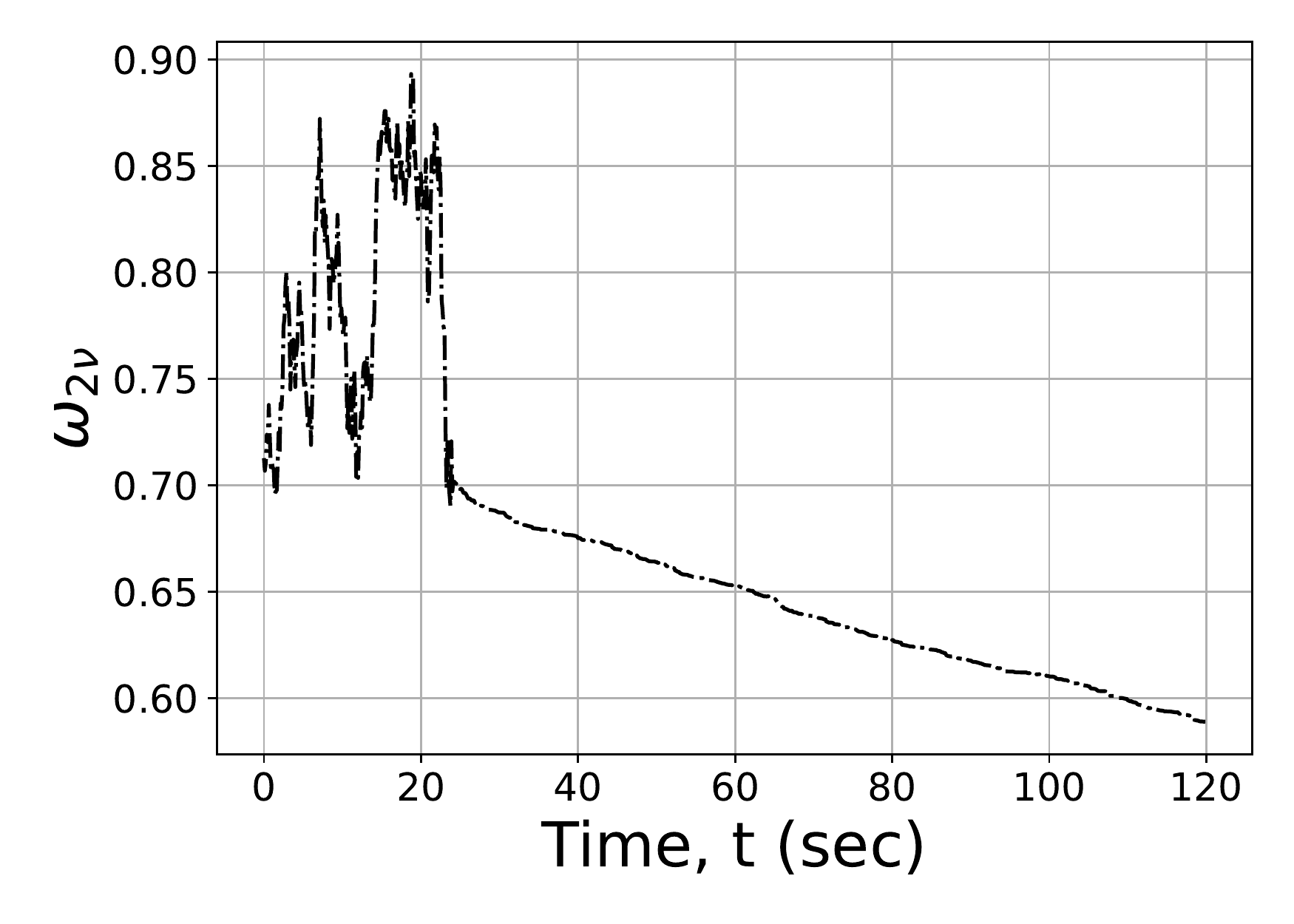}%
  }
     \hfill\mbox{}     
  \caption{Adaptation of the wrist actor gains for experiment~4.}
  \label{fig:actor_adaptation_j4_noise}
\end{figure}

\clearpage
\newpage
\bibliographystyle{elsarticle-num}
\bibliography{paper}

\end{document}